\documentclass[twocolumn]{aastex61}
\usepackage{comment}
\usepackage{amsmath,amssymb}
\usepackage{graphicx}
\usepackage{textcomp}
\usepackage{booktabs}
\usepackage{bm}

\newcommand\aastex{AAS\TeX}

\newcommand\ltsima{$\; \buildrel <\over\sim \;$}
\newcommand\simlt{\lower.5ex\hbox{\ltsima}}
\newcommand\gtsima{$\; \buildrel >\over\sim \;$}
\newcommand\simgt{\lower.5ex\hbox{\gtsima}}

\shorttitle{\aastex\ Terrestrial mass and Neptune mass free-floating planets}
\shortauthors{Koshimoto et al.}

\defcitealias{sumi2023}{S23}
\begin{document}

\title{Terrestrial and Neptune mass Free-Floating Planet candidates
from the MOA-II 9-year Galactic Bulge survey}

\correspondingauthor{Naoki Koshimoto}
\email{koshimoto.work@gmail.com}
\author[0000-0003-2302-9562]{Naoki koshimoto}
\affiliation{Code 667, NASA Goddard Space Flight Center, Greenbelt, MD 20771, USA}
\affiliation{Department of Astronomy, University of Maryland, College Park, MD 20742, USA}
\affiliation{Center for Research and Exploration in Space Science and Technology, NASA/GSFC, Greenbelt, MD 20771, USA}
\affiliation{Department of Earth and Space Science, Graduate School of Science, Osaka University, Toyonaka, Osaka 560-0043, Japan}
\author[0000-0002-4035-5012]{Takahiro Sumi}
\affiliation{Department of Earth and Space Science, Graduate School of Science, Osaka University, Toyonaka, Osaka 560-0043, Japan}
\author{David P.~Bennett}
\affiliation{Code 667, NASA Goddard Space Flight Center, Greenbelt, MD 20771, USA}
\affiliation{Department of Astronomy, University of Maryland, College Park, MD 20742, USA}
\affiliation{Center for Research and Exploration in Space Science and Technology, NASA/GSFC, Greenbelt, MD 20771, USA}
\author{Valerio Bozza}
\affiliation{Dipartimento di Fisica `E.R. Caianiello', Universit\'a di Salerno, Via Giovanni Paolo 132, Fisciano I-84084, Italy}
\affiliation{Istituto Nazionale di Fisica Nucleare, Sezione di Napoli, Via Cintia, Napoli I-80126, Italy}
\author[0000-0001-7016-1692]{Przemek Mr{\'o}z}
\affiliation{Astronomical Observatory, University of Warsaw Al. Ujazdowskie 4, 00-478 Warszawa, Poland}
\author[0000-0001-5207-5619]{Andrzej Udalski}
\affiliation{Astronomical Observatory, University of Warsaw Al. Ujazdowskie 4, 00-478 Warszawa, Poland}
\author{Nicholas J. Rattenbury}
\affiliation{Department of Physics, University of Auckland, Private Bag 92019, Auckland, New Zealand}

\nocollaboration

\author{Fumio Abe}
\affiliation{Institute for Space-Earth Environmental Research, Nagoya University, Nagoya 464-8601, Japan}
\author{Richard Barry}
\affiliation{Code 667, NASA Goddard Space Flight Center, Greenbelt, MD 20771, USA}
\author{Aparna Bhattacharya}
\affiliation{Code 667, NASA Goddard Space Flight Center, Greenbelt, MD 20771, USA}
\affiliation{Department of Astronomy, University of Maryland, College Park, MD 20742, USA}
\affiliation{Center for Research and Exploration in Space Science and Technology, NASA/GSFC, Greenbelt, MD 20771, USA}
\author{Ian A. Bond}
\affiliation{Institute of Natural and Mathematical Sciences, Massey University, Auckland 0745, New Zealand}
\author{Hirosane Fujii}
\affiliation{Institute for Space-Earth Environmental Research, Nagoya University, Nagoya 464-8601, Japan}
\author{Akihiko Fukui}
\affiliation{Department of Earth and Planetary Science, Graduate School of Science, The University of Tokyo, 7-3-1 Hongo, Bunkyo-ku, Tokyo 113-0033, Japan}
\affiliation{Instituto de Astrof\'isica de Canarias, V\'ia L\'actea s/n, E-38205 La Laguna, Tenerife, Spain}
\author{Ryusei Hamada}
\affiliation{Department of Earth and Space Science, Graduate School of Science, Osaka University, Toyonaka, Osaka 560-0043, Japan}
\author{Yuki Hirao}
\affiliation{Department of Earth and Space Science, Graduate School of Science, Osaka University, Toyonaka, Osaka 560-0043, Japan}
\author{Stela Ishitani Silva}
\affiliation{Department of Physics, The Catholic University of America, Washington, DC 20064, USA}
\affiliation{Code 667, NASA Goddard Space Flight Center, Greenbelt, MD 20771, USA}
\affiliation{Center for Research and Exploration in Space Science and Technology, NASA/GSFC, Greenbelt, MD 20771, USA}
\author{Yoshitaka Itow}
\affiliation{Institute for Space-Earth Environmental Research, Nagoya University, Nagoya 464-8601, Japan}
\author{Rintaro Kirikawa}
\affiliation{Department of Earth and Space Science, Graduate School of Science, Osaka University, Toyonaka, Osaka 560-0043, Japan}
\author{Iona Kondo}
\affiliation{Department of Earth and Space Science, Graduate School of Science, Osaka University, Toyonaka, Osaka 560-0043, Japan}
\author{Yutaka Matsubara}
\affiliation{Institute for Space-Earth Environmental Research, Nagoya University, Nagoya 464-8601, Japan}
\author{Shota Miyazaki}
\affiliation{Department of Earth and Space Science, Graduate School of Science, Osaka University, Toyonaka, Osaka 560-0043, Japan}
\author{Yasushi Muraki}
\affiliation{Institute for Space-Earth Environmental Research, Nagoya University, Nagoya 464-8601, Japan}
\author{Greg Olmschenk}
\affiliation{Code 667, NASA Goddard Space Flight Center, Greenbelt, MD 20771, USA}
\author{Cl\'ement Ranc}
\affiliation{Sorbonne Universit\'e, CNRS, UMR 7095, Institut d'Astrophysique de Paris, 98 bis bd Arago, 75014 Paris, France}
\author{Yuki Satoh}
\affiliation{Department of Earth and Space Science, Graduate School of Science, Osaka University, Toyonaka, Osaka 560-0043, Japan}
\author{Daisuke Suzuki}
\affiliation{Department of Earth and Space Science, Graduate School of Science, Osaka University, Toyonaka, Osaka 560-0043, Japan}
\author{Mio Tomoyoshi}
\affiliation{Department of Earth and Space Science, Graduate School of Science, Osaka University, Toyonaka, Osaka 560-0043, Japan}
\author{Paul J. Tristram}
\affiliation{University of Canterbury Mt.\ John Observatory, P.O. Box 56, Lake Tekapo 8770, New Zealand}
\author{Aikaterini Vandorou}
\affiliation{Code 667, NASA Goddard Space Flight Center, Greenbelt, MD 20771, USA}
\affiliation{Department of Astronomy, University of Maryland, College Park, MD 20742, USA}
\affiliation{Center for Research and Exploration in Space Science and Technology, NASA/GSFC, Greenbelt, MD 20771, USA}
\author{Hibiki Yama}
\affiliation{Department of Earth and Space Science, Graduate School of Science, Osaka University, Toyonaka, Osaka 560-0043, Japan}
\author{Kansuke Yamashita}
\affiliation{Department of Earth and Space Science, Graduate School of Science, Osaka University, Toyonaka, Osaka 560-0043, Japan}

\collaboration{(MOA collaboration)}



\begin{abstract}
We report the discoveries of low-mass free-floating planet (FFP) candidates from the analysis of 2006-2014 MOA-II 
Galactic bulge survey data. 
In this dataset, we found 6,111 microlensing candidates and identified a statistical sample consisting of 3,535 high quality single lens events with Einstein radius crossing times in the range $0.057 < t_{\rm E}/{\rm days} < 757$, including 13 events that show clear finite source effects with angular Einstein radii of $0.90<\theta_{\rm E}/{\rm \mu as} <332.54$.
Two of the 12 events with $t_{\rm E} < 1\,$day have significant finite source effects, and one event, 
MOA-9y-5919, with $t_{\rm E}=0.057\pm 0.016$ days and $\theta_{\rm E}= 0.90 \pm 0.14$\,$\mu$as, is the second terrestrial mass FFP candidate to date. 
A Bayesian analysis indicates a lens mass of $0.75^{+1.23}_{-0.46}$ $M_\earth$ for this event.
The low detection efficiency for short duration events implies a large population of low-mass FFPs. 
The microlensing detection efficiency for low-mass planet events depends on both the Einstein radius crossing times and the angular Einstein radii, so we have used image-level simulations to determine the detection efficiency dependence
on both $t_{\rm E}$ and $\theta_{\rm E}$. This allows us to use a Galactic model to simulate the $t_{\rm E}$ and $\theta_{\rm E}$ distribution of events produced
by the known stellar populations and models of the FFP distribution that are fit to the data.
Methods like this will be needed for the more precise FFP demographics determinations from 
Nancy Grace Roman Space Telescope data.

\end{abstract}


\section{Introduction} 
\label{sec:intro}

Gravitational microlensing enables us to study a variety of objects \citep{pac86} with masses ranging from  that of
exoplanets \citep{mao1991, sumi2011, suzuki2016, Mroz17} to black holes \citep{sah22, lam22, mro22}.
This is because the Einstein radius crossing time (or Einstein timescale) $t_{\rm E}$, the only quantity that can be measured in all events, takes measurable values ranging from minutes to years for lens masses $M$ of exoplanets to black holes:
\begin{align}
  t_{\rm E} &= \frac{\theta_{\rm E}}{\mu_{\rm rel}} = \frac{\sqrt{\kappa M \pi_{\rm rel} }}{\mu_{\rm rel}} \notag\\
  &\simeq 28\,{\rm days} \left( \frac{M}{1 \, M_\sun} \right)^{1/2} \left( \frac{\pi_{\rm rel}}{18 \, {\rm \mu as}} \right)^{1/2} \left( \frac{\mu_{\rm rel}}{5 \, {\rm mas/yr}} \right)^{-1} ,
  \label{eq:te}
\end{align}
where $\mu_{\rm rel}$ is the lens-source relative proper motion, $\theta_{\rm E}$ is the angular Einstein radius given by $\theta_{\rm E} = \sqrt{\kappa M \pi_{\rm rel}}$,
$\kappa$ is a constant given by $\kappa =4G/(c^2 {\rm au})= 8.144 {\rm mas}/M_\odot$, and $\pi_{\rm rel}$ is the lens-source relative parallax given by
$\pi_{\rm rel}=\pi_{\rm l}^{-1}- \pi_{\rm s}^{-1}=$ au $(D_{\rm l}^{-1}-D_{\rm s}^{-1})$
with the observer-lens distance $D_{\rm l}$ and the observer-source distance $D_{\rm s}$. 
Because microlensing is observed as a time variation of the light of a magnified background source star, $t_{\rm E}$ is measurable even if the lens object is dark.

Currently, three survey groups, the Microlensing Observations in Astrophysics collaboration \citep[MOA,][]{bon01, sumi03},
the Optical Gravitational Lensing Experiment \citep[OGLE,][]{uda94, uda15},
and the Korea Microlensing Telescope Network \citep[KMTNet,][]{kmtnet2010, kim16},
are conducting wide-field high-cadence surveys toward the Galactic bulge.
Because the lens mass $M$ is given by
 \begin{equation}
  \label{eq:M}
M=\frac{t_{\rm E}^2\mu_{\rm rel}^2}{\kappa \pi_{\rm rel}}=
5M_\earth  \left(  \frac{t_{\rm E}}{0.1{\rm d}} \right)^2
 \left(  \frac{\pi_{\rm rel}}{18\mu \rm as} \right)^{-1}
 \left(  \frac{\mu_{\rm rel}}{5 \rm mas/yr} \right)^2,
\end{equation}
survey groups that observe their target fields with up to 10-15 minutes cadence are sensitive to free-floating planets (FFPs) even with terrestrial masses.
However, $\pi_{\rm rel}$ and $\mu_{\rm rel}$ in Eq. (\ref{eq:M}) are highly uncertain.
Therefore, even though the mass can be estimated by Bayesian analysis with prior stellar density and velocity distributions of our Galaxy, the uncertainty of the mass estimate is also large.

In the cases where the projected lens trajectory passes close to the source star disk, we can measure the angular Einstein radius $\theta_{\rm E}$ in addition to $t_{\rm E}$ by utilizing the finite source effect, in which
the angular size of the source star affects the light curve.  Such events are called the FSPL (finite-source and point-lens) events.
With $\theta_{\rm E}$, the lens mass is given by
 \begin{equation}
  \label{eq:M_thetaE_pirel}
M=\frac{\theta_{\rm E}^2}{\kappa \pi_{\rm rel}}  =
5 M_\earth  \left(  \frac{\theta_{\rm E}} {1.5\mu \rm as} \right)^2
\left(  \frac{\pi_{\rm rel}}{18\mu \rm as} \right)^{-1},
\end{equation}
and no longer depends on the lens-source relative proper motion $\mu_{\rm rel}$.
Although Eq. (\ref{eq:M_thetaE_pirel}) still has one uncertain parameter $\pi_{\rm rel}$, the angular Einstein radius $\theta_{\rm E}$ gives us an inferred mass of the lens with
much less uncertainty than that by $t_{\rm E}$ solely.

Seven short FSPL events have been discovered to date \citep{Mroz18,Mroz19b,Mroz20b,Mroz20c,Kim2021,Ryu2021}.
The measured angular Einstein radii are $\theta_{\rm E}<10\,\mu$as, which implies that the lenses are most likely to have planetary mass.
All the sources of these events are red giants. This is likely (partly intentional) selection bias because their angular radii, i.e., cross-section,
are significantly larger than the main sequence stars.

Of these seven, OGLE-2016-BLG-1928 is the shortest FSPL event with $t_{\rm E}=0.0288_{-0.0016} ^{+0.0024}$ days and also has the smallest angular Einstein radius $\theta_{\rm E}=  0.842 \pm 0.064 \, \mu$as \citep{Mroz20b}.
The lens, OGLE-2016-BLG-1928L, is currently the only terrestrial mass FFP candidate and the first evidence of such a population.

This paper presents the systematic analysis of the 9-year MOA-II survey toward the Galactic bulge in 2006 -- 2014, and reports discoveries of a terrestrial mass ($\theta_{\rm E} \sim 0.9 \, \mu$as) and
a Neptune mass ($\theta_{\rm E} \sim 5 \mu$as) pair of FFP candidates with $\theta_{\rm E}$ measurements.
The terrestrial mass FFP candidate, MOA-9y-5919, could have the second smallest angular Einstein radius measured so far. 
Our analysis is an extended study of \citet{sumi2011} who analyzed the MOA-II data in 2006 -- 2007, and first suggested the existence of an FFP population.
Our analysis also includes the data observed in 2006 -- 2007, but differs from \citet{sumi2011} in that we have removed systematic trends that were found in the baseline correlated with the seeing and airmass \citep{bennett2012}.
There is a companion paper \citep[][hereafter \citetalias{sumi2023}]{sumi2023} that presents a statistical analysis of this sample and derives the mass function for the FFP population.
We present the calculation of the detection efficiency using a new method that takes into account the finite source effect. This method is important for a statistical analysis of short timescale events in which the finite source effect affects the detection efficiency, such as the measurement of the FFP mass function presented in \citetalias{sumi2023}.

This paper is organized as follows.
We describe our observations in section \S\,\ref{sec:Observation}.
The data analysis is presented in section \S\,\ref{sec:analysis}.
Section \S\,\ref{sec:select} describes the selection of microlensing events. Section \S\,\ref{sec:shortevents} describes short timescale events discovered with $t_{\rm E} < 1~{\rm day}$ and refines fits for them.
We analyze FSPL events in the sample including two FFP candidates in section \S\,\ref{sec:FSPL}.
We present our detection efficiency calculation that takes into account the finite source effect in section \S\,\ref{sec:efficiency}. In section \S\, \ref{sec:eff_thetaE}, we calculate the detection efficiency for FSPL events.
The discussion and conclusions are presented in section \S\,\ref{sec:discussionAndSummary}.

%



\begin{deluxetable*}{lrrrrrrrrrrr}
\tabletypesize{\scriptsize}
\tablecaption{MOA-II Galactic bulge fields with the central coordinates, cadences in 2006-07 and 2008-14,
the number of observed frames  ($N_{\rm f}$) and the used frames ($N_{\rm f,use}$), 
the number of all microlensing events ($N_{\rm ev}$) and ones selected by Criteria CR1 ($N_{\rm ev,CR1}$) and CR2 ($N_{\rm ev,CR2}$) .
 \label{tbl:fld}}
\tablewidth{0pt}
\tablehead{
    \colhead{Field} & \colhead{$\rm R.A.$} & \colhead{$\rm Dec.$} & \colhead{$l$} & \colhead{$b$} & 
    \multicolumn{2}{c}{Cadence} &
    \colhead{$N_{\rm f}$\tablenotemark{a}}  & \colhead{$N_{\rm f,use}$\tablenotemark{b}}  & \colhead{$N_{\rm ev}$\tablenotemark{c}} & \colhead{$N_{\rm ev,CR1}$} & \colhead{$N_{\rm ev,CR2}$} \\
        \colhead{} & \colhead{} & \colhead{} & \colhead{} & \colhead{} & 
    \colhead{06-07}  & \colhead{08-14} &
    \colhead{}  & \colhead{}  & \colhead{} & \colhead{} \\
    \colhead{} & \colhead{(2000)} & \colhead{(2000)} & \colhead{($^\circ$)}  & \colhead{($^\circ$)} 
     & \colhead{$\rm(min.)$}   & \colhead{($\rm min.$)}  & \colhead{} & \colhead{} 
}
\startdata
 gb1 & 17:47:31.41 & -34:14:31.09 & -4.3284 & -3.0982 &   50 &   47 &   11065 &   10488 &     316 &     194 &     193\\
 gb2 & 17:54:01.41 & -34:29:31.09 & -3.8600 & -4.3800 &   50 &   47 &   10968 &   10367 &     221 &     134 &     134\\
 gb3 & 17:54:01.41 & -32:44:31.09 & -2.3440 & -3.4997 &   50 &   15 &   21741 &   20662 &     348 &     197 &     196\\
 gb4 & 17:54:01.41 & -30:59:31.09 & -0.8308 & -2.6169 &   50 &   15 &   23034 &   22061 &     597 &     309 &     308\\
 gb5 & 17:54:01.41 & -29:14:31.09 &  0.6803 & -1.7323 &   10 &   15 &   33263 &   31659 &    1029 &     496 &     493\\
 gb6 & 17:54:01.41 & -27:29:31.09 &  2.1900 & -0.8465 &   50 & 1day &    3776 &    3561 &      63 &       0 &       0\\
 gb7 & 18:00:01.41 & -32:44:31.09 & -1.7141 & -4.5938 &   50 &   93 &    7116 &    6761 &     165 &      99 &      99\\
 gb8 & 18:00:01.41 & -30:59:31.09 & -0.1875 & -3.7309 &   50 &   47 &   11047 &   10570 &     296 &     186 &     186\\
 gb9 & 18:00:01.41 & -29:14:31.09 &  1.3360 & -2.8654 &   10 &   15 &   31006 &   29341 &     736 &     469 &     466\\
gb10 & 18:00:01.41 & -27:29:31.09 &  2.8572 & -1.9979 &   50 &   15 &   20667 &   19739 &     479 &     284 &     283\\
gb11 & 18:06:01.41 & -32:44:31.09 & -1.0979 & -5.6961 &   50 &   93 &    6160 &    5859 &      70 &      47 &      46\\
gb12 & 18:06:01.41 & -30:59:31.09 &  0.4422 & -4.8530 &   50 &   93 &    6597 &    6250 &     132 &      86 &      86\\
gb13 & 18:06:01.41 & -29:14:31.09 &  1.9784 & -4.0064 &   50 &   47 &    9388 &    8813 &     281 &     188 &     188\\
gb14 & 18:06:01.41 & -27:29:31.09 &  3.5114 & -3.1569 &   50 &   15 &   19977 &   18905 &     422 &     258 &     255\\
gb15 & 18:06:01.41 & -25:44:31.09 &  5.0419 & -2.3052 &   50 &   93 &    6935 &    6587 &     156 &      82 &      82\\
gb16 & 18:12:01.41 & -29:14:31.09 &  2.6079 & -5.1550 &   50 &   93 &    6140 &    5799 &     155 &     100 &      99\\
gb17 & 18:12:01.41 & -27:29:31.09 &  4.1530 & -4.3234 &   50 &   47 &    9075 &    8556 &     201 &     138 &     138\\
gb18 & 18:12:01.41 & -25:44:31.09 &  5.6946 & -3.4887 &   50 &   47 &    8695 &    8259 &     162 &     104 &     104\\
gb19 & 18:18:01.41 & -25:29:31.09 &  6.5571 & -4.5619 &   50 &   93 &    5275 &    5024 &     116 &      83 &      82\\
gb20 & 18:18:01.41 & -23:44:31.09 &  8.1062 & -3.7401 &   50 &   93 &    5114 &    4850 &      95 &      67 &      65\\
gb21 & 18:18:01.41 & -21:59:31.09 &  9.6523 & -2.9155 &   50 &   93 &    4960 &    4702 &      61 &      33 &      32\\
gb22 & 18:36:25.41 & -23:53:31.09 &  9.9063 & -7.5509 &   50 & 1day &    3391 &    3219 &      10 &       0 &       0\\
Total &          -- &          -- &      -- &      -- &   -- &   -- &  263491 &  248936 &    6111 &    3554 &    3535\\
\enddata

\tablenotetext{a}{Number of observed frames, i.e., exposures.}
\tablenotetext{b}{Maximum number of used frames among 10 chips.}
\tablenotetext{c}{Number of all microlensing event candidates including ones that did not pass either CR1 or CR2 criteria.}
\end{deluxetable*}

\section{Observations} 
\label{sec:Observation}
The data used in this analysis were taken during the 2006-2014 seasons of the MOA-II  
high cadence photometric survey toward the Galactic bulge.
MOA-II uses the 1.8-m MOA-II telescope located at the University of Canterbury's Mount\ John
Observatory in New Zealand. The telescope is equipped with a wide field camera, MOA-cam3 \citep{sako2008}, which consists of ten
2k $\times$ 4k pixel CCDs with $15\,\mu$m pixels. With the pixel scale
of 0.58 arcsec/pixel scale, this gives a
2.18 deg$^2$ field of view (FOV).
The median seeing for this data set is $2.0''$. 
The images were mainly taken through the custom MOA-Red wide-band filter, which
is equivalent to the sum of the standard Kron/Cousins $R$ and $I$-bands.
Although $V$-band observations are occasionally conducted, we do not include them in this analysis.

The central coordinates of the 22 fields of the MOA-II Galactic bulge survey and the cadences are listed in Table \ref{tbl:fld}.
In the 2006-2007 seasons, two fields, gb5 and gb9, were most densely sampled with a 10 minute cadence, 
and the 19 other fields were sampled with a 50 minute cadence.
In 2008-2014 seasons, six fields, gb5, gb9, gb10, gb4, gb3 and gb14,   
were densely sampled with a 15 minute cadence, 
six fields, gb1, gb2, gb8, gb13, gb17 and gb18,      
were sampled with a 47 minute cadence,
eight fields, gb7, gb11, gb12, gb15, gb16, gb19, gb20 and gb21    
were sampled with a 93 minute cadence,
and two fields, gb6 and gb22, 
were sampled with a 1 day cadence.

The number of frames, i.e., exposures in each field, $N_{\rm f}$ is given in  Table~\ref{tbl:fld}.
The number of frames actually used in the light curves differs chip by chip even in the same field because of 
CCD chip hardware failure, partial cloud and analysis failure owing to low signal to noise etc.
The maximum number of used frames among 10 chips in each field, $N_{\rm f, use}$, is also shown in Table~\ref{tbl:fld}.
The used dataset consists of 2,489,362 CCD images in total, which corresponds to 248,936 effective exposures.
The total duration of the dataset is 3146 days over the period HJD = 2453824--2456970.

The use of high cadence observation is to detect very short timescale events with 
$t_{\rm E} < 1$ day, which are expected due to lensing by
free-floating planets \citep{sumi2011,Mroz17,Mroz20b}, primodial blackholes \citep{Niikura2019a, Niikura2019b}
and/or short planetary anomalies
in the light curves of stellar microlensing events \citep{mao1991,sumi2010,bennett_rev,gaudi_rev, Kondo2019, Hirao2020,YounKil2020}.

High cadence observations are also important for improving the accuracy to which lensing parameters can be determined via light curve fitting. This is 
important for the accurate measurement of the microlensing timescale distribution, event rate and optical depth.

The Optical Gravitational Lensing Experiment \cite[OGLE;][]{uda15} also
conducts a microlensing survey toward the Galactic bulge, using the 1.3 m Warsaw telescope
at the Las Campanas Observatory in Chile.
The fourth phase of OGLE, OGLE-IV, started its high cadence survey observations in 
2010 with a 1.4 deg$^2$ FOV mosaic CCD camera. OGLE observes bulge fields
with cadences ranging from one observation every 20 minutes for 3 central
fields to fewer than one observation every night for the outer bulge fields.
Most observations are taken in the standard Kron-Cousin $I$-band with 
occasional observations in the Johnson $V$-band.
OGLE-IV issues $\sim 2000$ microlensing event alerts in real time 
each year.\footnote{\url{http://www.astrouw.edu.pl/\~ogle/ogle4/ews/ews.html}}
During 2001--2009, OGLE was operating its third phase survey, OGLE-III, using a 0.35 deg$^2$ camera.

\section{Data analysis} \label{sec:analysis}

The analysis method used here is similar to what was used by \cite{sumi2011,sumi2013},
but includes a correction of systematic errors.
The observed images were reduced with MOA's implementation \citep{bon01} of 
the difference image analysis (DIA) method \citep{tom96,ala98,ala00}. 
In the DIA, a high quality, good seeing, reference image is subtracted from each observed
image after matching the seeing, photometric scaling and position. 
This method generally provides more precise photometry in the very crowded Galactic bulge fields
than point spread function (PSF)-fitting routines, such as DOPHOT \citep{sch93}.
Each field consists of 10 chips and each chip is divided into eight 1024$\times$1024 pixel subfields 
during the DIA process.

In the MOA photometric light curve produced by DIA, 
we found that there were systematic errors 
that correlate with the seeing and airmass which causes positional shift,
 i.e., differential refraction, and  absorption, i.e., differential extinction, of stars.
The systematic trends owing to the relative proper motion of the source, 
lens and/or nearby stars can be modeled as
a linear function of time. 
To correct for these systematic trends in each event light curve, 
we used the baseline portions of the light curves and fitted a polynomial model in the same 
manner as \cite{bennett2012} and \cite{sumi2016a}. The model is given by the following equation;
\begin{align}
F_{\rm add} &= a_0 + a_1 ~{\rm JD} + a_2 ~ {\rm airmass} + a_3 ~{\rm airmass}^2 \notag\\
&+ a_4 ~{\rm seeing} + a_5 ~{\rm seeing}^2 + a_6 ~\tan {z} \cos {\phi} \notag\\
&+ a_7 ~\tan {z} \sin{\phi} + a_8 ~{\rm airmass}\tan{z} \cos{\phi} ~{\rm seeing} \notag\\
 &+ a_9 ~{\rm airmass} \tan{z}\sin{\phi} ~{\rm seeing},
\end{align}
where the elevation angle ($z$) and parallactic angle ($\phi$) of the target were included to correct for differential refraction.
$F_{\rm add}$ represents the additional flux for the correction, and the corrected flux is obtained by adding $F_{\rm add}$ to the original flux.
For each event, the correction was calculated using the light curve excluding the region of microlensing magnification, and the correction was applied to the full light curve.

This de-trending improved the fitting $\chi^2$ significantly in the baseline for many events, 
which indicates that the systematics have been reduced. 
This correction is important to have confidence in the light curve fitting parameters.
This is one of the major improvements from the previous analysis \citep{sumi2011,sumi2013}
in addition to the extension of the survey duration.

The DIA light curve photometry values are given as flux values which are scaled to the MOA reference images.
The instrumental magnitudes of the MOA reference images were
calibrated to the Kron/Cousins $I$-band by cross-referencing the MOA-II DOPHOT catalog
 to the OGLE-III photometry map of the Galactic bulge \citep{uda11}.

The OGLE data were reduced with the OGLE DIA \citep{wozniak2000}
 photometry pipeline \citep{uda15}. 
 In this analysis, we use data from OGLE-III and OGLE-IV.

\section{Microlensing event selection}
\label{sec:select}

In this work, we distinguished and selected single lens microlensing events from periodic variable stars,
other astrophysical phenomena such as cataclysmic variables (CVs), fast-moving stars including asteroids, and non-astrophysical artifacts due to dusts on the CCD detectors or leakages from saturated stars.

The observed flux during gravitational microlensing of a point source by a single point lens (PSPL) is represented by (\citealt{pac86}):
\begin{equation}
  \label{eq:ft}
F(t) = f_s A(t) + f_b,
\end{equation}
where $f_s$ is the unamplified source flux and $f_b$ is the total background flux.
The time variation of the magnification $A(t)$ is given by
\begin{equation}
  \label{eq:amp-u}
  A(t)= \frac{u^2(t)+2}{u(t)\sqrt{u^2(t)+4}},
\end{equation}
where $u (t)$ is the projected angular separation of the lens and source
in units of the angular Einstein radius $\theta_{\rm E}$.
The time variation of $u (t)$ is given by
\begin{equation}
  \label{eq:u}
  u(t)=\sqrt{u_{\rm 0}^2 + \left( \frac{t-t_{0}}{t_{\rm E}} \right)^2},
\end{equation}
where $u_{\rm 0}$ is the minimum impact parameter in units of
$\theta_{\rm E}$ and $t_{0}$ is the time of maximum magnification.

To model FSPL events, the angular size of the source star needs to be taken into account
by introducing an additional parameter, 
\begin{equation}
  \label{eq:rho}
  \rho= \frac{\theta_*}{\theta_{\rm E}},
\end{equation}
where
 $\theta_*$ is the angular radius of the source.
By combining  $\theta_*$ estimated from the source color and magnitude, one can obtain $\theta_{\rm E}$ and
the proper motion of the lens $\mu_{\rm rel}=\theta_{\rm E}/t_{\rm E}$ \citep{gou92,gou94a,nem94,wit94}.

The distribution of binary lens events is out of the scope of this paper.
In short, we selected light curves with a single instantaneous
brightening episode and a flat constant baseline, which can be fitted well
with a point-lens microlensing model.


\begin{deluxetable*}{lll}
\tablecaption{Event Selection Criteria \label{tbl:criteria}}
\tablewidth{0pt}
\tablehead{
\colhead{level}  & \colhead{criteria} & \colhead{comments}  \\
}
\startdata  
CR1 &  & \\
cut0 & $N_{\rm continue,8}\ge3$         & Number of continuous detection within 8 days from previous detection \\
       & $\sigma_{\rm x,y}  \le 1$  if $S_{\rm IM,max}\ge 3.5$    & Require small Standard deviation of (x, y) coordinates of objects, $\sigma_{\rm x,y}$, \\
       & $\sigma_{\rm x,y}  \le 0.8$ if $S_{\rm IM,max}<3.5$  &  depending on S/N on the image, $S_{\rm IM,max}$. Rejecting moving objects\\
\hline
cut1 &$N_{\rm data}  \ge 1,000$                                                    & Number of data points\\  
&$N_{\rm data}/N_{\rm f, use}  \ge 0.2$ if $S_{\rm max} <30$            & Require $\ge$20\% of data points for low S/N events\\  
&$N_{\rm out}  \ge 500$                        & Number of data points outside of the 1400-day window\\  
%
&$\Sigma_{\rm S, max} \ge 65$                 & Total significance of consecutive points with $S_i> 3$\\ 
& $\Sigma_{\rm S,max} \ge  75$ if $\chi^2_{\rm out}/dof >  3$       & Stricter requirement  on $\Sigma_{\rm S,max}$ for scattered light curves\\  
&$N_{\rm bump,in} < 20$ if $S_{\rm max} <12$    & remove scattered noisy light curves with low S/N\\  
&$N_{\rm bump,out} < 1$  if $S_{\rm max} <15$    & remove scattered noisy light curves with low S/N. Also reject repeating CV.\\  
&$N_{\rm bump,out} < 2$  if $\Sigma_{\rm S, max} <250$   & remove scattered noisy light curves with low S/N. Also reject repeating CV.\\   
\hline
cut2 
& $3824 \le t_0 \le 6970$ JD$'$              & Peak should be within observational period\\  
& $u_0 \le 1$                            & The minimum impact parameter\\  
& $ 0.05 \le t_{\rm E} \le 1,000$ days      & Einstein radius crossing timescale\\
& $ ISTAT \ge 2$                                     & Full covariance matrix in Minuit minimization\\  
& $N_{t_{\rm E}} \ge 5$                           & Number of data points in $|t-t_0| \le t_{\rm E}$ \\  
& $N_{t_{\rm E,n}} \ge 1$                    & Number of data points in $ -t_{\rm E} \le t-t_0 \le 0 $ \\  
& $N_{t_{\rm E2,n}} \ge 2$                   & Number of data points in $ -2t_{\rm E} \le t-t_0 \le 0 $ \\  
& $\sigma_{t_{\rm E}}  < 40$ days       & Error in $t_{\rm E}$\\    
& $\sigma_{t_{\rm E}}/t_{\rm E} \le 0.6$ OR $\sigma_{t_{\rm E}} \le 0.7$ days & Error in $t_{\rm E}$ is less than 60\% or 0.7days\\  
& $\sigma_{t_{\rm E}}/t_{\rm E} \le 0.6$ OR $\Sigma_{\rm S, max} \ge 210$ & Error in $t_{\rm E}$ is less than 60\% for low S/N\\  
& $\chi^2/{\rm dof} \le 2.5$                      & reduced $\chi^2$ for all data\\  
& $\chi^2_1/{\rm dof} \le 2.5$                  & reduced $\chi^2$ for $|t-t_0| \le t_{\rm E}$ \\  
& $\chi^2_1/{\rm dof} \le S_{\rm max}/12$   & reduced $\chi^2$ for $|t-t_0| \le t_{\rm E}$ requirement proportional to the S/N\\  
& $\chi^2_1/{\rm dof} \le 0.9 + \chi^2/{\rm dof} $      & reduced $\chi^2$ for $|t-t_0| \le t_{\rm E}$ should not be too bad relative to  $\chi^2/{\rm dof}$ \\  
& $\chi^2_2/{\rm dof} \le 1.5 + \chi^2/{\rm dof} $      & reduced $\chi^2$ for $|t-t_0| \le 2t_{\rm E}$ should not be too bad relative to  $\chi^2/{\rm dof}$ \\  
& $ 10 \le I_s \le 21.4$                             & Apparent $I$-band source magnitude\\ 
& $I_{\rm c} - I_{\rm s} \le 0.6$  if  ($I_{\rm c} - I_{\rm s})/\sigma_I \ge 7$                    & remove if source is significantly brighter than cataloged star\\  
& $\Sigma_{\rm S, max} \ge 100 N_{1\sigma} -1200 $ & Stricter requirement  on $\Sigma_{\rm S,max}$ for light curves with systematic residuals\\ 
\hline
cut3
&
same as cut2 but for FSPL model \\
\hline
 & & Additional criteria for CR2\\
\hline
CR2 & $\sigma_{t_{\rm E}}/t_{\rm E} \le 0.5$ OR $\sigma_{t_{\rm E}} \le 0.2$ days & Error in $t_{\rm E}$ is less than 50\% or 0.2days\\
%
%
%
\enddata
\end{deluxetable*}

\subsection{Selection criteria} \label{sec:SC}

We use a similar analysis pipeline and microlensing event selection 
criteria to those used in \citet{sumi2011}, and the details are summarized 
in their Supplementary Information. However, we made several improvements to 
 optimize our method for the extended dataset used in this work.
In our dataset, there is an increased number of artifacts because of the increased number of image frames.
On the other hand, the extended baseline helps to (i) distinguish the long timescale events from long variables
and (ii) reject repeating flare stars.
We empirically defined the following selection criteria to maximize the number of
microlensing candidates and discard all non-microlensing light curves.
All criteria are summarized in Table \ref{tbl:criteria}.

\begin{itemize}
\item[$(1)$]
Cut-0:
We conducted a blind search on the subtracted images rather than a limited search 
on the pre-identified stars on the reference images. On subtracted images, 
we detect variable objects by using a custom implementation
of the IRAF task DAOFIND \citep{daophot} with the modification
that both positive and negative PSF profiles are searched for simultaneously. 

This algorithm finds peaks with a signal to 
noise ratio (S/N) of $S_{\rm IM}> 2.7$ on difference images and then
applies several additional criteria to avoid the detection of 
spurious variations that are not associated with stellar variability,
such as cosmic ray hits, satellite 
tracks and electrons leaked from the saturated images of bright  stars.
Here we slightly modified these criteria from the previous work to optimize our method when using our new dataset.
Furthermore, in this analysis,
we applied the PSF fitting at the detected objects on difference images and used their $\chi^2$ values 
as one of the criteria to reduce spurious detections.
Here we used the PSF function derived by DOPHOT on the reference images, which was then 
convolved by the kernel to match the seeing, scale, and PSF shape variation on each observed
subframe. 
We used the kernels which are derived in DIA process.

Lists of variable objects are created by using
the positions of detected objects in the first frame.
Then, in each new frame time sequentially analyzed, the positions of detected objects are checked against those in the list of variable objects.
When no object is cross-referenced within 2~pixels, the object is
classified as new and added to the list of variable objects with its position.
If the object has previously been detected within 2~pixels, 
the number of detections for this object, $N_{\rm detect}$, is 
incremented and $N_{\rm detect} \ge 3$ is required to pass.
If the detection in the new frame has a higher S/N, 
then the position of the object in the list is replaced 
by the new position.
The maximum value of $S_{\rm IM}$ among the frames is recorded
as $S_{\rm IM, max}$ for each object.
At this stage, we found 5,791,159 objects.
In this work, we further require that these detections should be continuous and without a significant time gap
because some types of artifacts tend to be not correlated in time.
Each detection should be within 8 days from the previous detection for a variable, $N_{\rm continue,8}$, to be
incremented. We required $N_{\rm continue,8} \ge 3$ for an event to pass this cut.
As a result, 2,409,061 variable objects were detected at this stage
of the analysis, including a number of image artifacts of various types to be removed by subsequent criteria.

The $(x,y)$ coordinates of the detected peak with the highest S/N are adopted as 
the final coordinate for the corresponding object. 
We found many moving objects, asteroids, satellites, and dust specks on the CCD chip in the sample.
These tend to have large standard deviations of (x, y) coordinates, $\sigma_{\rm x,y}$. 
We required  $\sigma_{\rm x,y}\le 1$ pixel and $\le0.8$ pixel for S/N of
$S_{\rm IM, max}\ge 3.5$ and $<3.5$, respectively,   
to reject these moving objects.

\item[(2)]
Cut-1:
Light curves of the candidates passing Cut-0 were then created by using
PSF fitting photometry on the difference images. 
Here, DOPHOT PSF functions on the reference images are used rather than the empirical numerical PSF in the previous work \citep{sumi2011}, 
so that the flux scale can be linked with the DOPHOT catalog of the reference images easily for precise calibration.
Here the data points which failed the PSF fitting due to various reasons, such as saturated pixels, dead pixels, 
satellite track, cosmic ray hits, etc, are removed. 
We retain light curves only if the number of 
the data points ($N_{\rm data}$) is more than 1,000.

It is known that in a stellar crowded region like the Galactic bulge, the error bar estimates from the photometry
code provide only an approximate description of the photometric uncertainty for
each measurement.
The photometric error bars were multiplied by a normalization factor that standardizes the distribution of residuals of the constant fits to non-variable stars in each subfield.

To find the bump and define the baseline flux of the light curve, 
firstly, we place a 1400-day moving window on each light curve.
Note that the window size is increased from 120-day in the previous work thanks to the longer baselines of the time series used in this work.
The window moves from the beginning to the end of the light curve with a step size of 50 days.
In each window position, we fit the light curves outside of the window to get an average baseline flux of $F_{\rm base}$ 
and $\chi^2_{\rm out}/{\rm dof}$.
We require the number of data points in baseline, $N_{\rm out}$ to be more than 500.

We then search for positive light curve ``bumps" inside the 1400-day window relative to the baseline.
The actual scatter of the light curves depends upon the
spatial distribution of stars in the immediate vicinity of the target and/or low level variabilities of these stars including the target itself.
We therefore define a significance of each data point relative to the baseline taking the scatters of the baseline into account as, 
$S_{i} = \left( F_i - F_{\rm base}  \right)/ (\sigma_i \sqrt{\chi^2_{\rm out}/{\rm dof} })$,
where $\sigma_i$ is the error bar of the $i$th measurement of flux $F_i$.

We then define a ``bump" as a brightening episode with 
more than 3 consecutive measurements with excess flux $S_i> 3$.

We define a statistic $\Sigma_{\rm S}= \Sigma_i S_i$ summed over consecutive points with $S_i  > 3$ and
require $\Sigma_{\rm S} \ge 10$.
The number of bumps, $N_{\rm bump,in}$ and $N_{\rm bump,out}$ are counted inside and outside of the window, respectively.
The bump with the highest $\Sigma_{\rm S}$ inside the window is defined as the primary bump.
The maximum value of $\Sigma_{\rm S}$ and $S_i$ of this primary bump among the moving window positions
are defined as $\Sigma_{\rm S, max}$ and $S_{\rm max}$, respectively.
We require $\ge$20\% of data points used, i.e., 
$N_{\rm data}/N_{\rm f, use}  \ge 0.2$ for low S/N events with $S_{\rm max} <30$.

There are 549,445 light curves that satisfy tentative looser criteria of $\Sigma_{\rm S, max} \geq 40$ or
 $\Sigma_{\rm S, max} \ge  75$ and $\chi^2_{\rm out}/{\rm dof} >  3$. 
With these light curves, we moved on to cut-2 for a trial run. Here, a looser version of the current cut-2 criteria was used to select light curves. 
We then visually inspected 
tens of thousands light curves with their best fit models in order of 
higher $\Sigma_{\rm S, max}$ and smaller $\chi^2_{\rm out}/{\rm dof}$ 
until the frequency of plausible events appeared to be almost zero. 
During this process, we found 6,111 microlensing candidates.
Note that although this sample contains the most of microlensing candidates in this dataset,
this whole sample is not statistically complete with certain criteria.

As a result,  we increased the limit to $\Sigma_{\rm S,max} \geq 65$.
We also placed the upper limit of  $N_{\rm bump,in}$ and $N_{\rm bump,out}$ depending on the S/N, i.e.,
$\Sigma_{\rm S, max}$ and $S_{\rm max}$, to remove scattered and noisy light curves, 
low level variable stars, and repeating flare stars.
All criteria in cut-1 are summarized in Table \ref{tbl:criteria}. 
There are 67,242 light curves remained after applying cut-1.


\begin{deluxetable*}{llcccrrrccc}
\tabletypesize{\scriptsize}
\tablecaption{Short timescale event candidates with $t_{\rm E}<1$ day with RA, Dec., Alerted ID, catalog star's $I$-band magnitude, number of data points and passed criteria.
\label{tbl:candlistShort1}
}
\tablewidth{0pt}
\tablehead{ 
\colhead{ID} &
\colhead{internal ID} &
\colhead{R.A.} & 
\colhead{Dec.} &
\colhead{ID$_{\rm alert}$} &
\colhead{$I_{\rm c}$} &
\colhead{$N_{\rm data}$} &
\colhead{criteria} \\
\colhead{} &
\colhead{(field-chip-subfield-ID)} &
\colhead{(2000)} & 
\colhead{(2000)} &
\colhead{} &
\colhead{(mag)} &
\colhead{} &
\colhead{} 
}
\startdata
 MOA-9y-537 &  gb2-10-7-248887 & 17:54:18.974 & -33:50:27.24  &           -- &  18.85 $\pm$  0.09 & 10265 & CR2\\
 MOA-9y-570 &   gb3-2-1-129006 & 17:52:19.998 & -32:26:31.67  & 2009-BLG-115 &  18.26 $\pm$  0.05 & 20210 & CR1\\
 MOA-9y-600 &   gb3-2-4-455860 & 17:50:58.418 & -32:23:11.30  &           -- &  17.29 $\pm$  0.02 & 20331 & CR2\\
 MOA-9y-671 &    gb3-4-2-82374 & 17:52:29.716 & -33:10:03.01  & 2009-BLG-206 &  17.45 $\pm$  0.04 & 20426 & CR2\\
 MOA-9y-770 &    gb3-7-6-65303 & 17:55:16.892 & -33:08:35.69  &           -- &  16.00 $\pm$  0.01 & 20438 & CR2\\
MOA-9y-1173 &   gb4-5-6-114001 & 17:52:41.125 & -31:33:50.59  &           -- &  17.86 $\pm$  0.07 & 21831 & CR2\\
MOA-9y-2175 &   gb5-8-0-185381 & 17:56:37.038 & -29:04:52.67  &           -- &  16.58 $\pm$  0.01 & 30099 & CR2\\
MOA-9y-2202 &   gb5-8-1-542070 & 17:56:05.269 & -29:11:29.62  & 2014-BLG-215 &  18.49 $\pm$  0.09 & 31191 & CR1\\
MOA-9y-3945 &     gb10-5-1-431 & 17:57:52.940 & -28:16:56.55  &           -- &  16.70 $\pm$  0.05 & 19350 & CR2\\
MOA-9y-5057 &   gb14-8-3-66703 & 18:06:26.706 & -27:26:44.97  &           -- &  18.34 $\pm$  0.05 & 18114 & CR2\\
MOA-9y-5919 &   gb19-7-7-39836 & 18:18:41.318 & -25:57:15.65  &           -- &  17.07 $\pm$  0.01 &  4940 & CR2\\
MOA-9y-6057 &   gb21-3-3-11851 & 18:17:40.655 & -22:01:30.52  &           -- &  18.25 $\pm$  0.04 &  4575 & CR2\\
\enddata
\tablecomments{The list of all microlensing event candidates is available in the electronic version. 
\\}
\end{deluxetable*}

\begin{deluxetable*}{lrccccccccccccccccccccc}
\tabletypesize{\scriptsize}
\tablecaption{Results of refined FSPL fits for the short timescale event candidates of $t_{\rm E}<1$ day.
\label{tbl:candlistShort}
}
\tablewidth{0pt}
\tablehead{
\colhead{ID} & \colhead{$\Delta \chi^2\tablenotemark{a}$} &  \colhead{$t_0$}        & &   \multicolumn{2}{c}{$t_{\rm E}$} & &  \multicolumn{2}{c}{$u_0$}      & & \multicolumn{2}{c}{$\rho$}      & & \colhead{$I_{\rm s}$}    \\
             &                                    & \colhead{$(\rm HJD')$}  & &   \multicolumn{2}{c}{$(\rm day)$} & &   \multicolumn{2}{c}{}          & & \multicolumn{2}{c}{}            & & \colhead{${(\rm mag)}$}  \\
             &                                    & \colhead{best}          & & \colhead{best} & \colhead{mean}   & & \colhead{best} & \colhead{mean} & & \colhead{best} & \colhead{mean} & & \colhead{best}
}                                                                                                                                                                                               
\startdata                                                                                                                                                                                      
 MOA-9y-537 &      0.2                           & 6845.279                 & &  0.326     & 0.390 $\pm$ 0.101    & & 0.478      & 0.337 $\pm$ 0.114 & &  0.488  &   0.172 $\pm$ 0.147   & &    18.50 \\
 MOA-9y-570 &      0.9                           & 4919.257                 & &  0.688     & 0.809 $\pm$ 0.280    & & 0.380      & 0.253 $\pm$ 0.102 & &  0.405  &   0.135 $\pm$ 0.126   & &    18.26 \\
 MOA-9y-600 &      0.9                           & 6359.091                 & &  0.803     & 0.536 $\pm$ 0.451    & & 0.001      & 0.031 $\pm$ 0.043 & &  0.036  &   0.225 $\pm$ 0.408   & &    21.48 \\
 MOA-9y-671 &      0.0                           & 4955.068                 & &  0.765     & 0.765 $\pm$ 0.053    & & 0.440      & 0.449 $\pm$ 0.056 & &  0.173  &   0.177 $\pm$ 0.115   & &    17.49 \\
 MOA-9y-770 &    525.8                           & 4647.043                 & &  0.319     & 0.315 $\pm$ 0.017    & & 0.101      & 0.208 $\pm$ 0.130 & &  1.054  &   1.084 $\pm$ 0.070   & &    16.17 \\
MOA-9y-1173 &      2.9                           & 4945.128                 & &  0.195     & 0.236 $\pm$ 0.264    & & 0.044      & 0.061 $\pm$ 0.037 & &  0.075  &   0.087 $\pm$ 0.041   & &    21.80 \\
MOA-9y-2175 &      0.0                           & 4581.306                 & &  0.755     & 0.725 $\pm$ 0.105    & & 0.359      & 0.428 $\pm$ 0.132 & &  0.117  &   0.237 $\pm$ 0.177   & &    17.72 \\
MOA-9y-2202 &      0.2                           & 6771.610                 & &  0.757     & 0.957 $\pm$ 0.316    & & 0.380      & 0.142 $\pm$ 0.107 & &  0.589  &   0.123 $\pm$ 0.141   & &    18.31 \\
MOA-9y-3945 &      0.6                           & 3910.751                 & &  0.903     & 0.737 $\pm$ 0.225    & & 0.238      & 0.323 $\pm$ 0.195 & &  0.365  &   0.134 $\pm$ 0.177   & &    19.58 \\
MOA-9y-5057 &      0.2                           & 5025.062                 & &  0.260     & 0.307 $\pm$ 0.066    & & 0.694      & 0.434 $\pm$ 0.109 & &  0.755  &   0.186 $\pm$ 0.144   & &    18.01 \\
MOA-9y-5919 &     35.0                           & 4601.092                 & &  0.066     & 0.057 $\pm$ 0.016    & & 0.031      & 0.572 $\pm$ 0.436 & &  1.018  &   1.399 $\pm$ 0.460   & &    18.58 \\
MOA-9y-6057 &      0.1                           & 3923.229                 & &  0.168     & 0.222 $\pm$ 0.057    & & 0.469      & 0.255 $\pm$ 0.120 & &  0.559  &   0.356 $\pm$ 0.200   & &    17.89 \\
 \enddata
\tablenotetext{a}{$\Delta\chi^2=\chi^2 - \chi^2_{\rm FS}$.}
\tablecomments{Best columns show values of the best-fit models. Mean columns show the mean and standard deviation values of the posterior distribution of the MCMC, where a prior of
$0.8 < \mu_{\rm rel}/{\rm mas \, yr^{-1}} < 20.0$ was additionally applied to derive those values. }
\end{deluxetable*}

\item[(3)]
Cut-2: 
We fit the light curves that passed the cut-1 criteria with the PSPL model given by Eq.(\ref{eq:ft}).

For the fitting, we used the MIGRAD minimization algorithm in MINUIT package \citep{James1994}.
To get an accurate distribution of the microlensing timescale,
we require the full covariance matrix calculated in MINUIT minimization, i.e., $ISTAT\ge 2$.
The parameter errors are determined using the MINOS procedure of the MINUIT package, 
except in cases where MINOS failed. In those cases, the error bars 
from the MIGRAD procedure are used.

We select only events with the peak time within the survey duration $3824 \le t_0 \le 6970$ JD$'$ where
JD'=HJD$-2450000$ and a minimum impact parameter of $u_0<1.0$.  The $I$-band source
magnitudes are required to be $10 \le I_s \le 21.4$ mag and  not significantly brighter than cataloged star
on the reference images.
We select only events with a timescale of $ 0.05 \le t_{\rm E} \le 1,000$ days because
the events whose parameters severely degenerate or are not due to microlensing 
tend to have very small or larger $t_{\rm E}$ values outside of this range.
The errors in $t_{\rm E}$, $\sigma_{t_{\rm E}}$, should be less than 40 days, 
which can effectively reject artifacts with long-term variability and/or systematics.
We also require $\sigma_{t_{\rm E}}$ to be either $\leq 0.6 \, t_{\rm E}$ or $\leq 0.7 \, {\rm days}$ for the nominal criteria (CR1).
We also test stricter criteria (CR2) which require $\sigma_{t_{\rm E}} $ to be either $\leq 0.5 \, t_{\rm E}$ or $\leq 0.2 \, {\rm days}$,
to see the effect of the choice of the selection criteria.

One of the main mimics of microlensing is a CV- or flare-type brightening which 
shows a fast rise and slow decline, in which usually only the decline phase is observed.
To differentiate these from microlensing events, we require the number of data points $N_{t_{\rm E}} \ge 5$  
during $|t-t_0| \le t_{\rm E}$.
Furthermore, we also require at least one data point during the rising phase, i.e.,
$N_{t_{\rm E,n}} \ge 1$ during $ -t_{\rm E} \le t-t_0 \le 0 $ and two data points, i.e., 
$N_{t_{\rm E2,n}} \ge 2$ during $ -2t_{\rm E} \le t-t_0 \le 0 $.

We also require $\chi^2/{\rm dof} \le 2.5$ for the entire light curve and
$\chi^2_1/{\rm dof} \le 2.5$ during $|t-t_0| \le t_{\rm E}$.
To remove many low S/N artifacts, we further apply the upper limit for 
$\chi^2_1/{\rm dof}$ and $\chi^2_2/{\rm dof}$ depending on $S_{\rm max}$ and overall $\chi^2/{\rm dof}$, such that
,
 $\chi^2_1/{\rm dof} \leq S_{\rm max}/12$,  $\chi^2_1/{\rm dof} \leq 0.9 + \chi^2/{\rm dof}$, 
and $\chi^2_2/{\rm dof} \leq 1.5 + \chi^2/{\rm dof}$, as shown in Table \ref{tbl:criteria}, 
where $\chi^2_2/{\rm dof}$ is defined for the light curve during $|t-t_0| \le 2t_{\rm E}$.

Events with systematic residuals from the best fit model are also rejected. 
This cut depends on the significance of the microlensing signal. 
We defined $N_{1\sigma}$ as the maximum number of consecutive measurements 
that are scattered from the best fit model with excess flux more than 1-$\sigma$. 
We require that $\Sigma_{\rm S, max} \ge 100 N_{1\sigma} -1200 $.

\item[(4)]
Cut-3: 
The light curves are also fit using the FSPL model that considers the finite source effect parameterized by $\rho$ given by Eq. (\ref{eq:rho}).
We use the \citet{Bozza2018} algorithm to calculate the magnification by FSPL. 
 The source angular radius $\theta_{*}$ is calculated by using the relation between
the limb-darkened stellar angular diameter, $\theta_{\rm LD}$, $(V-I)$ and $I$ (private communication, \citet{Boyajian2014}, see \citealt{Fukui2015}).
Here we estimated the $(V-I)_{\rm s}$ color and error by taking the mean and standard deviation, respectively, of stars on the 
color magnitude diagram (CMD) at the magnitude of the best fit $I_{\rm s}$, assuming that the source is in the bulge \citep{ben08}.
We use MOA's CMD combined with {\it Hubble Space Telescope} (HST)'s CMD \citep{hol98} for bright and faint stars respectively.
Extinction and reddening are corrected by using the position of the red clump giants (RCGs) in the CMD in each subfield.

Then we derived the angular Einstein radius, $\theta_{\rm E}=\theta_{*} /\rho$ and
the lens-source relative proper motion,  $\mu_{\rm rel}=\theta_{\rm E}/t_{\rm E}$.
We found that many non-microlensing light curves tend to have better $\chi^2$ values for FSPL models compared to PSPL models.
However, most of these provide unphysically small $\mu_{\rm rel}$ values of less than $\sim$0.8 mas$\,$yr$^{-1}$.
We adopt the FSPL results if $\chi^2$ is improved by more than 20 and 50 over those when using a PSPL model
with $0.8 < \mu_{\rm rel} \leq 0.9$ mas$\,$yr$^{-1}$ and $\mu_{\rm rel} >0.9$ mas$\,$yr$^{-1}$, respectively.

Then, the cut-2 criteria are applied for the results of the FSPL fit parameters.
Although we identified 18 FSPL events  visually in all candidates,  only 13 events passed 
all of our selection criteria.

Note that although KMTNet's sample of giant-source events \citep{Kim2021, Gould2022} contains events with super red giant source stars that have an extremely large source size ($\theta_{*} > 10~\mu$as), there are no such events in our sample
because such bright stars saturate in MOA at $I\sim14$ mag.
This is one of the reasons that the number of FSPL events are relatively small compared to the KMTNet survey.

\end{itemize}

\begin{figure*}[htb]
\centering
  \begin{tabular}{@{}cccc@{}}
  \includegraphics[width=8cm,keepaspectratio]{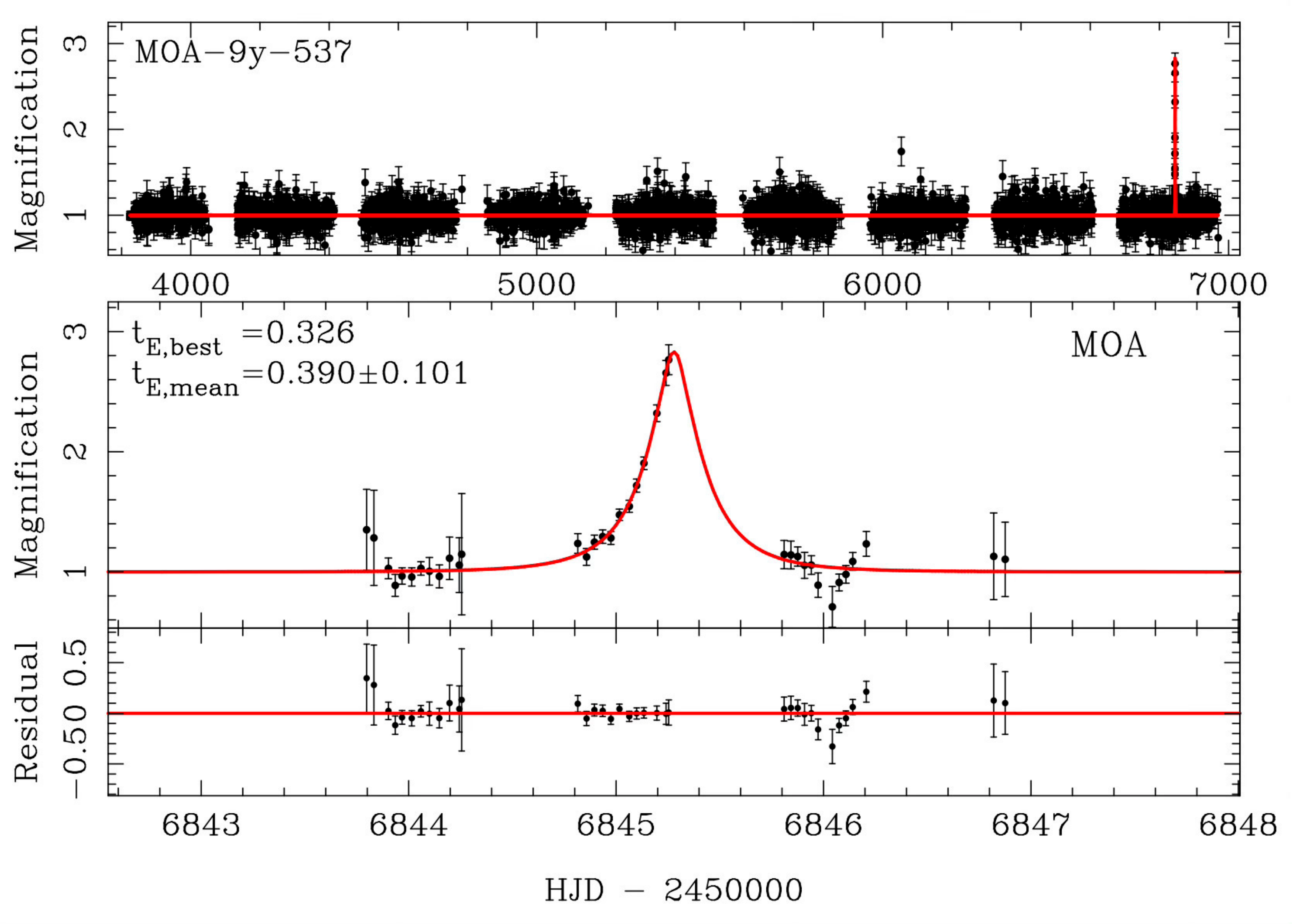} &
  \includegraphics[width=8cm,keepaspectratio]{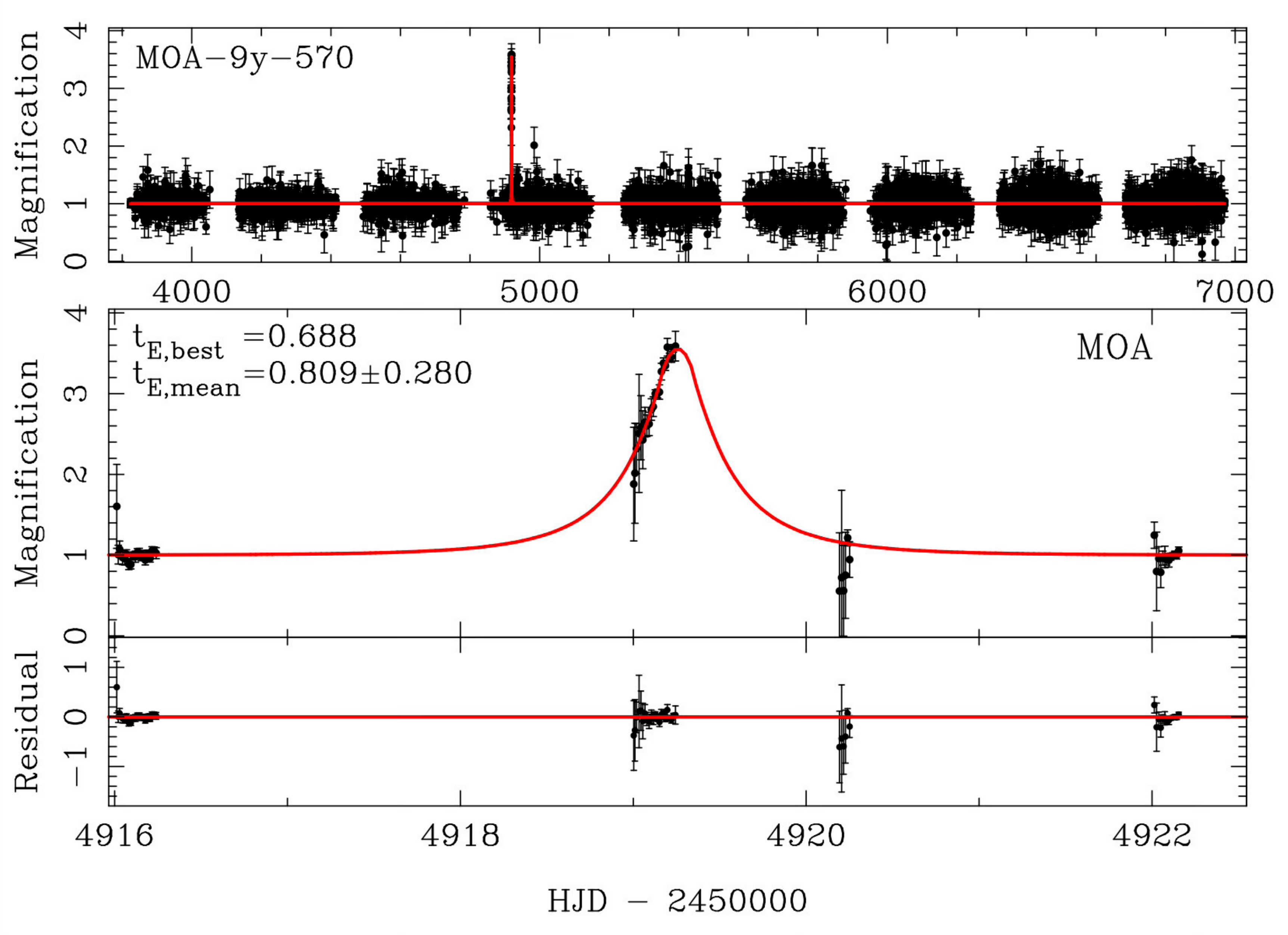} \\
  \includegraphics[width=8cm,keepaspectratio]{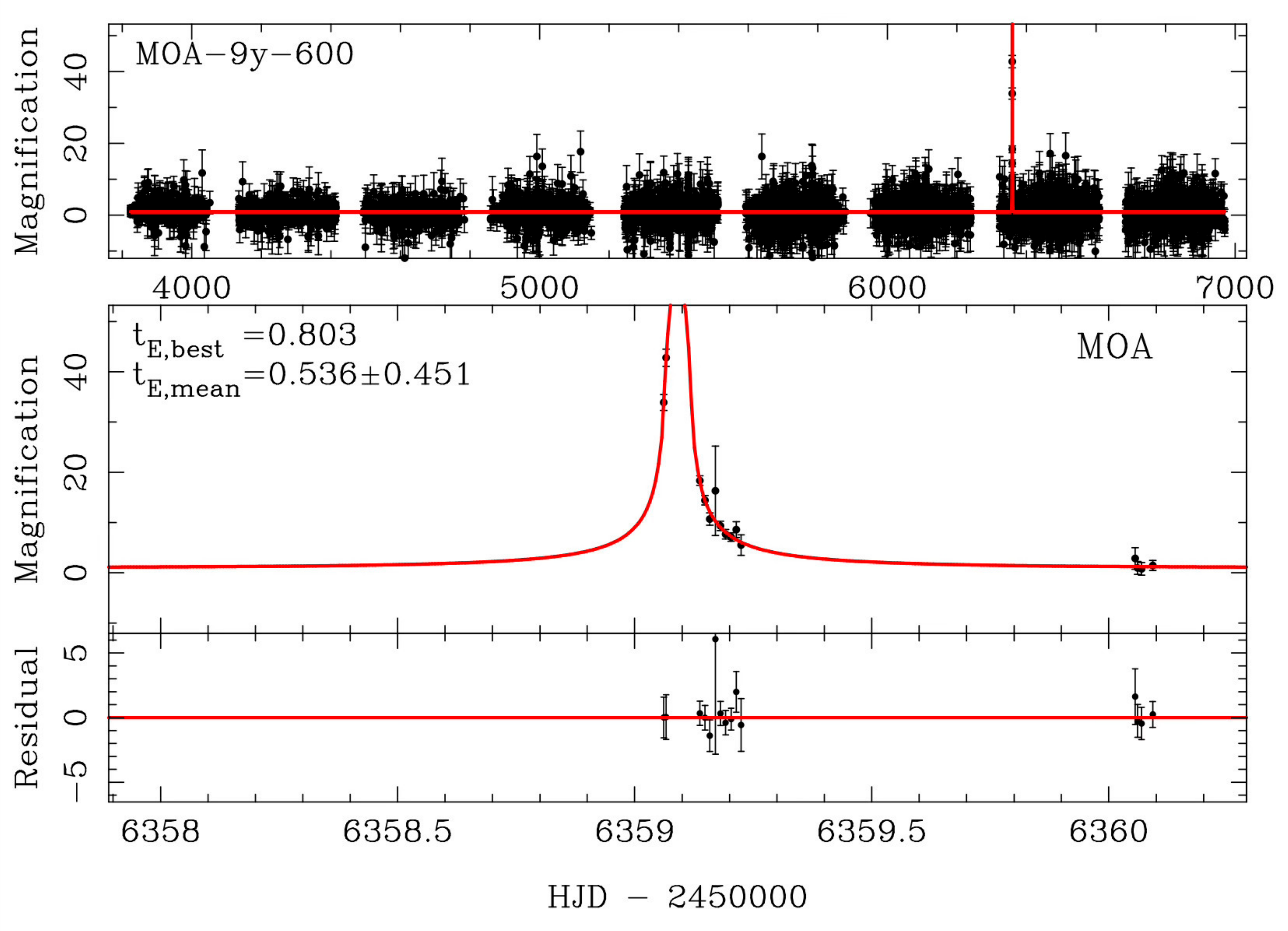} &
  \includegraphics[width=8cm,keepaspectratio]{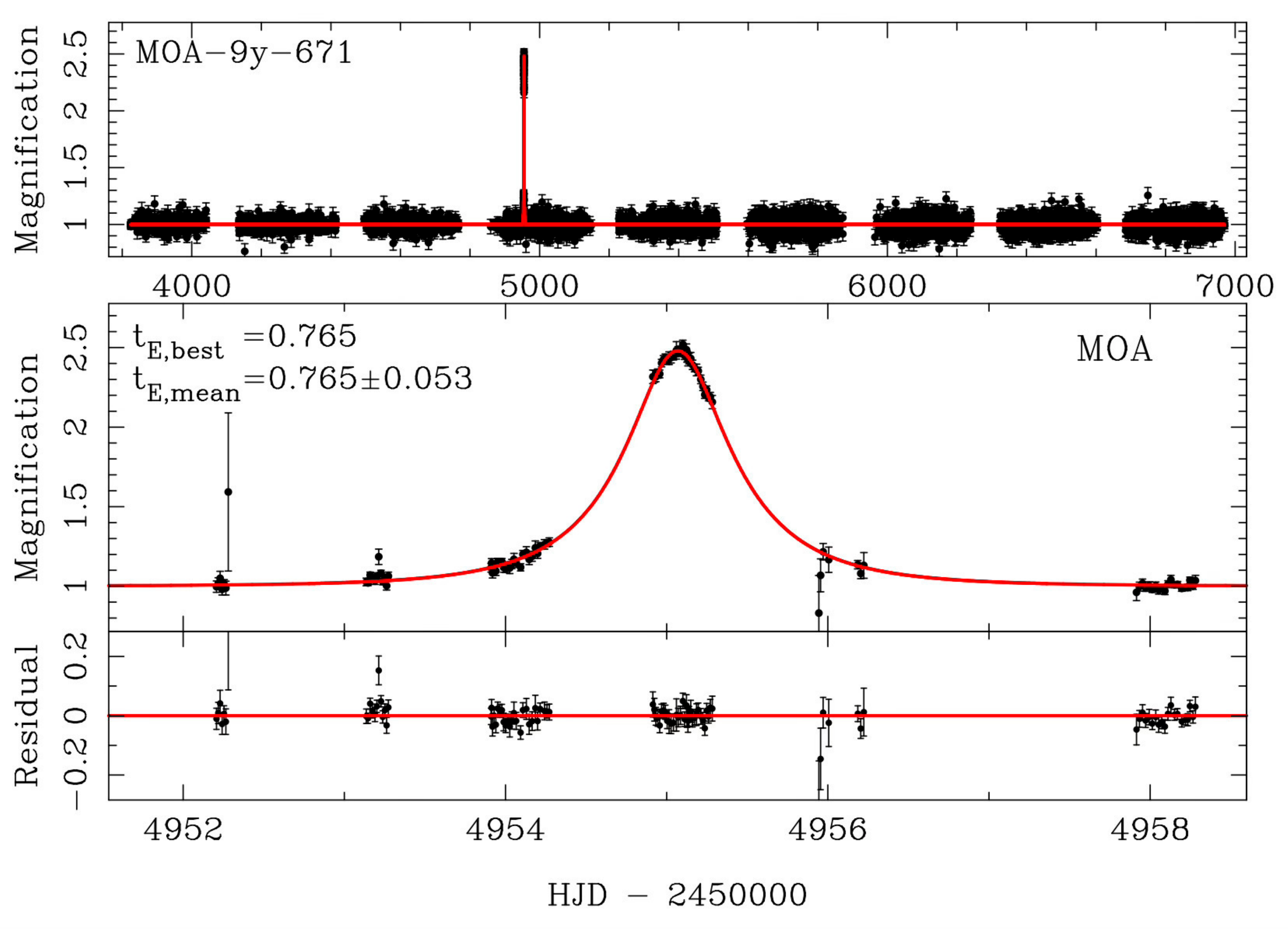} \\
  \includegraphics[width=8cm,keepaspectratio]{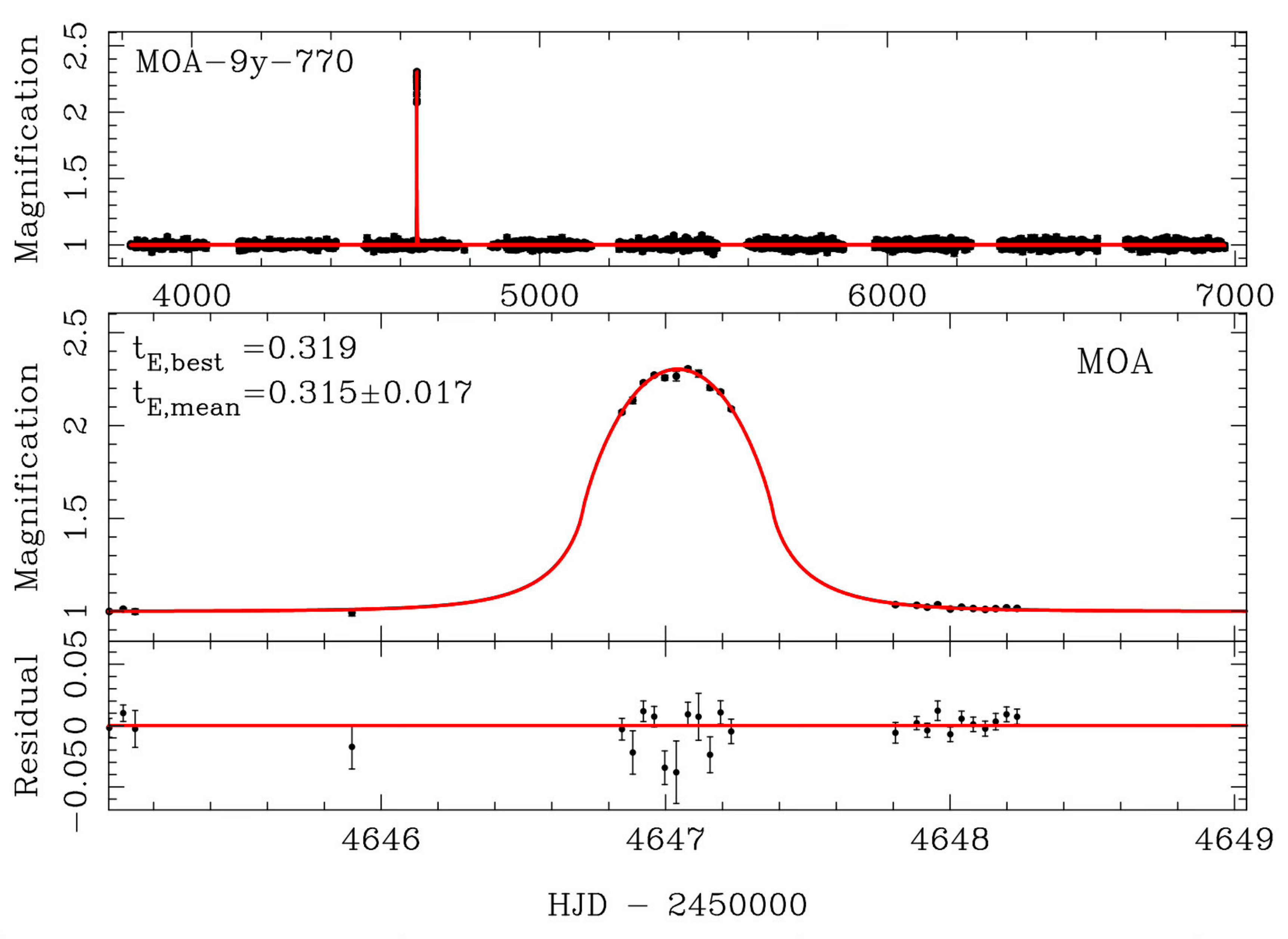} &
  \includegraphics[width=8cm,keepaspectratio]{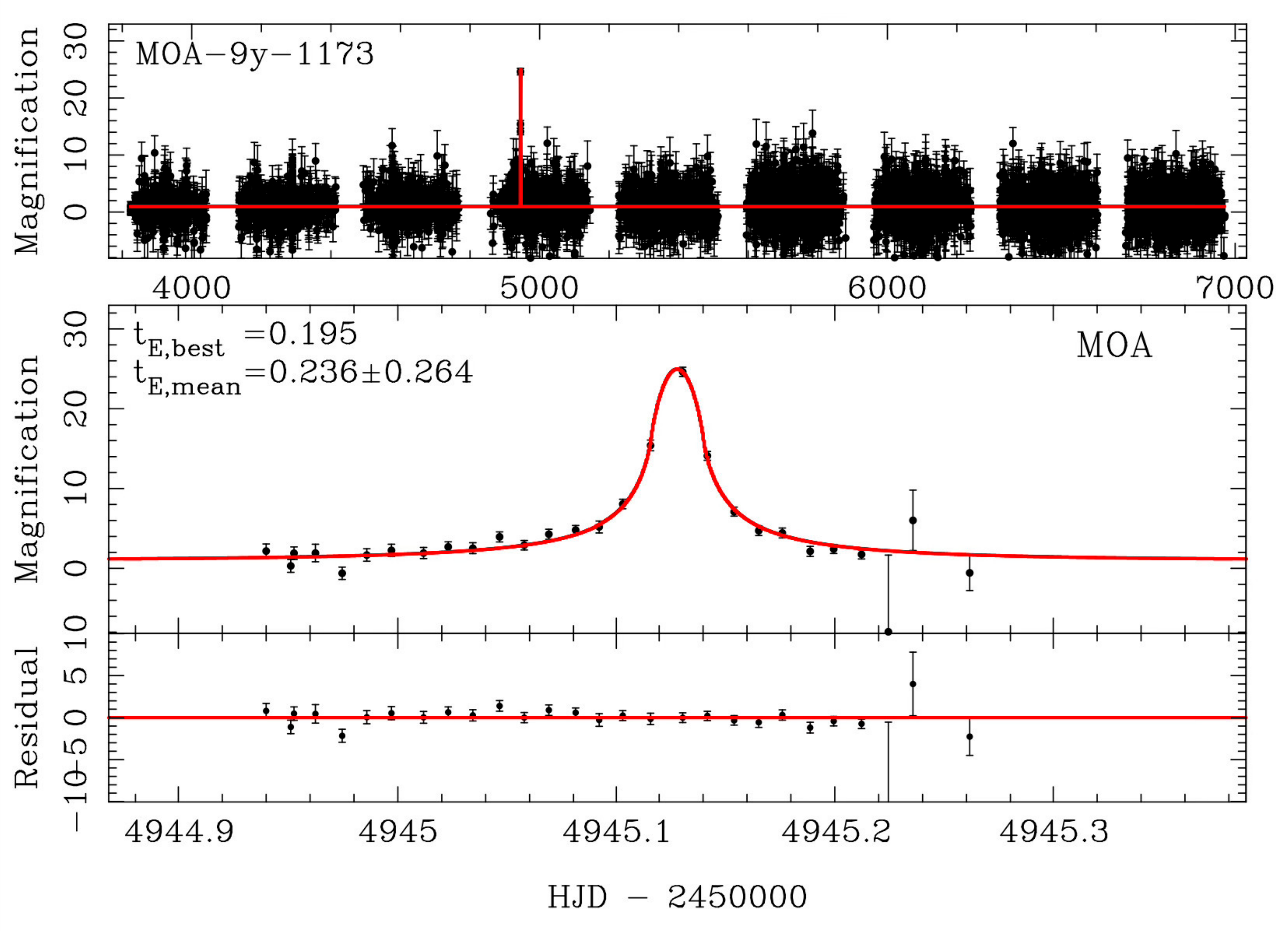} \\
  \includegraphics[width=8cm,keepaspectratio]{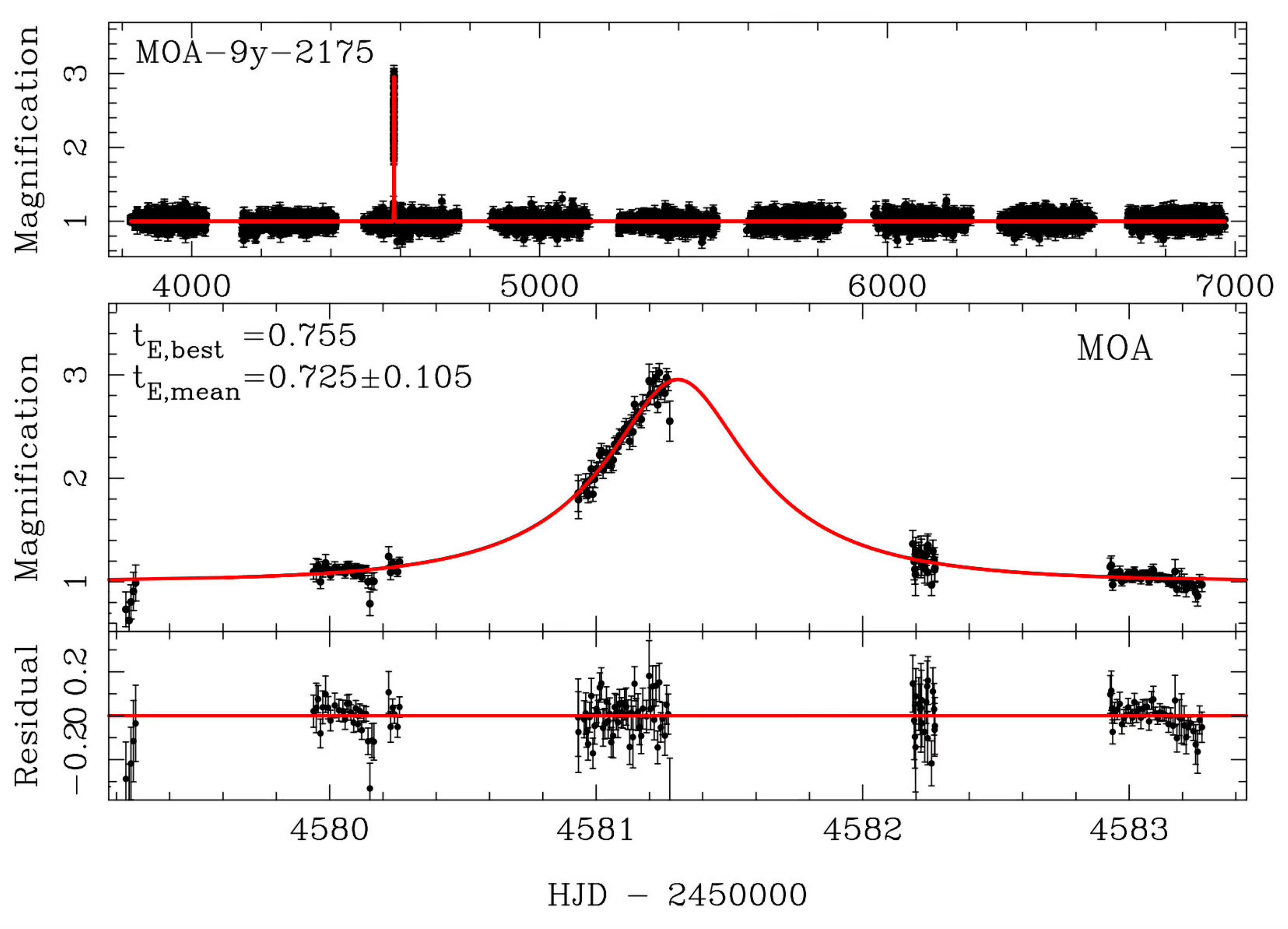} &
  \includegraphics[width=8cm,keepaspectratio]{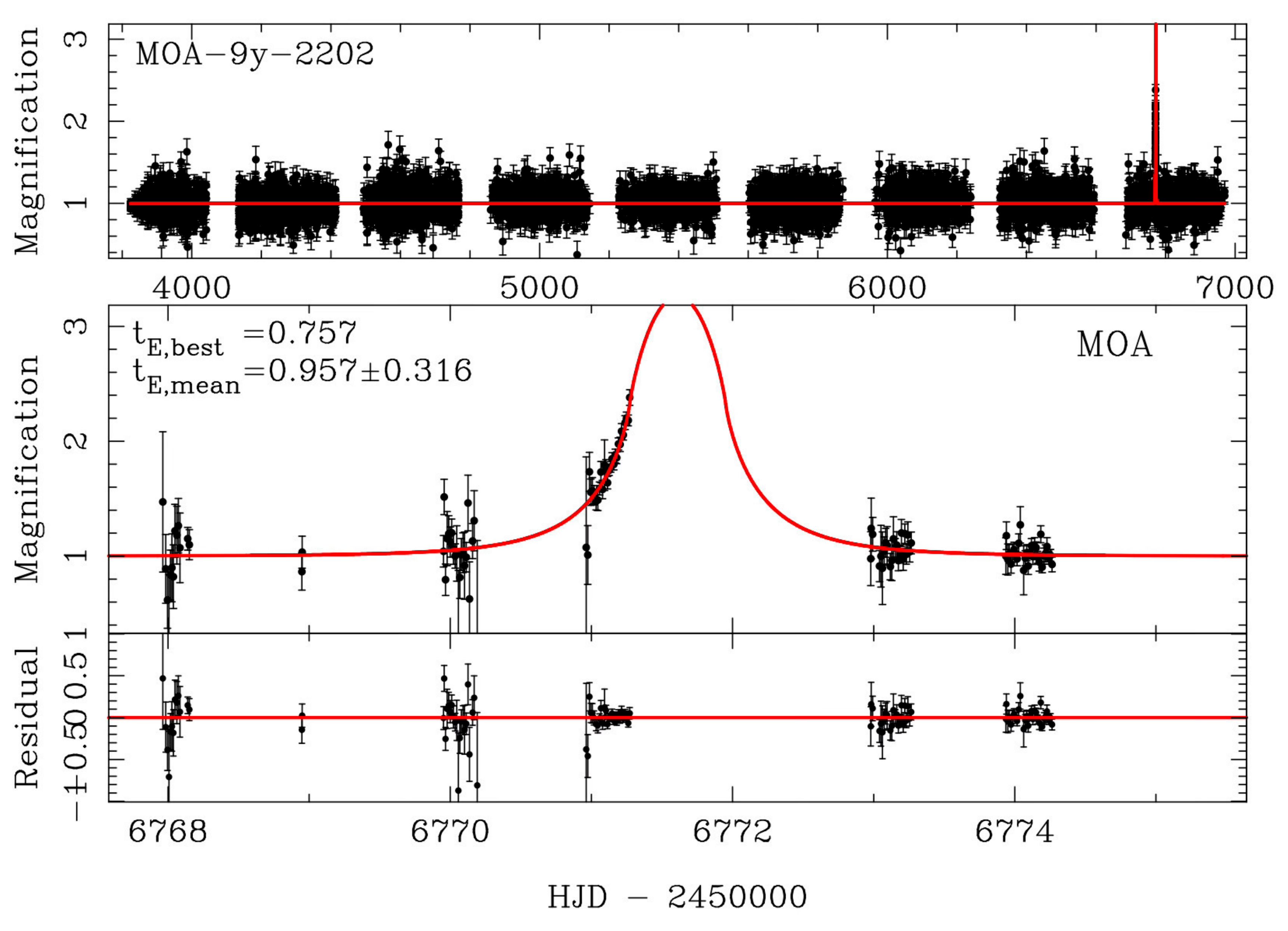}
  \end{tabular}
\caption{
  \label{fig:lightcurve}
Light curves of the short timescale microlensing candidates with $t_{\rm E}< 1$ day.
}
\end{figure*}


\begin{figure*}
 \addtocounter{figure}{-1} 
\begin{center}
\includegraphics[width=8.2cm,keepaspectratio]{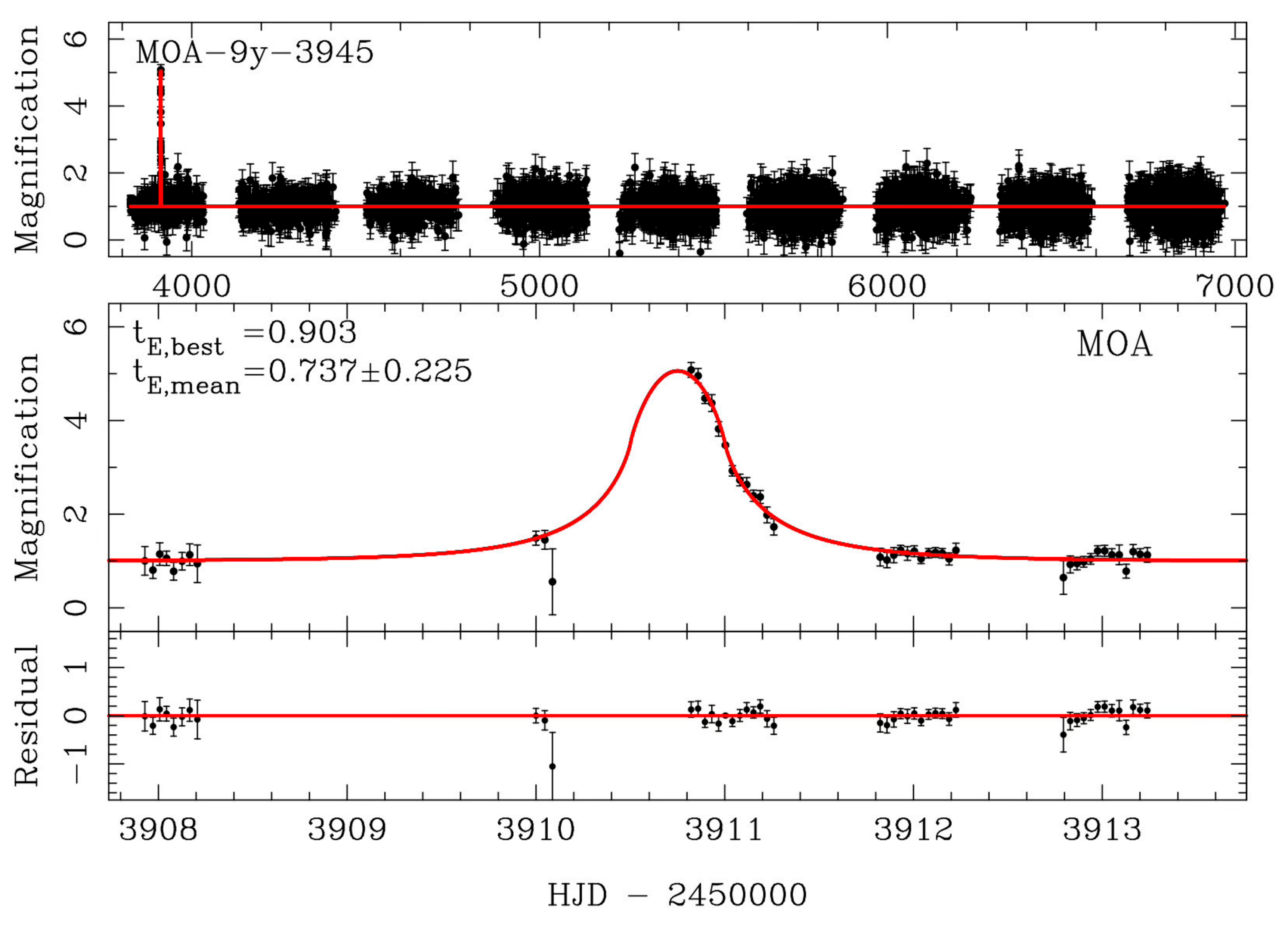}
\includegraphics[width=8.2cm,keepaspectratio]{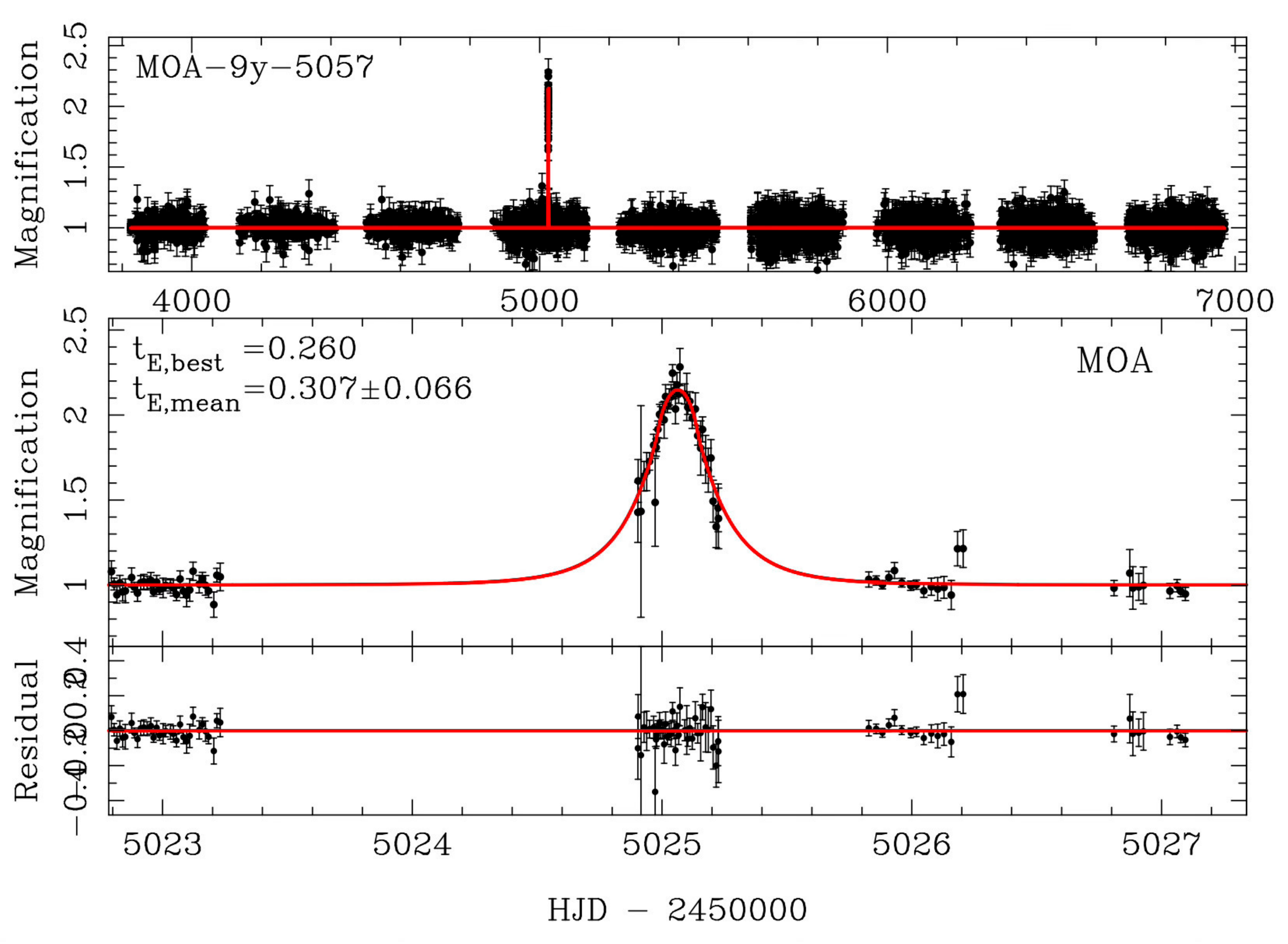}
\includegraphics[width=8.2cm,keepaspectratio]{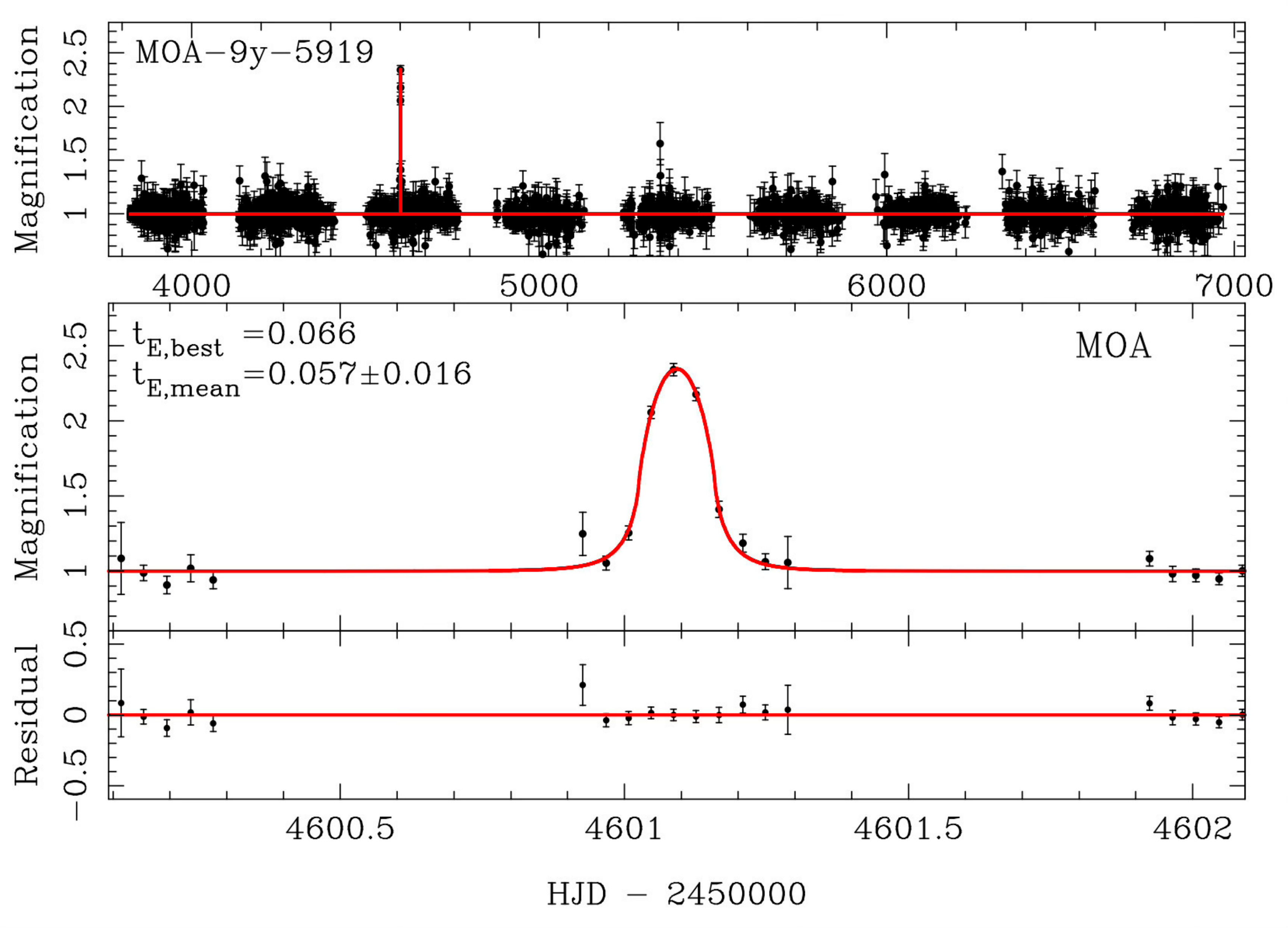}
\includegraphics[width=8.2cm,keepaspectratio]{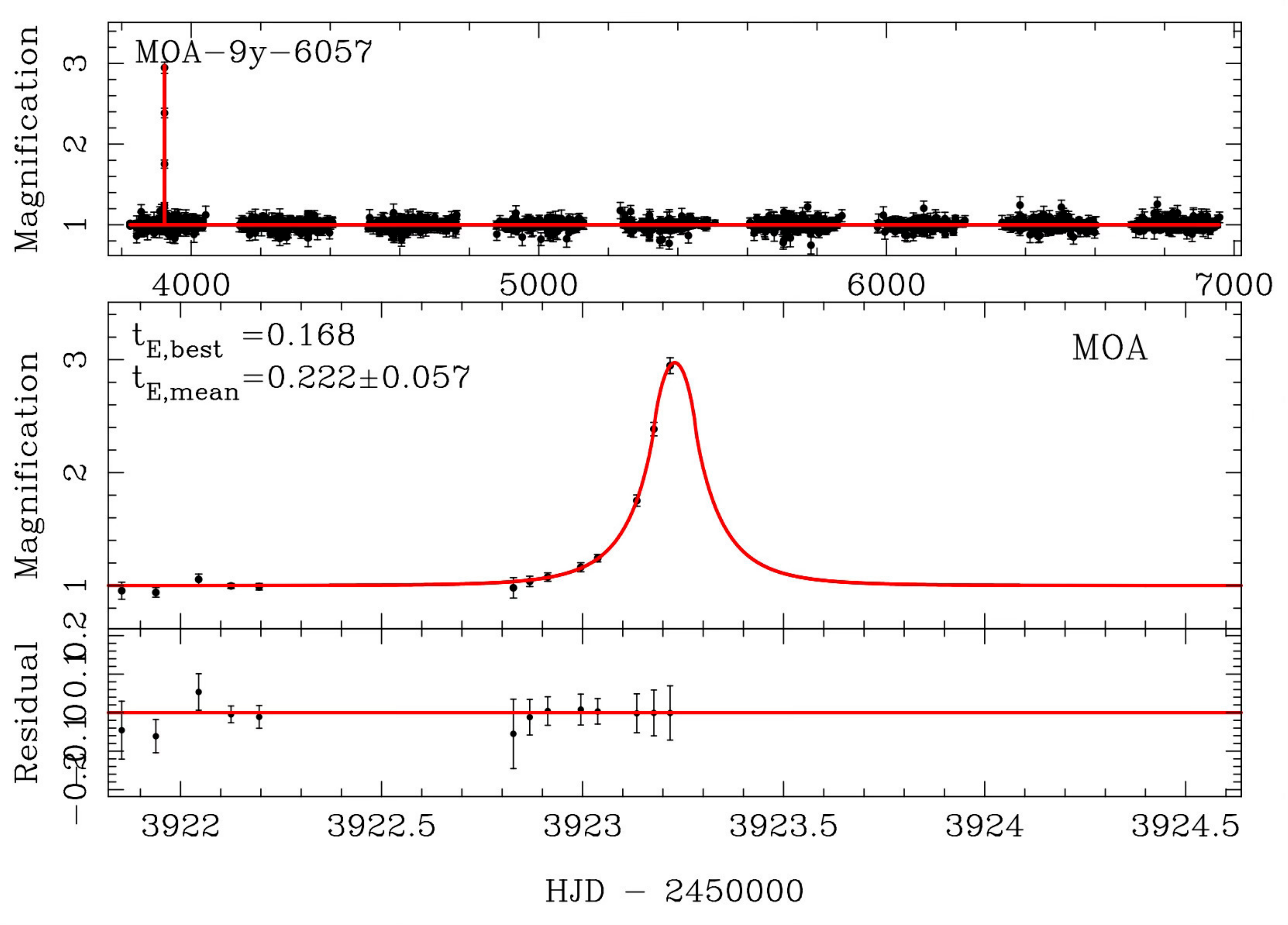}
\caption{
  \label{fig:lightcurve2}
  Continued.
  Light curves of the short timescale microlensing candidates with $t_{\rm E}< 1$ day continued.
}
\end{center}
\end{figure*}
\begin{figure*}
\begin{center}
\includegraphics[width=8.5cm,keepaspectratio]{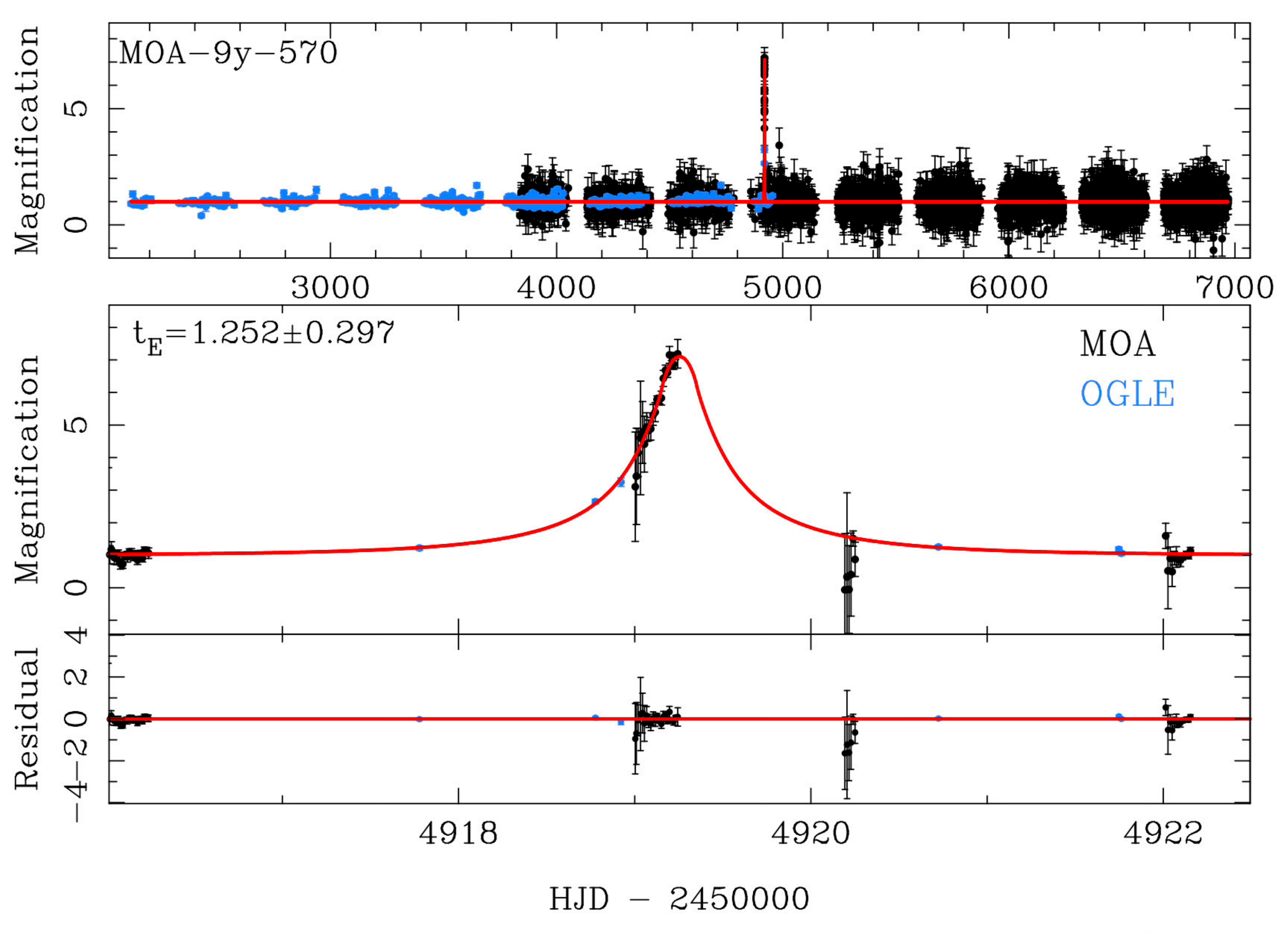}
\includegraphics[width=8.5cm,keepaspectratio]{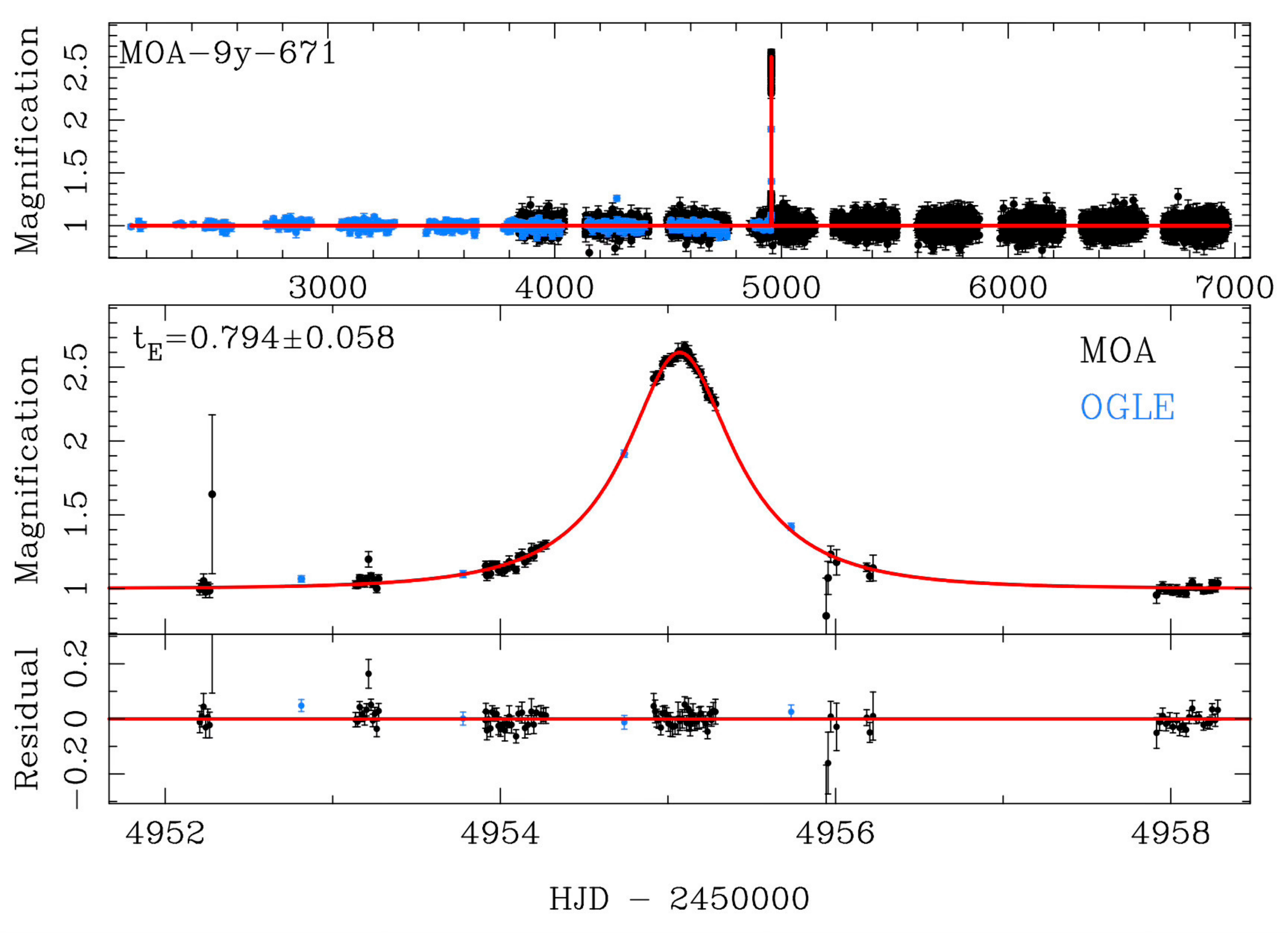}
\includegraphics[width=8.5cm,keepaspectratio]{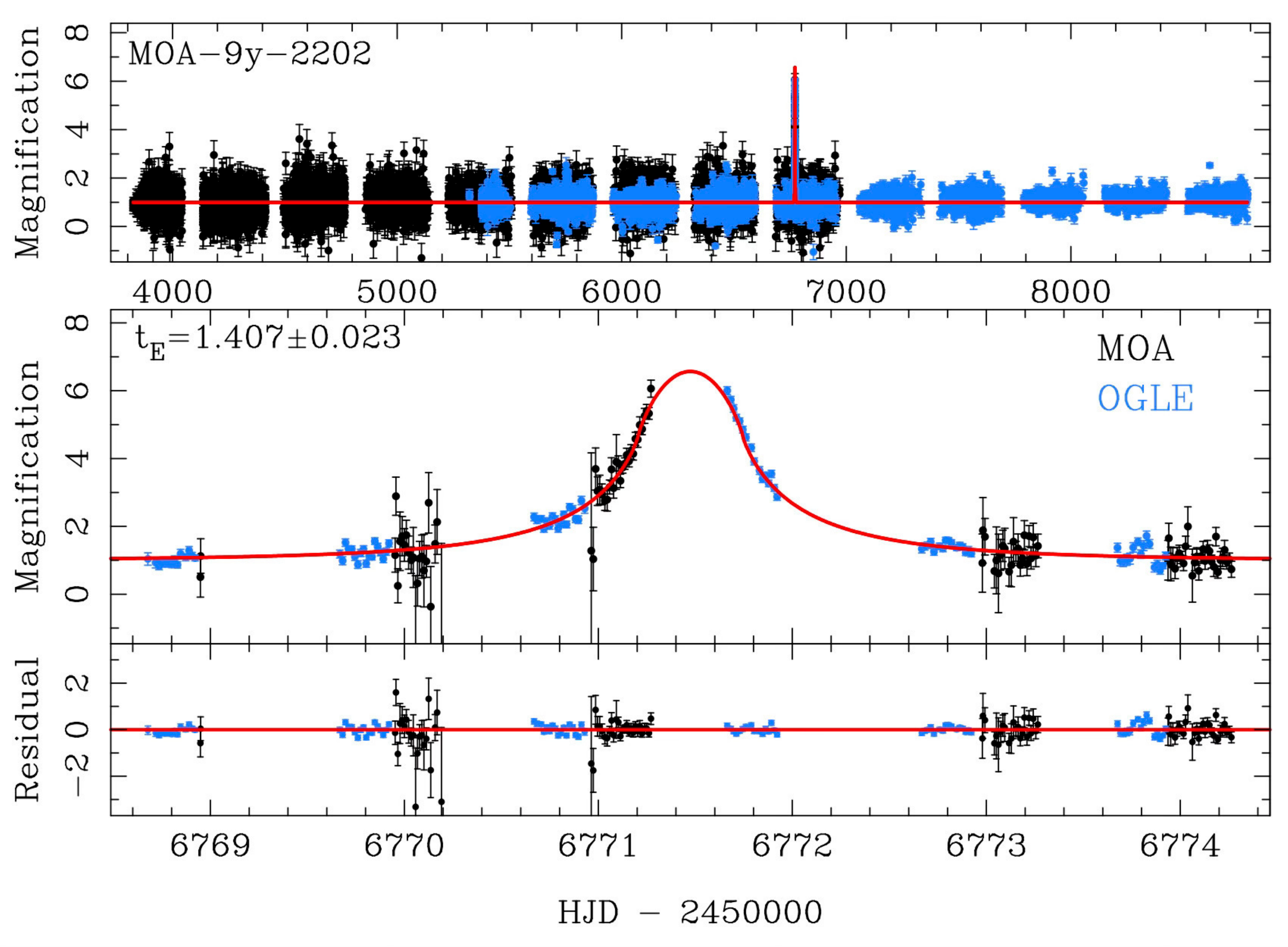}
\includegraphics[width=8.5cm,keepaspectratio]{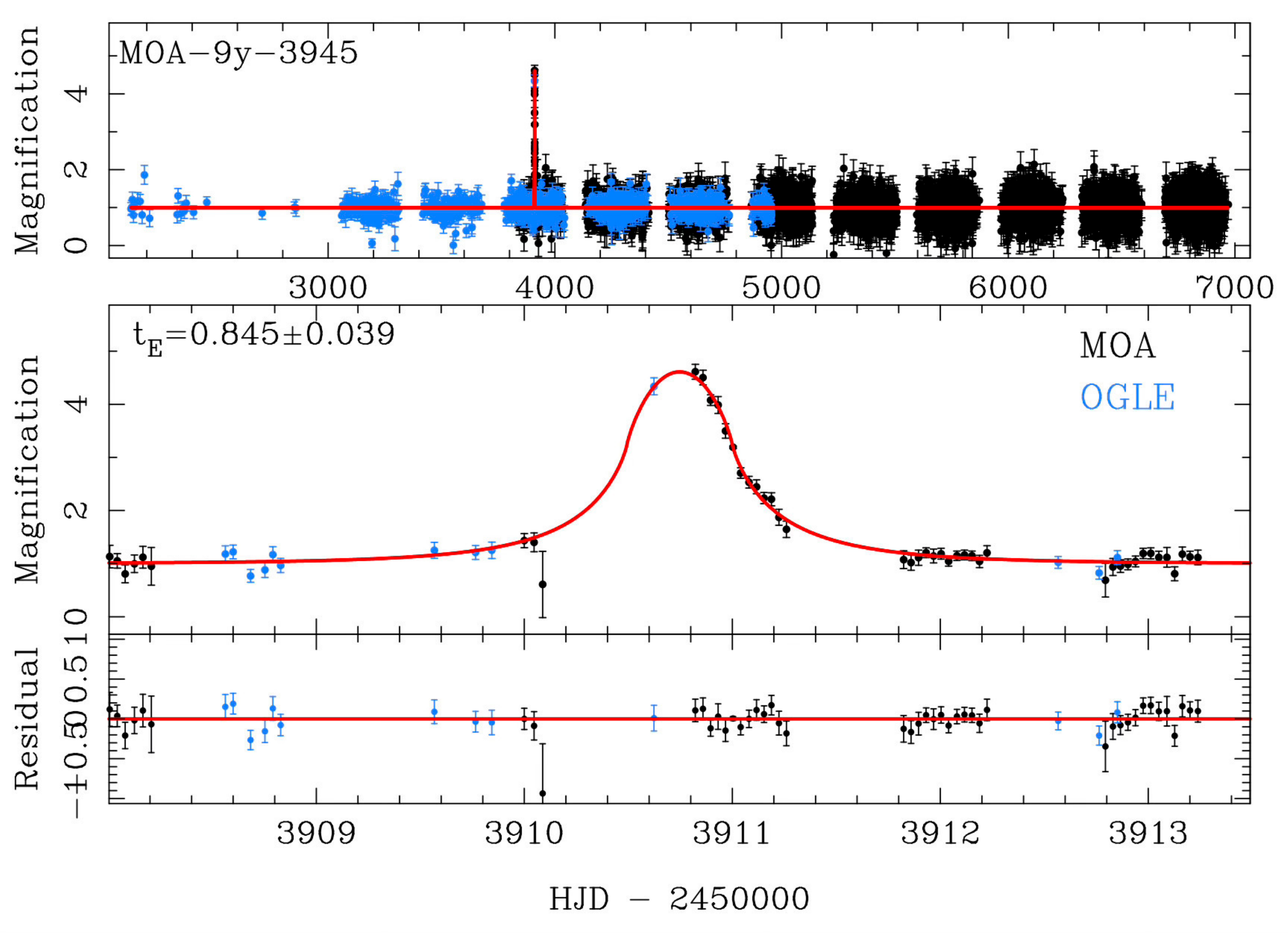}
\caption{
  \label{fig:lightcurveOGLE}
Light curves of the 4 short events with OGLE data (blue) during magnification 
(MOA-9y-570, MOA-9y-671, MOA-9y-2202 (OGLE-2014-BLG-0617), MOA-9y-3945).
The red lines indicate the best fit model by using both MOA and OGLE data.
}
\end{center}
\end{figure*}


\subsection{Parallax}
\label{sec:Parallax}

We found 66 candidates with likely microlensing parallax signals which is the long term distortion in 
a light curve due to the orbital motion of the Earth \citep{an02, smi02}.
Most of these events failed to pass the criteria due to the bad fit and/or unphysical parameters when fitting either PSPL or FSPL models.
However, 17 events with weak possible parallax signals survived in the final sample because
their signals are too weak to distinguish between PSPL and FSPL. 
These may not be even real parallax signals because long-term low-level systematics or small levels of source star variability
may resemble a parallax signal. Thus we classified these as uncertain parallax events.
Further careful analysis is needed to treat these parallax events for a statistical study 
on the long timescale events which possibly includes black hole lens events.

In this work, we included these uncertain parallax candidates in the final sample
because (1) these may not be parallax, (2) the effect on the $t_{\rm E}$ values is small,
(3) they do not affect the short timescale event distribution at all and 
(4) their number is negligible (only $<0.5$\%) even for long timescale events.

\subsection{binary}
\label{sec:binary}
We identified 581 binary lens candidates in all 6,111 microlensing candidates by visual inspection.  
Although this analysis is intended to sample only single lens microlensing events,
45 possible binary lens candidates remain among the final candidates.
These possible binary events have relatively weak signatures and it is difficult to distinguish these
from a noisy single lens event by using the numerical selection criteria.

The fraction of those possible binaries is relatively small, $<1.3$\%, compared to the total number of the sample.
The timescale of these binary candidates assuming a single lens model ranges over $11 < t_{\rm E} < 120$ days, 
where the number of other single lens events is large enough to neglect these binary candidates.
We confirmed that our final results for the mass function parameters presented in \citetalias{sumi2023} do not change when we include these 45 binary candidates.  
In the following analysis, we rejected these binary candidates.

\subsection{Final sample}
\label{sec:finalsample}
In order to determine the detection efficiency in our simulation, we need to  determine and correct for source star extinction and reddening \citep{sumi2011,sumi2013}.
We use red clump giants (RCGs) as standard candles for this purpose.  We determine the extinction and reddening towards each subfield in which the CMD shows a clear red clump, and correct for these effects in both our sample and simulation.
In the most of the subfields of gb6 and gb22, and some subfields in other fields, totaling about 12\% of all area, 
a clear RCG population could not be identified in the CMD.
We exclude fields gb6 and gb22 and any subfields of other fields in which RCG are not clearly identified.

After these relatively strict cuts, 3,554 and 3,535 objects remained as microlensing candidates after applying criteria CR1 and CR2, 
respectively, among the all visually identified 6,111 candidates.
We visually confirmed that there is no obvious non-microlensing events in the final sample.
Applying these strict criteria ensures that $t_{\rm E}$ is well constrained
for each event and that there is no significant contamination by
misclassified events. 
The number of all candidates $N_{\rm ev}$ and selected candidates by CR1, $N_{\rm ev, CR1}$, 
 and by CR2, $N_{\rm ev, CR2}$, in each field are listed in Table \ref{tbl:fld}.
The light curve data of all the 6111 events are publicly available via NASA Exoplanet Archive \footnote{\url{https://exoplanetarchive.ipac.caltech.edu/}}\citep{ake13}.

\section{Short events}
\label{sec:shortevents}

In the final sample, there are 12 and 10 short time scale events with $t_{\rm E} <1$ day for CR1 and CR2, respectively.
We show the light curves of these short events in Figure \ref{fig:lightcurve}.
The clear short single instantaneous magnification can be seen in the 9-year constant baseline.
The major source of false positives among the short events is flare stars and dwarf novae.
We confirmed that there is no other transient event in the 9-year baseline for all these candidates.
 The flares are usually associated with spotted stars showing a low amplitude and short periodic variability.
There is no periodic variability in the light curve of these objects, indicating that these are not spotted/flaring stars.

We found counterpart objects in the OGLE database for 11 short events. 
OGLE has 4, 3, 66, and 4 data points during the period of magnification in 4 short events,
MOA-9y-570, MOA-9y-671, MOA-9y-2202 (OGLE-2014-BLG-0617) and MOA-9y-3945, 
which confirmed the magnification found by MOA light curve as shown in Figure \ref{fig:lightcurveOGLE}.
For these four events, the combined fit with MOA and OGLE by MINUIT gives the following $t_{\rm E}$ values:
$t_{\rm E} = 1.252 \pm 0.297$\,days ($0.809 \pm 0.280$\,days),
$t_{\rm E} = 0.794 \pm 0.058$\,days  ($0.765 \pm 0.053$\,days),
$t_{\rm E} = 1.407 \pm 0.023$\,days  ($0.957 \pm 0.316$\,days), and
$t_{\rm E} = 0.845 \pm 0.039$\,days  ($0.737 \pm 0.225$\,days), respectively, where the mean and 
standard deviation of $t_{\rm E}$ from the MCMC calculations using only MOA data (see Section \ref{sec:ref_fit}) are 
shown in parenthesis.
Although adding the OGLE data changed the $t_{\rm E}$ values somewhat for MOA-9y-570 and MOA-9y-2202, the differences are not statistically significant. These two events have relatively large uncertainties for the $t_{\rm E}$ values due 
to the sparse coverage of the magnified part of the light curve. 
The $\chi^2$ for the MOA data in the best fit models increases by only 1.6 for MOA-9y-570 and 3.0 for MOA-9y-2202, when the OGLE data are added to the fit. These $\chi^2$ increases indicate relative probabilities of 0.45 and 0.22, respectively, assuming normal distributions of the data. This also indicates that the best fit joint MOA plus OGLE models are consistent with the MOA data alone. Note that MOA-9y-570 and MOA-9y-2202 are rejected by the stricter selection criteria CR2.
\citetalias{sumi2023} obtained consistent mass functions for both CR1 and CR2 samples, and whether these two events are included or not does not affect their conclusion.
We also confirmed that there is no flare nor periodic variability in the OGLE baseline during 7-10 years for all 11 candidates.

We list 12 short microlensing candidates with ID numbers, coordinates (R.A., Dec.)(2000), 
the corresponding MOA real-time Alert ID, I-band magnitude of Dophot catalog star and 
the number of data points in Table \ref{tbl:candlistShort1}.
The complete lists  including all the visually identified 6,111 candidates are available online.
Two of these short events show a finite source effect, as described and further analyzed in \S \ref{sec:FSPL}.

\begin{figure*}
\begin{center}
\includegraphics[angle=0,width=18cm,keepaspectratio]{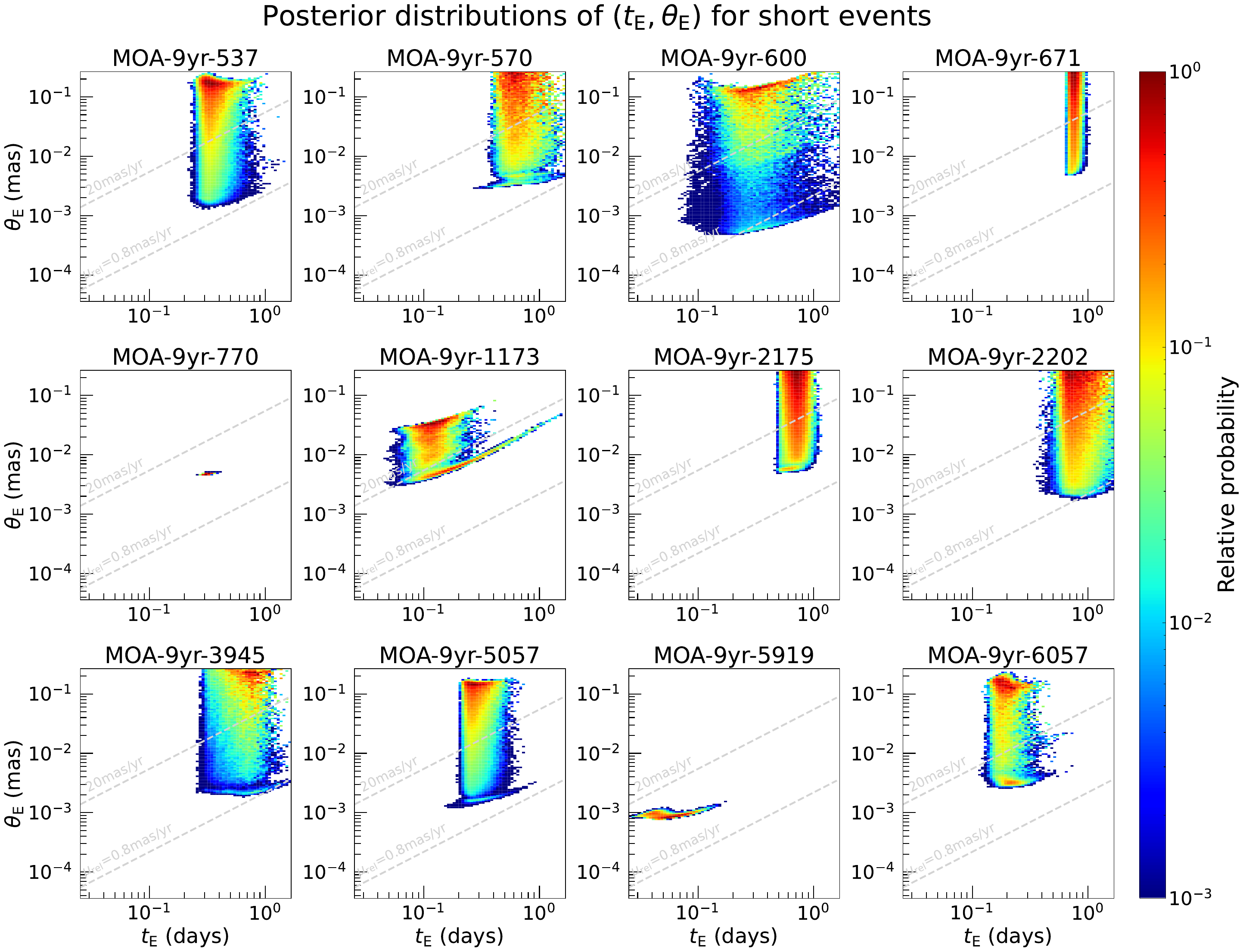}
\caption{
  \label{fig:MCMCtEthE}
Marginalized posterior distributions of $(t_{\rm E}, \theta_{\rm E})$ calculated using the MCMC method for the 12 short timescale events with the best-fit $t_{\rm E}$ values smaller than 1 day. The prior was corrected so that it becomes uniform in ($\log t_{\rm E}$, $\log \theta_{\rm E}$) although the MCMC was performed with the uniform prior in ($1/t_{\rm E}$, $t_*$). The two gray dashed lines indicate relative proper motion values of $\mu_{\rm rel} = 0.8$ mas/yr and $\mu_{\rm rel} = 20$ mas/yr, between which most events reside considering the structure of our Galaxy.
}
\end{center}
\end{figure*}

\subsection{Short events in 2006-2007 sample}
\label{sec:shortevents_sumi2011}

\cite{sumi2011} reported 10 short events with $t_{\rm E}<2$ days using their 2006-2007 dataset which is included in this work.
Only 5 of those events survived CR1 in this work because the fitting results  have changed due to the re-reduction of the images and light curves,
especially, the de-correlation to remove color-dependent differential refraction systematic errors.
Note that all events except ip-1 and ip-6 had only modest changes in parameters with new $t_{\rm E}$ values within 1-2 $\sigma$ of the values in \cite{sumi2011}. The best-fit $t_{\rm E}$ values decreased for ip-3, ip-4, ip-5 and ip-9, and increased for ip-2, ip-7, ip-8 and ip-10.
Two events (ip-1, ip-6) failed to pass our criteria because they have too faint best-fit source magnitudes with the new light curve data to meet our requirement of $I_s \leq 21.4$.




Two events (ip-2, ip-7) were excluded owing to their large values of $\sigma_{t_{\rm E}}/t_{\rm E}$.
One event (ip-3) failed to pass due to its impact parameter, $u_0=1.01$, exceeding the threshold value of 1.0 for $u_0$.
Furthermore, the event (ip-5) failed to pass the  CR2  criteria where $\sigma_{t_{\rm E}}/t_{\rm E}>0.5$ due to its sightly smaller $t_{\rm E}$ and larger $\sigma_{t_{\rm E}}$.

On the other hand, two events, MOA-9y-6057 ($t_{\rm E}=0.17$ days) and MOA-9y-3036 ($t_{\rm E}=1.52$ days) are newly found with $t_{\rm E}<2$ days in the 2006-2007 dataset. 
In total, the 10 events of \citet{sumi2011} decreased to 7 and 6 events after applying CR1 and CR2, respectively, using our new dataset.
As a result, the excess in the $t_{\rm E}$ distribution at $t_{\rm E}=0.5-2$ days is not significant anymore, however, an even shorter event MOA-9y-6057 is added.

\subsection{Refined FSPL fit for short events} \label{sec:ref_fit}

We refined the FSPL fits for 12 short timescale events 
with updated limb darkening parameters $u_{\rm MOA,Red}$ (see section \ref{sec:FSPL}) 
using the image centered, ray-shooting light curve modeling 
method of \citet{bennett96} and \citet{bennett-himag}, which is now known as 
\texttt{eesunhong}\footnote{\url{https://github.com/golmschenk/eesunhong}}, after co-author
Sun Hong Rhie \citep{rhie_obit}.
This modeling effort was conducted on the full light curve parameter set to find
the refined best fit models and the posterior ($t_{\rm E}$, $\theta_{\rm E}$) probability distribution for each event 
using the Markov Chain Monte Carlo (MCMC) method \citep{metrop}.
This is done because the ($t_{\rm E}$, $\theta_{\rm E}$) probability distributions for the short events
are needed to determine the mass function of planetary mass objects, 
which is the aim of our companion paper \citetalias{sumi2023}.

Although FSPL models are adopted for only two short events in the cut-3 of the event selection, 
we conduct FSPL fitting for all 12 events, because we need to determine which $\theta_{\rm E}$ values are consistent
with each short event. 
Even for events without significant finite source effects, we can put an upper limit on the source size parameter $\rho$, which corresponds to a lower limit on $\theta_{\rm E}$. This is especially useful for short timescale events, because they 
very sensitive to finite source effects.
In fact, Figure \ref{fig:MCMCtEthE} indicates
that an important range of $\theta_{\rm E}$ values can be excluded for MOA-9y-537, MOA-9y-570, MOA-9y-671,
MOA-9y-2175, MOA-9y-5057, MOA-9y-6057, and especially MOA-9y-1173.
If we ignored the fact that small $\theta_{\rm E}$ values are excluded for these events, it would bias our results toward
small $\theta_{\rm E}$ by including the $\theta_{\rm E}$ measurements for events MOA-9y-770 and MOA-9y-5919.
Thus, we must use the constraints on $\theta_{\rm E}$ for all the short events, even though most of them have large
uncertainties.

 In our light curve modeling, we constrained the source star to be fainter than the brightness of
 the catalog star ($I_{\rm c}$) at the position of the event, with a flux equivalent to $I_{\rm bk} = 19.0 \pm 0.3$ added to
 represent the unresolved stellar background.
 Table \ref{tbl:candlistShort1} lists the catalog star magnitudes, $I_{\rm c}$.

The error bars in the light curve data were re-normalized to give $\chi^2/{\rm dof} \sim 1$ for 
the best fit model, in order to improve the error estimate for the light curve parameters.
The MCMC calculations were conducted with uniform priors in $1/t_{\rm E}$ and the source radius crossing 
time, $t_*$, which is used instead of $\rho = t_*/ t_{\rm E}$ by the \citet{bennett-himag} modeling code because $t_*$
is usually more tightly constrained by the light curve data than $\rho$.
Since most of the events do not show a significant finite source effect, their allowed $\theta_{\rm E}$ can range over 
an order of magnitude or more, with limits imposed priors on proper motion and the lack of significant finite source effects.
In such cases, the prior distributions assumed for $\theta_{\rm E}$ and $t_{\rm E}$ can be important. So, we 
apply a uniform prior in $(\log t_{\rm E}, \log \theta_{\rm E})$, and using the following conversion.
Let $p_0 (x)$ be the prior probability density function of a parameter $x$, 
then the prior of $(\log t_{\rm E}, \log \theta_{\rm E})$ is given by 
\begin{align}
p_0(\log t_{\rm E}, \log \theta_{\rm E}) &= p_0(1/t_{\rm E}, t_*) \left|\frac{\partial (1/t_{\rm E}, t_*)}{\partial (\log t_{\rm E}, \log \theta_{\rm E})}\right| \notag\\
&\simeq p_0(1/t_{\rm E}, t_*) \, \rho \, (\ln 10)^2 ,
\label{eq:rho0}
\end{align}
where we assumed that $\theta_*$ depends almost solely on the source brightness, which depends on
$\log t_{\rm E}$ with little dependence on $\log \theta_{\rm E}$. This yields the approximation on the second
line of equation~\ref{eq:rho0}.
This assumption was confirmed to be reasonable for the short events by examining correlations between the parameters 
in the posterior distributions.
Because $p_0(1/t_{\rm E}, t_*)$ is constant, the prior can be converted so that it becomes uniform in $(\log t_{\rm E}, \log \theta_{\rm E})$ by weighting each MCMC link by $1/\rho$. All the MCMC results presented in this paper have used this conversion.

Table \ref{tbl:candlistShort} shows the refined FSPL fit results for the short timescale events, and Figure \ref{fig:MCMCtEthE} shows the resulted posterior distribution of ($t_{\rm E}$, $\theta_{\rm E}$) for each event marginalized over the other light curve parameters. 
The figure shows that in most events, the areas of high posterior probability are located at $\mu_{\rm rel} > 20$ mas/yr, which is unlikely considering the structure of our Galaxy \citep[e.g., see Fig. 1 of][]{kos21b}. 
Therefore, we additionally applied a prior of $0.8 < \mu_{\rm rel}/({\rm mas \, yr^{-1}}) < 20$ to derive the mean and standard deviation values shown in Table \ref{tbl:candlistShort}.
Combining the restrictions from the light curve and the $\mu_{\rm rel}$ prior enabled us to determine the $\rho$ value for MOA-9y-1173 moderately well.
The use of these sharp $\mu_{\rm rel}$ cuts is a rather crude way of imposing a prior distribution on $\mu_{\rm rel}$, since the true $\mu_{\rm rel}$ distribution is a smooth function. So, our companion paper \citetalias{sumi2023} applies the Galactic prior on the $\mu_{\rm rel}$ distribution based on the Galactic model of \citet{Koshimoto2021}.

For MOA-9y-570, MOA-9y-600, MOA-9y-1173, and MOA-9y-2202, the mean and standard deviation of $t_{\rm E}$ in Table \ref{tbl:candlistShort} appear to be inconsistent with the classification of the event selection, i.e., CR1 or CR2,  in Table \ref{tbl:candlistShort1} in terms of the  criteria on $\sigma_{t_{\rm E}}/t_{\rm E}$.
This is because of differences between the refined fits and the fits used for our selection process. 
The refined fits used the FSPL model for all short events whereas the PSPL model fits were adopted during the event selection
unless finite source effects are significantly detected. During the selection process, the parameter errors are determined
using the MINOS procedure of
the MINUIT package, except in cases where MINOS failed. In those cases, the error bars from the MIGRAD procedure were used.
The refined fits used the more robust MCMC method to determine the error bars.
The photometric error bars for each light curve were re-normalized to give $\chi^2/{\rm dof} \sim 1$ for the refined fits. 
The constraints on the source magnitude from the catalog star magnitude and the 
$\mu_{\rm rel}$ cut were applied only to the refined fits. 
The selection process models are used to define the selected sample of events and to determine the detection efficiencies.
The refined fits are needed to determine the range of $t_{\rm E}$ and $\theta_{\rm E}$ values that are consistent with the 
data in order to determine the constraints that the data impose upon the FFP population.

\section{FSPL events}
\label{sec:FSPL}
There are 13 FSPL candidates in the final sample as listed in Table \ref{tbl:candlistFS1}, including two short events MOA-9y-770 and MOA-9y-5919.
The original FSPL fit during the event selection was done with a tentative limb darkening 
coefficient of $u_{\rm MOA\text{-}Red}=0.566$ which corresponds to a G2 type star in the MOA-Red wide band \citep{Claret2011}. 
For the selected events, we fit the light curves again with updated limb darkening coefficients 
estimated by taking the source color into account with the procedures described below.
Note that the changes in the parameters are negligible. 

The final best-fit parameters are shown in Table \ref{tbl:candlistFS2}.
In this sample, we estimated the angular Einstein radius, $\theta_{\rm E}$, and
the lens-source relative proper motion,  $\mu_{\rm rel}$, as follows.

\begin{deluxetable*}{llcccrrrccc}
\tabletypesize{\scriptsize}
\tablecaption{FSPL candidates with RA, Dec., Alerted ID, catalog star's
$I$-band magnitude, number of data points and passed criteria.
\label{tbl:candlistFS1}
}
\tablewidth{0pt}
\tablehead{
\colhead{ID} &
\colhead{internal ID} &
\colhead{R.A.} & 
\colhead{Dec.} &
\colhead{ID$_{\rm alert}$} &
\colhead{$I_{\rm c}$} &
\colhead{$N_{\rm data}$} &
\colhead{criteria} \\
\colhead{} &
\colhead{(field-chip-sub-ID)} &
\colhead{(2000)} & 
\colhead{(2000)} &
\colhead{} &
\colhead{(mag)} &
\colhead{} &
\colhead{} 
}
\startdata
  MOA-9y-81 &   gb1-3-2-117560 & 17:46:17.838 & -34:20:24.70  & 2011-BLG-093 &  16.60 $\pm$  0.02 & 10239 & CR2\\
 MOA-9y-707 &   gb3-5-5-398397 & 17:52:07.344 & -33:24:19.91  & 2013-BLG-611 &  18.27 $\pm$  0.04 & 20495 & CR2\\
 MOA-9y-770 &    gb3-7-6-65303 & 17:55:16.892 & -33:08:35.69  &           -- &  16.00 $\pm$  0.01 & 20438 & CR2\\
MOA-9y-1117 &   gb4-4-4-329819 & 17:50:55.994 & -31:19:39.12  & 2014-BLG-425 &    ---  & 21568 & CR2\\
MOA-9y-1248 &    gb4-7-3-59884 & 17:54:14.854 & -31:11:02.67  & 2007-BLG-233 &  16.51 $\pm$  0.02 & 20801 & CR2\\
MOA-9y-1772 &   gb5-4-3-477919 & 17:53:58.399 & -29:44:56.05  & 2009-BLG-411 &  15.82 $\pm$  0.02 &  9464 & CR2\\
MOA-9y-2881 &   gb8-5-4-211663 & 17:57:01.624 & -31:38:42.64  & 2013-BLG-145 &    ---  &  9061 & CR2\\
MOA-9y-3312 &   gb9-4-0-331071 & 17:57:08.881 & -29:44:58.28  & 2010-BLG-523 &  17.10 $\pm$  0.06 & 28577 & CR2\\
MOA-9y-3430 &    gb9-5-5-58496 & 17:57:47.616 & -29:50:46.67  &           -- &  18.28 $\pm$  0.07 & 29000 & CR2\\
MOA-9y-3888 &   gb10-4-1-78451 & 17:58:29.239 & -27:59:21.90  & 2008-BLG-241 &  17.28 $\pm$  0.10 & 19343 & CR2\\
MOA-9y-5175 &   gb15-3-2-26189 & 18:05:00.407 & -25:47:03.72  & 2007-BLG-176 &  17.85 $\pm$  0.04 &  6523 & CR2\\
MOA-9y-5238 &   gb15-7-0-92708 & 18:08:49.977 & -25:57:04.30  & 2010-BLG-311 &  19.25 $\pm$  0.05 &  6148 & CR2\\
MOA-9y-5919 &   gb19-7-7-39836 & 18:18:41.318 & -25:57:15.65  &           -- &  17.07 $\pm$  0.01 &  4940 & CR2\\
\hline
MOA-9y-1944\tablenotemark{$a$} & gb5-6-0-416936 & 17:56:25.942 & -29:54:04.90  & 2012-BLG-403 &  17.68 $\pm$  0.03 & 35613 & ---\\
\enddata
\tablenotetext{a}{MOA-9y-1944 is not in the final sample. $N_{\rm data}$ includes 4632 OGLE data points.}
\tablecomments{The list of all microlensing event candidates is available in the electronic version.
\\}
\end{deluxetable*}

\begin{deluxetable*}{lrrrrrrr}
\tabletypesize{\scriptsize}
\tablecaption{Parameters for FSPL events.
\label{tbl:candlistFS2}
}
\tablewidth{0pt}  
\tablehead{
\colhead{ID} &
\colhead{$t_0$} &  
\colhead{$t_{\rm E}$} & 
\colhead{$u_0$} & 
\colhead{$\rho$} &
\colhead{$I_{\rm s}$} &
\colhead{$\chi^2$} &
\colhead{$\Delta\chi^2$\tablenotemark{a}}  \\
\colhead{} & 
\colhead{$(\rm HJD')$} &   
\colhead{$(\rm day)$} &  
\colhead{} & 
\colhead{} & 
\colhead{${(\rm mag)}$} &  
\colhead{}  &
\colhead{}   
}
\startdata 
  MOA-9y-81                  &     5678.5543 &   15.001 $\pm$  0.032 & 0.028253 $\pm$ 0.000102 &  0.05349 $\pm$ 0.00013 &    16.59 &     7807 &  22713.8\\
 MOA-9y-707                  &     6536.7334 &   21.226 $\pm$  0.442 & 0.002520 $\pm$ 0.000104 &  0.00568 $\pm$ 0.00011 &    21.08 &    16207 &    181.2\\
 MOA-9y-770\tablenotemark{b} &     4647.0426 &    0.315 $\pm$  0.017 & 0.207823 $\pm$ 0.130471 &  1.08449 $\pm$ 0.07021 &    16.17 &    21686 &    525.8\\
MOA-9y-1117                  &     6887.5787 &   60.202 $\pm$  0.394 & 0.008510 $\pm$ 0.000355 &  0.00937 $\pm$ 0.00064 &    18.61 &    21302 &     23.6\\
MOA-9y-1248                  &     4289.2592 &   15.279 $\pm$  0.063 & 0.000040 $\pm$ 0.012629 &  0.03669 $\pm$ 0.00113 &    16.45 &    23226 &     97.1\\
MOA-9y-1772                  &     5052.5466 &   10.551 $\pm$  0.089 & 0.002064 $\pm$ 0.005415 &  0.02805 $\pm$ 0.00008 &    15.82 &     4050 &     61.6\\
MOA-9y-2881                  &     6367.0260 &    8.796 $\pm$  0.164 & 0.004062 $\pm$ 0.000491 &  0.00683 $\pm$ 0.00046 &    19.09 &    11721 &     94.4\\
MOA-9y-3312                  &     5432.6404 &   17.385 $\pm$  0.421 & 0.000985 $\pm$ 0.004596 &  0.00976 $\pm$ 0.00068 &    19.27 &    38055 &    649.3\\
MOA-9y-3430                  &     3951.9865 &   14.988 $\pm$  0.141 & 0.000502 $\pm$ 0.000045 &  0.00296 $\pm$ 0.00002 &    21.13 &    24996 &   4103.3\\
MOA-9y-3888                  &     4632.5647 &   16.748 $\pm$  0.098 & 0.000004 $\pm$ 0.000696 &  0.02049 $\pm$ 0.00049 &    17.53 &    26129 &    856.0\\
MOA-9y-5175                  &     4245.0575 &    9.090 $\pm$  0.116 & 0.025382 $\pm$ 0.002420 &  0.05556 $\pm$ 0.00097 &    17.92 &    16097 &   3710.8\\
MOA-9y-5238                  &     5365.1979 &   21.801 $\pm$  0.270 & 0.001350 $\pm$ 0.000019 &  0.00245 $\pm$ 0.00003 &    19.47 &     5290 &    890.2\\
MOA-9y-5919\tablenotemark{b} &     4601.0921 &    0.057 $\pm$  0.016 & 0.572225 $\pm$ 0.435984 &  1.39874 $\pm$ 0.45997 &    18.58 &     3729 &     35.0\\
\hline
MOA-9y-1944\tablenotemark{c} &    6098.0974  &   1.594  $\pm$  0.136 & 0.002866 $\pm$ 0.004371 &  0.00928 $\pm$ 0.00032 &    21.91 &     53693 & 194.0\\
 \enddata
\tablenotetext{a}{$\Delta\chi^2=\chi^2 - \chi^2_{\rm FS}$.}
\tablenotetext{b}{From the refined fits by MCMC described in Section \ref{sec:ref_fit}.}
\tablenotetext{c}{MOA-9y-1944 is not in the final sample. The values show the fitting results by using MOA and OGLE light curves.}
\tablecomments{The list of all microlensing event candidates is available in the electronic version.}
\end{deluxetable*} 

\begin{figure}
\begin{center}
\includegraphics[angle=0,width=8cm,keepaspectratio]{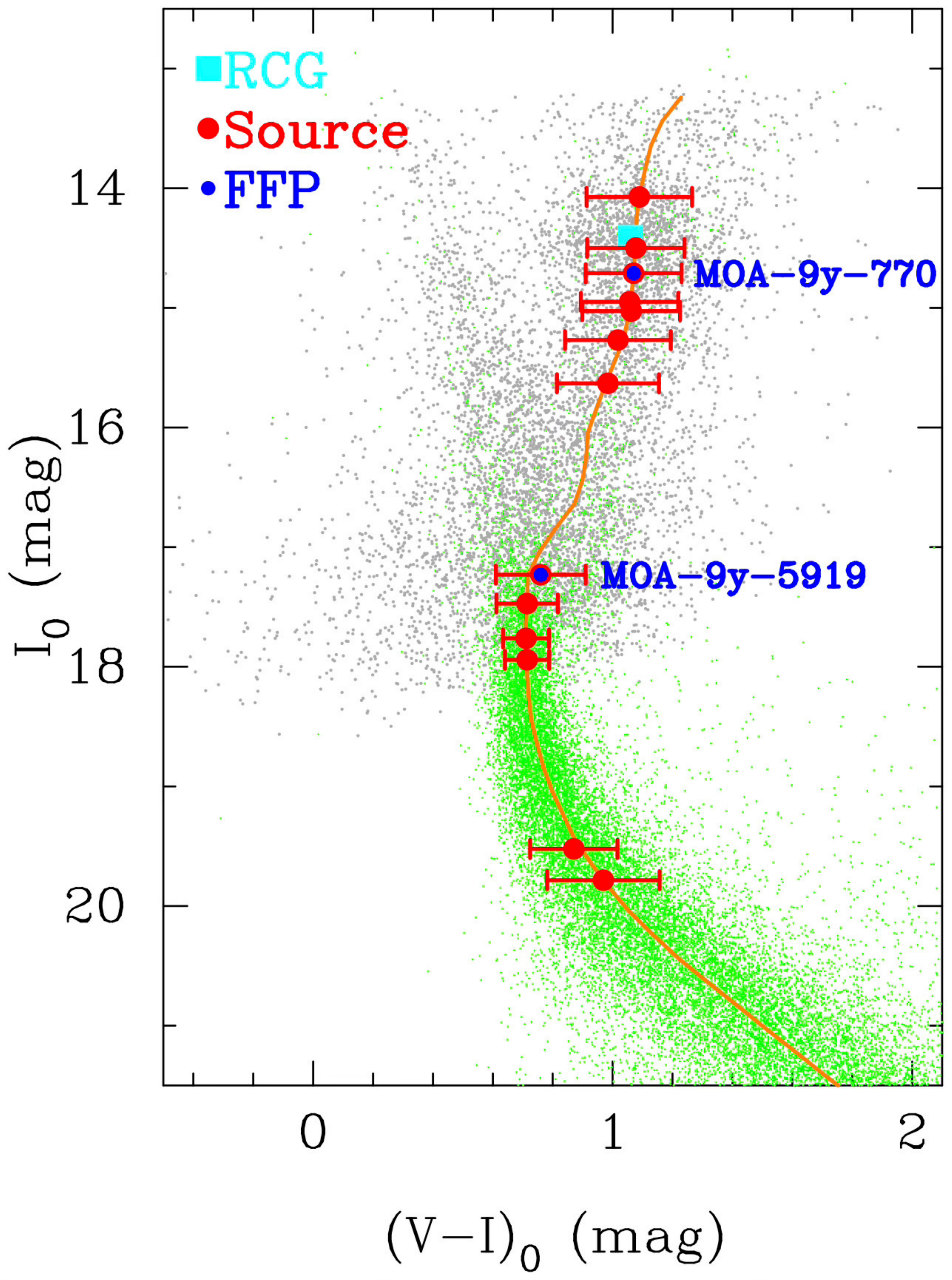}
\caption{
  \label{fig:CMDLDall}
The extinction-free CMD of a MOA subfield gb13-5-4, which combines the MOA data (black dots) and the HST data by \citet{hol98} (green dots). The orange curve is the isochrone matched to this subfield.
The cyan square is the RCG centroid. The red-filled circles with error bars are sources of the 13 FSPL events.
The blue-filled circles indicate the 2 FFP candidates.
}
\end{center}
\end{figure}

Because we do not have $V$-band observations during times of event magnification, 
we estimated the color of the sources by assuming 
the sources are main sequence stars or giants in the bulge, which 
is most likely correct.
Firstly, we determined an $I$ vs $(V-I)$ ``isochrone" sequence on the CMD at Baade's window (or MOA subfield gb13-5-4) that combines MOA's data for bright stars 
and {\it Hubble Space Telescope} (HST)'s data \citep{hol98} for faint stars, as shown in Figure \ref{fig:CMDLDall}.
The isochrone is corrected for extinction and reddening by comparing 
the RCG's positions on the CMD and de-reddened apparent magnitude of the RCG 
using the relation $I_{\rm RC,0}=14.3955-0.0239\times I+0.0122 \times |b|$ 
\citep[Eq. 2 of][]{Nataf2016} and the intrinsic color of $(V-I)_{\rm RC,0}  = 1.06$ \citep{Bensby2011,Bensby2013,Nataf2016}. 
The isochrone is shifted for each given sub-field using the difference of $I_{\rm RC,0}$ (i.e., the difference of distance modulus) from the Baade's window value.

The reddening free source colors $(V-I)_{\rm s,0}$ are estimated from 
the best fit extinction free source magnitude $I_{\rm s,0}$ by using the isochrone. 
The determined source positions are plotted together on the extinction-free CMD of gb13-5-4 in Figure \ref{fig:CMDLDall}.
The errors of $(V-I)_{\rm s,0}$ are defined by the standard deviation of 
the $(V-I)_{\rm 0}$ color of stars at given magnitude $I_{\rm s,0} \pm 0.5$ mag in the MOA+HST CMD.

The source angular radius $\theta_{*}$ is calculated by using the relation between
the limb-darkened stellar angular diameter, $\theta_{\rm LD}$, $(V-I)$ and $I$  (private communication, \citealt{Boyajian2014}, see \citealt{Fukui2015}).
Then, we estimated $\theta_{\rm E}=\theta_{*} /\rho$ and $\mu_{\rm rel}=\theta_{\rm E}/t_{\rm E}$.

The $(V-I,I)_{\rm s,0}$ and source angular radius $\theta_*$,  angular Eistein radius $\theta_{\rm E}$, 
relative lens-source proper motion $\mu_{\rm rel}$,  effective temperature  of source $T_{\rm eff}$, 
and limb darkening coefficient $u_{\rm MOA\text{-}Red}$ for 13 FSPL events are shown in  Table \ref{tbl:candlistFS3}.

\begin{deluxetable*}{lrrrrrrr}
\tabletypesize{\scriptsize}
\tablecaption{Derived parameters for FSPL events.
\label{tbl:candlistFS3}
}
\tablewidth{0pt}
\tablehead{ 
\colhead{ID} &
\colhead{$I_{\rm s,0}$} &  
\colhead{$(V-I)_{\rm s,0}$} &  
\colhead{$\theta_*$} & 
\colhead{$\theta_{\rm E}$} &
\colhead{$\mu_{\rm rel}$} & 
\colhead{$T_{\rm eff}$} &
\colhead{$u_{\rm MOA\text{-}Red}$}  \\
\colhead{} & 
\colhead{(mag)} &  
\colhead{(mag)} & 
\colhead{($\mu$as)} & 
\colhead{($\mu$as)} &
\colhead{(masyr$^{-1}$)} &
\colhead{(K)} & 
\colhead{}  
}
\startdata
  MOA-9y-81                  &      15.03 $\pm$   0.05 &   1.06 $\pm$   0.16 &   4.37 $\pm$   0.70 &  81.74 $\pm$  13.05 &     1.99 $\pm$   0.32 &   4771 $\pm$  298 & 0.6286\\
 MOA-9y-707                  &      19.52 $\pm$   0.05 &   0.87 $\pm$   0.15 &   0.46 $\pm$   0.07 &  80.61 $\pm$  11.64 &     1.39 $\pm$   0.20 &   5183 $\pm$  376 & 0.5866\\
 MOA-9y-770\tablenotemark{a} &      14.71 $\pm$   0.13 &   1.07 $\pm$   0.16 &   5.13 $\pm$   0.86 &   4.73 $\pm$   0.75 &     5.50 $\pm$   0.90 &   4753 $\pm$  290 & 0.6286\\ 
MOA-9y-1117                  &      15.63 $\pm$   0.06 &   0.98 $\pm$   0.17 &   3.07 $\pm$   0.51 & 327.82 $\pm$  58.98 &     1.99 $\pm$   0.36 &   4921 $\pm$  358 & 0.6044\\
MOA-9y-1248                  &      14.50 $\pm$   0.05 &   1.08 $\pm$   0.16 &   5.66 $\pm$   0.90 & 154.25 $\pm$  24.96 &     3.69 $\pm$   0.60 &   4743 $\pm$  288 & 0.6286\\
MOA-9y-1772                  &      14.07 $\pm$   0.05 &   1.09 $\pm$   0.18 &   6.97 $\pm$   1.20 & 248.32 $\pm$  42.65 &     8.60 $\pm$   1.48 &   4723 $\pm$  307 & 0.6286\\
MOA-9y-2881                  &      17.47 $\pm$   0.05 &   0.71 $\pm$   0.10 &   1.01 $\pm$   0.10 & 148.48 $\pm$  18.18 &     6.17 $\pm$   0.76 &   5643 $\pm$  352 & 0.5364\\
MOA-9y-3312                  &      17.76 $\pm$   0.05 &   0.71 $\pm$   0.08 &   0.88 $\pm$   0.07 &  90.53 $\pm$   9.49 &     1.90 $\pm$   0.20 &   5655 $\pm$  266 & 0.5364\\
MOA-9y-3430                  &      19.78 $\pm$   0.05 &   0.97 $\pm$   0.19 &   0.45 $\pm$   0.08 & 150.93 $\pm$  27.65 &     3.68 $\pm$   0.67 &   4953 $\pm$  407 & 0.6105\\
MOA-9y-3888                  &      14.95 $\pm$   0.05 &   1.06 $\pm$   0.16 &   4.51 $\pm$   0.72 & 219.95 $\pm$  35.35 &     4.80 $\pm$   0.77 &   4781 $\pm$  299 & 0.6286\\
MOA-9y-5175                  &      15.27 $\pm$   0.08 &   1.02 $\pm$   0.18 &   3.75 $\pm$   0.65 &  67.42 $\pm$  11.84 &     2.71 $\pm$   0.48 &   4854 $\pm$  349 & 0.6300\\
MOA-9y-5238                  &      17.94 $\pm$   0.06 &   0.71 $\pm$   0.07 &   0.82 $\pm$   0.06 & 332.54 $\pm$  25.69 &     5.57 $\pm$   0.44 &   5643 $\pm$  253 & 0.5364\\
MOA-9y-5919\tablenotemark{a} &      17.23 $\pm$   0.61 &   0.76 $\pm$   0.15 &   1.26 $\pm$   0.48 &   0.90 $\pm$   0.14 &     6.15 $\pm$   1.83 &   5499 $\pm$  446 & 0.5552\\ 
\hline
MOA-9y-1944\tablenotemark{$b$} &      20.14   $\pm$ 0.05 &    1.09 $\pm$  0.23 &   0.43  $\pm$ 0.10  &   46.10 $\pm$  10.50&  10.57 $\pm$   2.57 &   4715  $\pm$   405 & 0.6328\\
 \enddata
\tablenotetext{a}{From the refined fits by MCMC described in Section \ref{sec:ref_fit}.}
\tablenotetext{b}{MOA-9y-1944 is not in the final sample.}
\end{deluxetable*}

\begin{figure}
\begin{center}
\includegraphics[width=8.2cm,clip,keepaspectratio]{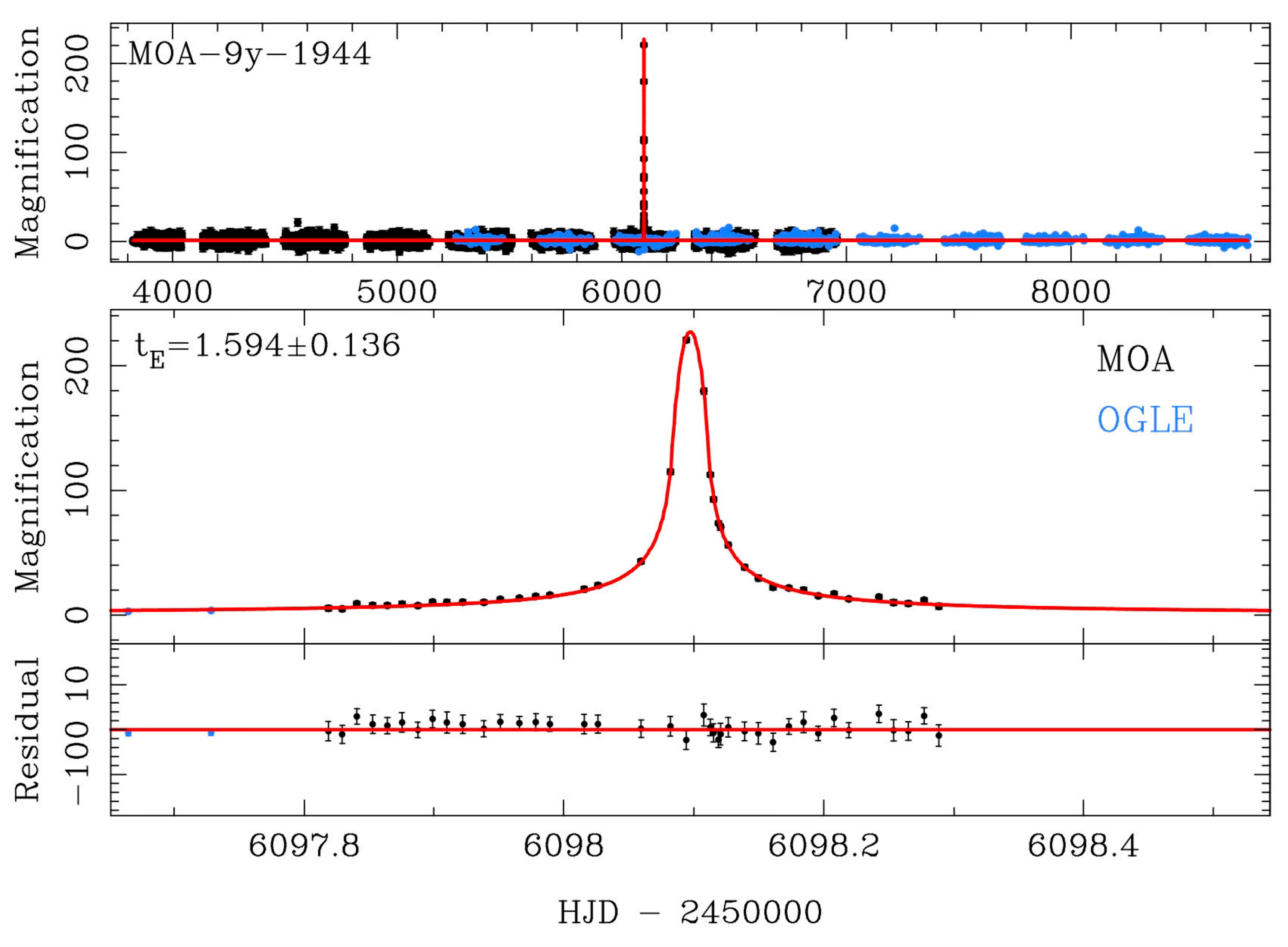}
\caption{
  \label{fig:lightcurveBD}
Light curves of the BD candidate MOA-9y-1944  
}
\end{center}
\end{figure}

\begin{figure}[htb]
\centering
  \includegraphics[scale=0.28,keepaspectratio]{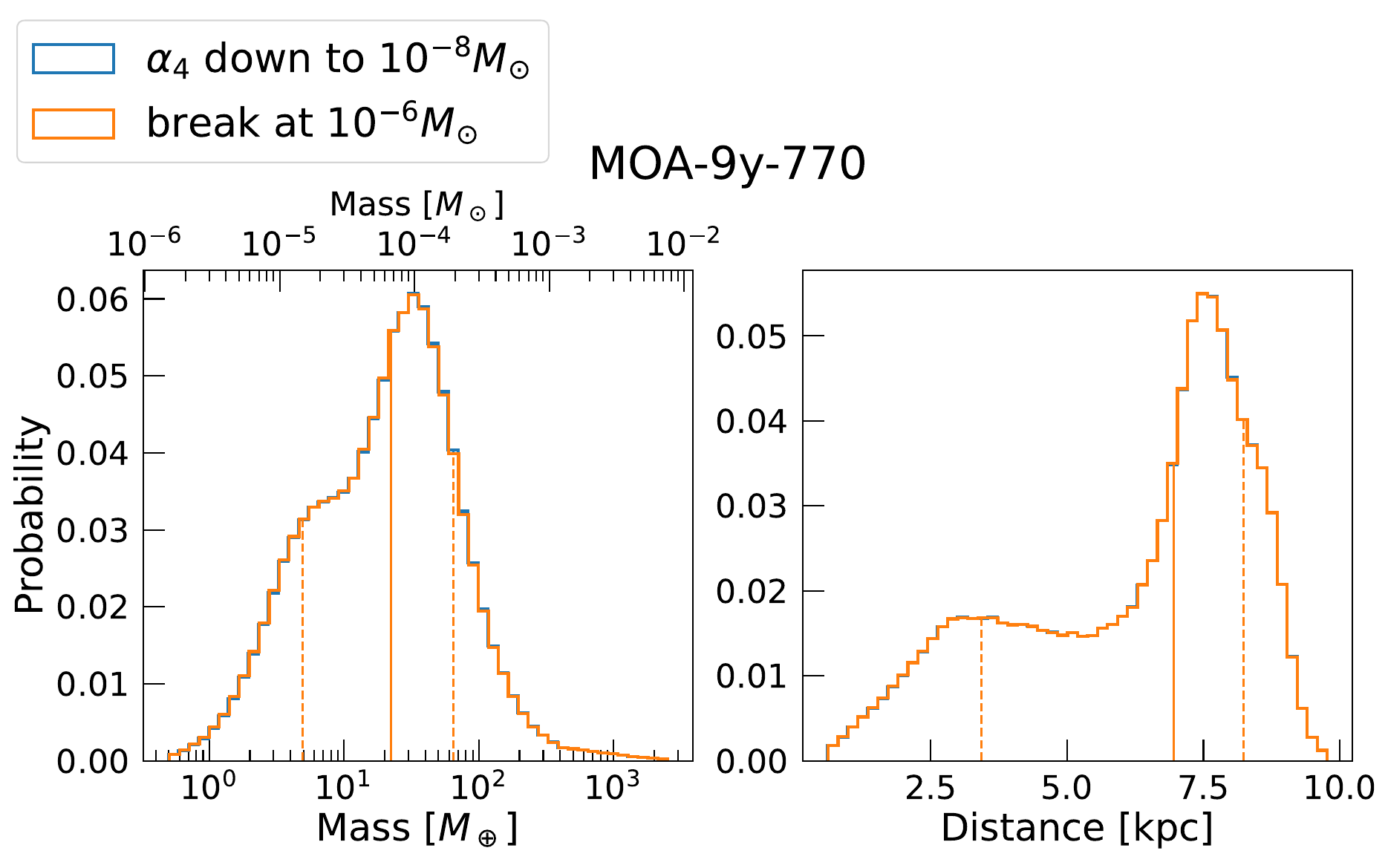} 
  \includegraphics[scale=0.28,keepaspectratio]{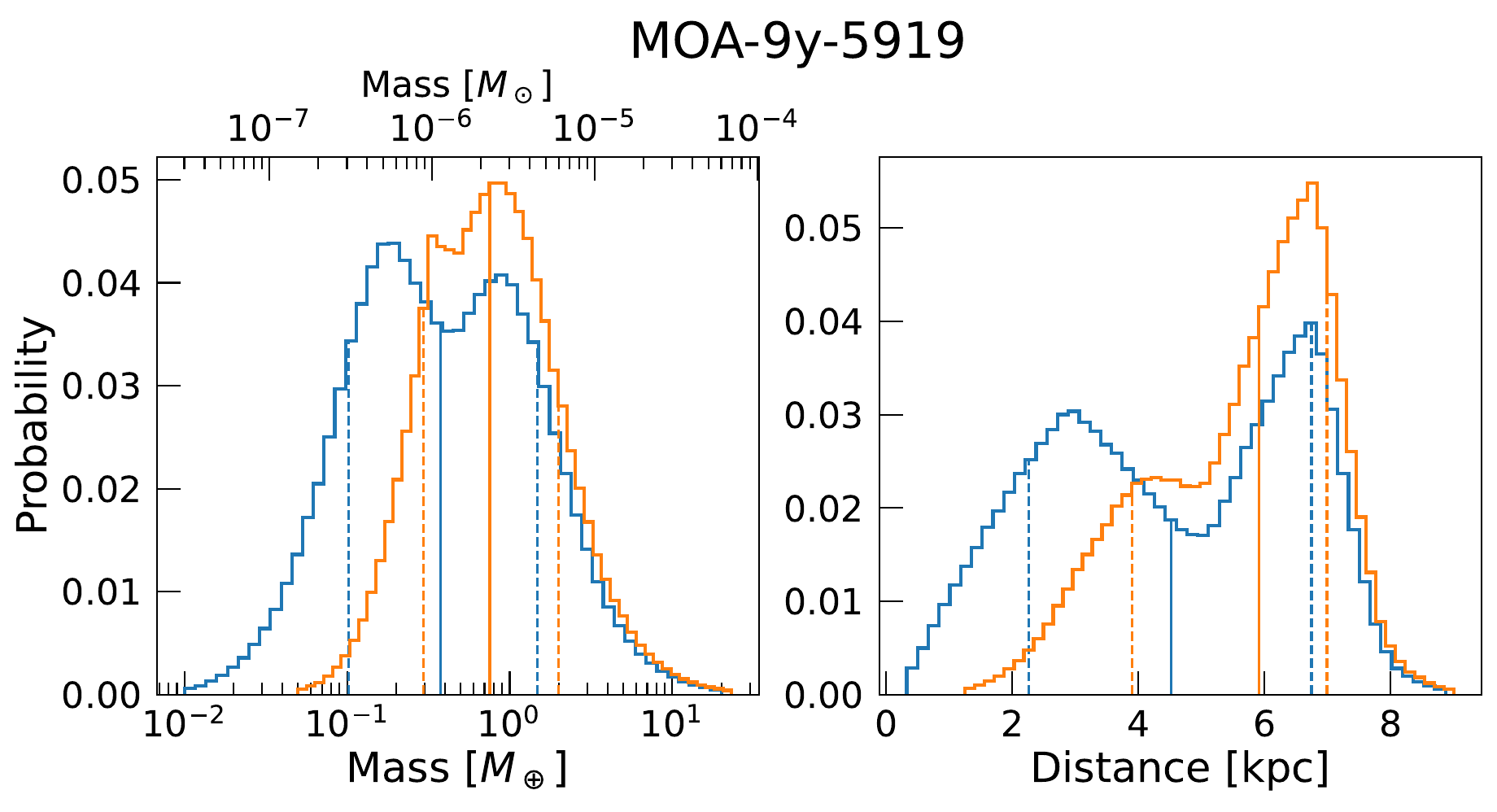}
  \includegraphics[scale=0.28,keepaspectratio]{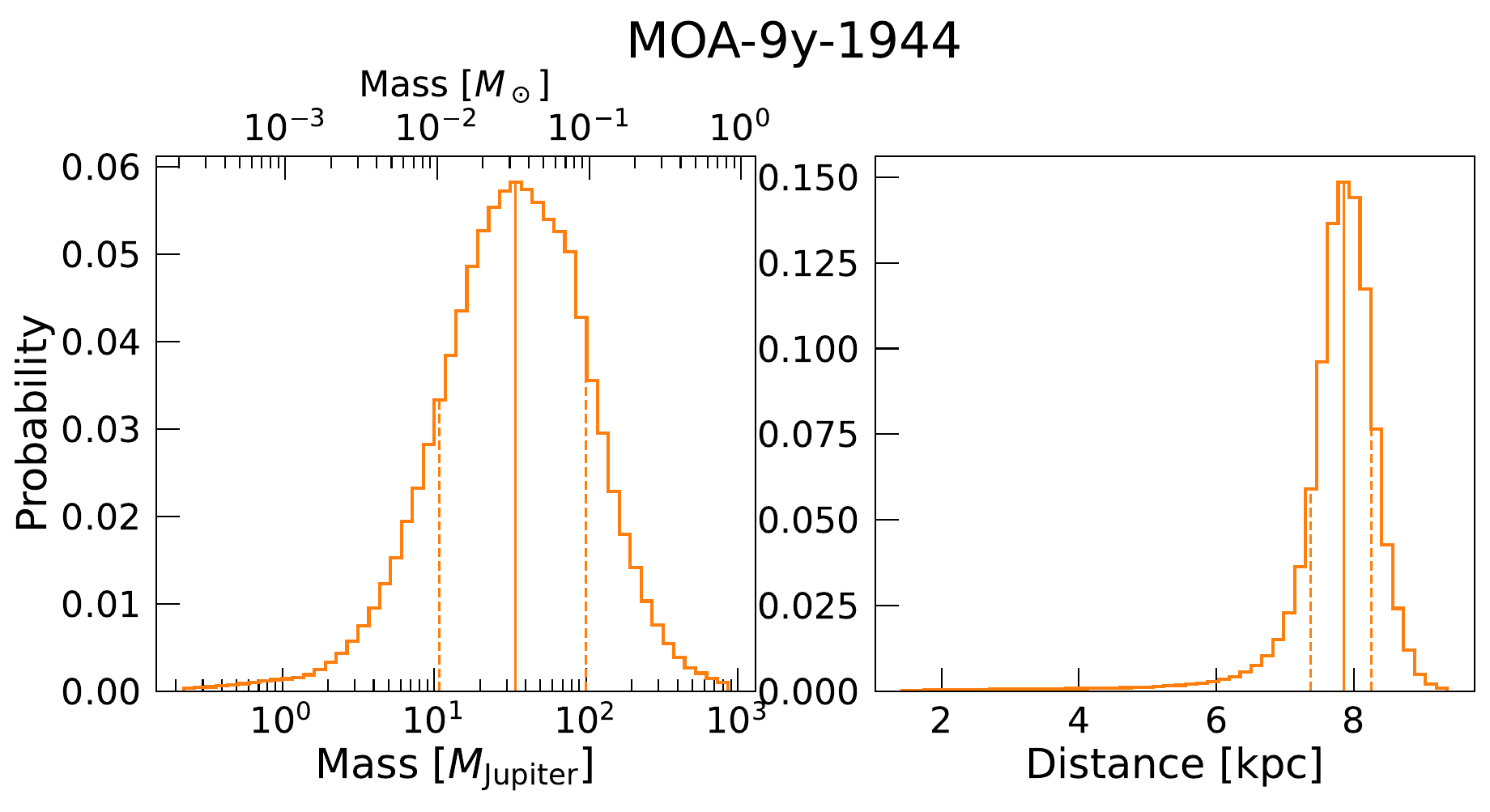}
\caption{
  \label{fig:Bayesian_FFPs}
Posterior mass and distance distributions of the two FFP and one brown dwarf candidates by a Bayesian analysis.
The blue histograms are for when the slope $\alpha_4$ is applied down to $10^{-8} M_{\odot}$ ($M_{\rm lo} = 0.0033 M_\earth$).
The orange histograms are for when the slope $\alpha_4$ is applied down to $10^{-6} M_{\odot}$ and $dN/d\log M = {\rm const.}$ is applied for $M < 10^{-6} M_{\odot}$ ($M_{\rm lo} = 0.33 M_\earth$).
}
\end{figure}


\begin{deluxetable}{lccccccccc}
\tabletypesize{\scriptsize}
\tablecaption{The posterior distributions of parameters by Bayesian analysis for short FSPL events and a BD candidate.
\label{tbl:Bayesian}
}
\tablewidth{0pt}
\tablehead{
\colhead{ID}  &        \multicolumn{2}{c}{Mass}          & & \multicolumn{2}{c}{Distance}   \\
              &        \multicolumn{2}{c}{$(M_\earth)$}         & & \multicolumn{2}{c}{(kpc)}     \\
\colhead{$M_{\rm lo} (M_\earth) = $} &  \colhead{$0.0033$} & \colhead{$0.33$}     & &    \colhead{$0.0033$}    & \colhead{$0.33$}
}
\startdata
 MOA-9y-770   &    \multicolumn{2}{c}{$22.3^{+42.2}_{-17.4}$}   & &    \multicolumn{2}{c}{$6.95^{+1.29}_{-3.52}$}   \\ 
MOA-9y-5919   & $0.37^{+1.11}_{-0.27}$ & $0.75^{+1.23}_{-0.46}$ & & $4.52^{+2.23}_{-2.25}$ & $5.92^{+1.07}_{-2.02}$ \\ 
\hline
MOA-9y-1944\tablenotemark{a} & \multicolumn{2}{c}{$0.033^{+0.062}_{-0.022} M_\sun$} & & \multicolumn{2}{c}{$7.85^{+0.40}_{-0.48}$} \\  
\enddata
\tablenotetext{a}{MOA-9y-1944 is not in the final sample.}
\tablecomments{Prior with $M_{\rm lo} = 0.0033 M_\earth$ applies $dN/d\log M \propto M^{-0.91}$ \citepalias{sumi2023} down to $3.3 \times 10^{-3} M_\earth$ while prior with $M_{\rm lo} = 0.33 M_\earth$ applies it down to $0.33 M_\earth$ and applies $dN/d\log M \propto {\rm const.}$ below.
Results for MOA-9y-770 and MOA-9y-1944 didn't vary among these two priors (see Fig. \ref{fig:Bayesian_FFPs}).
\\}
\end{deluxetable}

\subsection{Short FSPL events}
\label{sec:shortFSPL}
There are two short ($t_{\rm E}<1$ day) events in the final sample with a measured finite source effect,
MOA-9y-770 ($t_{\rm E}=0.315\pm 0.017$ days) and
MOA-9y-5919 ($t_{\rm E}=0.057\pm 0.016$ days).
The light curves of these events are shown in Fig. \ref{fig:lightcurve}.

We also report here our discovery of a brown dwarf candidate event MOA-9y-1944. Although the $t_{\rm E}$ value is slightly longer than 1 day and the source is too faint to pass our selection criteria, its angular Einstein radius value is close to the star/BD end of the Einstein desert \citep{Ryu2021, Gould2022}. The light curve is shown in Fig. \ref{fig:lightcurveBD}

\subsubsection{MOA-9y-770} 
Event MOA-9y-770 occurred in 2008 and a clear finite source effect was detected at the peak covered during one night of observation.
The timescale is short, $t_{\rm E}=0.315\pm 0.017$ days, and the ratio of source star size to $\theta_{\rm E}$ is very large, $\rho=1.08 \pm 0.07$.
The source is an RCG, as shown in Figure \ref{fig:CMDLDall}. 
The estimated source angular radius is $\theta_{*}= 5.13 \pm 0.86$\,$\mu$as, 
which results in a small angular Einstein radius of $\theta_{\rm E}= 4.73 \pm 0.75$\,$\mu$as.
This small $\theta_{\rm E}$ implies a very small lens mass.

To estimate the posterior distribution of the physical parameters of the lens, we performed a Bayesian analysis using the Galactic model of \citet{Koshimoto2021}.
We use their microlensing simulation tool \texttt{genulens}\footnote{\url{https://github.com/nkoshimoto/genulens}} \citep{kosran21} to sample many microlensing events toward the event direction, and calculate the posterior distribution by collecting $10^6$ simulated events that have $t_{\rm E}$, $\theta_{\rm E}$, and source magnitude and color values
consistent with the observed values.
For the parameters $t_{\rm E}$ and $\theta_{\rm E}$, we evaluate the consistency by comparing 
the values of the simulated events using a Gaussian distribution with a mean equal to the observed value and a standard deviation equal to the observed uncertainty. 
For the source magnitude and color, we use a uniform probability distribution with a width equal to three times the 
observed uncertainty to evaluate the consistency.
We ignored correlations among those parameters. More details are found in the usage document of \texttt{genulens}\footnote{\url{https://github.com/nkoshimoto/genulens/blob/main/Usage.pdf}}.

To apply the Galactic model of \citet{Koshimoto2021} for the FFP candidates, we extended their broken power law initial mass function as follows : 
\begin{equation}
 \frac{dN}{d\log M} \propto
  \left\{
   \begin{array}{ll}
     {M^{-1.32}} & (M > 0.86 M_\sun)\\
     {M^{-0.13}} & (0.08 M_\sun < M \leq 0.86 M_\sun) \\
     {M^{-\alpha_3}} & (M_3 < M \leq 0.08 M_\sun) \\
     {M^{-\alpha_4}} & (M_{\rm lo} < M \leq M_3) \\
     {\rm const.} & (10^{-8} M_\sun < M \leq M_{\rm lo}), \\
    \end{array}
    \right.
  \label{eq:MF}
\end{equation}
where the slope and break mass values above 0.08 $M_\sun$ are taken from the E+E$_{\rm X}$ model of \citet{Koshimoto2021}.
We use $\alpha_3 = -0.55$, $\alpha_4 = 0.92$, and $M_3 = 1.3 \times 10^{-3} \, M_\sun (\sim 1 M_{\rm Jupiter})$ taken from a tentative best-fit mass function to the $t_{\rm E}$ and $\theta_{\rm E}$ distribution of our sample, which is consistent with the final result presented in \citetalias{sumi2023}.
For the low mass break $M_{\rm lo}$, we applied two values, $M_{\rm lo} = 10^{-8} M_\sun$ or $0.0033 M_\earth$ and $M_{\rm lo} = 10^{-6} M_\sun$ or $0.33 M_\earth$, because it is uncertain to what extent the $\alpha_4$ slope continues below the sensitivity of our sample.
Note that the lowest mass of $10^{-8} M_\sun = 3.3 \times 10^{-3} M_\earth$ is lower than the mass of Mercury (i.e., the lowest mass planet in our solar system), $5.5 \times 10^{-2} M_\earth$, and close to but slightly higher than the mass of Eris (i.e., the most massive dwarf planet), $2.8 \times 10^{-3} M_\earth$.

The resultant median and 68\% intervals of the posterior distributions of the lens mass and distance for MOA-9y-770 are given in Table \ref{tbl:Bayesian} and the  posterior distributions are shown in Figure \ref{fig:Bayesian_FFPs}.
The estimated lens mass is $22.3^{+42.2}_{-17.4}$ $M_\earth$, which is close to Neptune's mass (17.2 $M_\earth$), and is located in the Galactic bulge ($D_{\rm L}=7.11^{+1.25}_{-3.49}$ kpc), regardless of the prior for $M_{\rm lo}$.

\subsubsection{MOA-9y-5919} 
Event MOA-9y-5919 occurred in 2008, and the interval when the magnification showed a clear finite source effect was well covered over the course of one night.
The timescale is the shortest in our sample, $t_{\rm E}=0.057\pm 0.016$ days and the ratio $\rho$ is large, $\rho=1.40 \pm 0.46$.
The source is a G7 turn-off star, as plotted on the CMD in Figure \ref{fig:CMDLDall}. 
The estimated source angular radius is $\theta_{*}=  1.26 \pm 0.48$\,$\mu$as, which leads to an angular Einstein radius of $\theta_{\rm E}= 0.90 \pm 0.14$\,$\mu$as, similar to the value of $\theta_{\rm E} \sim 0.84$ $\mu$as for OGLE-2016-BLG-1928, the shortest timescale event discovered to date \citep{Mroz20b}.

Our Bayesian analysis indicates that the lens, MOA-9y-5919L, has a terrestrial-mass regardless of the prior for $M_{\rm lo}$; $0.37^{+1.11}_{-0.27}$ $M_\earth$ (with $M_{\rm lo} = 0.0033 M_\earth$) or $0.75^{+1.23}_{-0.46}$ $M_\earth$ (with $M_{\rm lo} = 0.33 M_\earth$), as shown in Table \ref{tbl:Bayesian}. Thus, MOA-9y-5919L is the second terrestrial-mass FFP candidate discovered to date.
The posterior distributions are shown in Figure \ref{fig:Bayesian_FFPs}.
The mass distribution with $M_{\rm lo} = 0.0033 M_\earth$ (blue histogram) shows a non-negligible probability of $M$ even below Mercury's mass of $0.055 M_\earth$.
Nevertheless, we use results with $M_{\rm lo} = 0.33 M_\earth$ for our final results, to be conservative, as we have little sensitivity to planets below $0.33 M_\oplus$.
We confirmed the robustness of our conclusion that MOA-9y-5919L is most likely to have a terrestrial mass by repeating the same analysis but with mass function parameters that minimize the number of planetary-mass objects within the 
uncertainty range given by \citetalias{sumi2023}.

\subsubsection{MOA-9y-1944}  
MOA-9y-1944 is a brown dwarf (BD) candidate event that occurred in 2012 whose entire magnification part was well covered during one night.
The timescale is relatively short, $t_{\rm E} = 1.594  \pm  0.136 $ days and the value of the ratio $\rho$ is moderate, $\rho=0.00928 \pm 0.00032$.
The estimated source angular radius is $\theta_{*}=  0.43 \pm  0.10$\,$\mu$as, which leads to a value for $\theta_{\rm E}$ of $\theta_{\rm E}= 46.1 \pm   10.5$\,$\mu$as.
This value is distinctly larger than those of the two FFP candidates above, but the smallest among others, i.e., it is consistent with the lower edge of the star/BD population.
This is closer to the same edge of $\theta_{\rm E}= 30$\,$\mu$as as found by \cite{Gould2022}.
Note that this event is not in the final sample to be used for our statistical analysis because the source magnitude of $I_{\rm s}=21.9$ mag is fainter than the criteria threshold.
We showed this event as a reference to show the object near the gap of the Einstein desert \citep{Ryu2021, Gould2022} and also to show 
the usefulness of finding the events with not only a giant source, but also with dwarf sources.

Our Bayesian analysis indicates the lens mass is $0.033^{+0.062}_{-0.022}$   $M_\sun$, i.e., 
it is likely a brown dwarf in the Galactic bulge at $7.85^{+0.40}_{-0.48}$ kpc. The posterior distributions are shown in Figure \ref{fig:Bayesian_FFPs}.

\section{Detection Efficiency}
\label{sec:efficiency}

To be used for various statistical studies such as the measurement of the mass function by \citetalias{sumi2023}, we calculate the detection efficiency of the survey 
by conducting an image level simulation following \citet{sumi03, sumi2011}.
One major difference from the previous studies is that we consider the influence of the finite source effect on the detection efficiency in this work.
This makes the analysis more complicated because when the finite source effect is not negligible, the detection efficiency becomes a function of both $t_{\rm E}$ and $\theta_{\rm E}$, $\epsilon (t_{\rm E}, \theta_{\rm E})$.
Thus, the detection efficiency as a function of the timescale,
\begin{equation}
\tilde{\epsilon} (t_{\rm E} ; \Gamma) =  \int_{\theta_{\rm E}} \Gamma (\theta_{\rm E} | t_{\rm E})  \epsilon (t_{\rm E}, \theta_{\rm E}) d\theta_{\rm E},  \label{eq:eps_tE}
\end{equation}
depends on the event rate $\Gamma (t_{\rm E}, \theta_{\rm E})$ given by a Galactic model that includes the mass function of the lens objects, i.e., what we want to measure in \citetalias{sumi2023}.
Here, $P (A | B)$ is the probability of $A$ given $B$, and 
$\Gamma (\theta_{\rm E} | t_{\rm E})$ is thus the fraction of events with $\theta_{\rm E}$ among events with $t_{\rm E}$ in the model.
The true detection efficiency, $\epsilon (t_{\rm E}, \theta_{\rm E})$, depends on two variables, so if we want to express
the detection efficiency as a function of one variable, we must integrate over one of these variables. We, therefore, refer to 
$\tilde{\epsilon} (t_{\rm E} ; \Gamma)$ function as the ``integrated detection efficiency\rlap." However, the integrated 
detection efficiency depends on the event rate, $\Gamma (t_{\rm E}, \theta_{\rm E})$, which depends on the mass
function of lens objects. It is particularly sensitive to the FFP mass function because a large fraction of FFP microlensing
light curves shows significant finite source effects. Thus, the true two-dimensional nature of $\epsilon (t_{\rm E}, \theta_{\rm E})$
cannot be ignored in microlensing analyses of the FFP mass function.

The detection efficiencies depend on the (true) $\theta_{\rm E}$ value, especially 
in short events where the finite source effect can significantly change its amplitude 
and duration of magnification.
However, we note that selection criteria for $\epsilon (t_{\rm E}, \theta_{\rm E})$ do not 
require the measurement of $\theta_{\rm E}$ (see Table \ref{tbl:criteria}), 
which allows both PSPL and FSPL events to be detected.
On the other hand, we separately consider another detection efficiency for the FSPL events in Section \ref{sec:eff_thetaE} by adding a requirement of the $\theta_{\rm E}$ measurement to the selection criteria.

We first calculate the detection efficiency for events with $(t_{\rm E}, \theta_{\rm E})$, $\epsilon (t_{\rm E}, \theta_{\rm E})$, by an image level simulation in Section \ref{sec:eff_tEthE}.
Then, we calculate the integrated detection efficiency as a function of $t_{\rm E}$, $\tilde{\epsilon} (t_{\rm E} ; \Gamma)$, by integrating Eq. (\ref{eq:eps_tE}) for a given event rate $\Gamma (t_{\rm E}, \theta_{\rm E})$ in Section \ref{sec:eff_tE}.

\subsection{Image level simulation} 
\label{sec:eff_tEthE}
As described in Section \ref{sec:analysis}, our analysis has been conducted using 1024 pix $\times$ 1024 pix subframes as the smallest image unit.
We generated 40,000 artificial events in each subframe, i.e., 64M events in total,  
and embedded them at random positions between $0 \le x/{\rm pix} \le 2048$ and $0 \le y/{\rm pix} \le 4096$ in each CCD. 
The microlensing parameters are randomly assigned between $3824 \le t_0/{\rm JD}' \le 6970$,  
$0 \le u_0 \le 1.5$, and source magnitude of $ 14.2 \le I_{\rm s}/{\rm mag} \le 22$, uniformly.

The timescale $t_{\rm E}$ are randomly given with a log-uniform distribution between 0.02 and 1000 days for 12.5\% of the simulation, and 
between 0.02 - 10 days for the remained 87.5\% with a probability distribution proportional to $\log (t_{\rm E}/ 10\, {\rm days})^{-1}$.
The bias toward small $t_{\rm E}$ is because shorter events generally have smaller detection efficiencies and 
more simulations are needed to estimate the detection efficiency accurately enough.

\begin{figure*}
\begin{center}
\includegraphics[width=16cm]{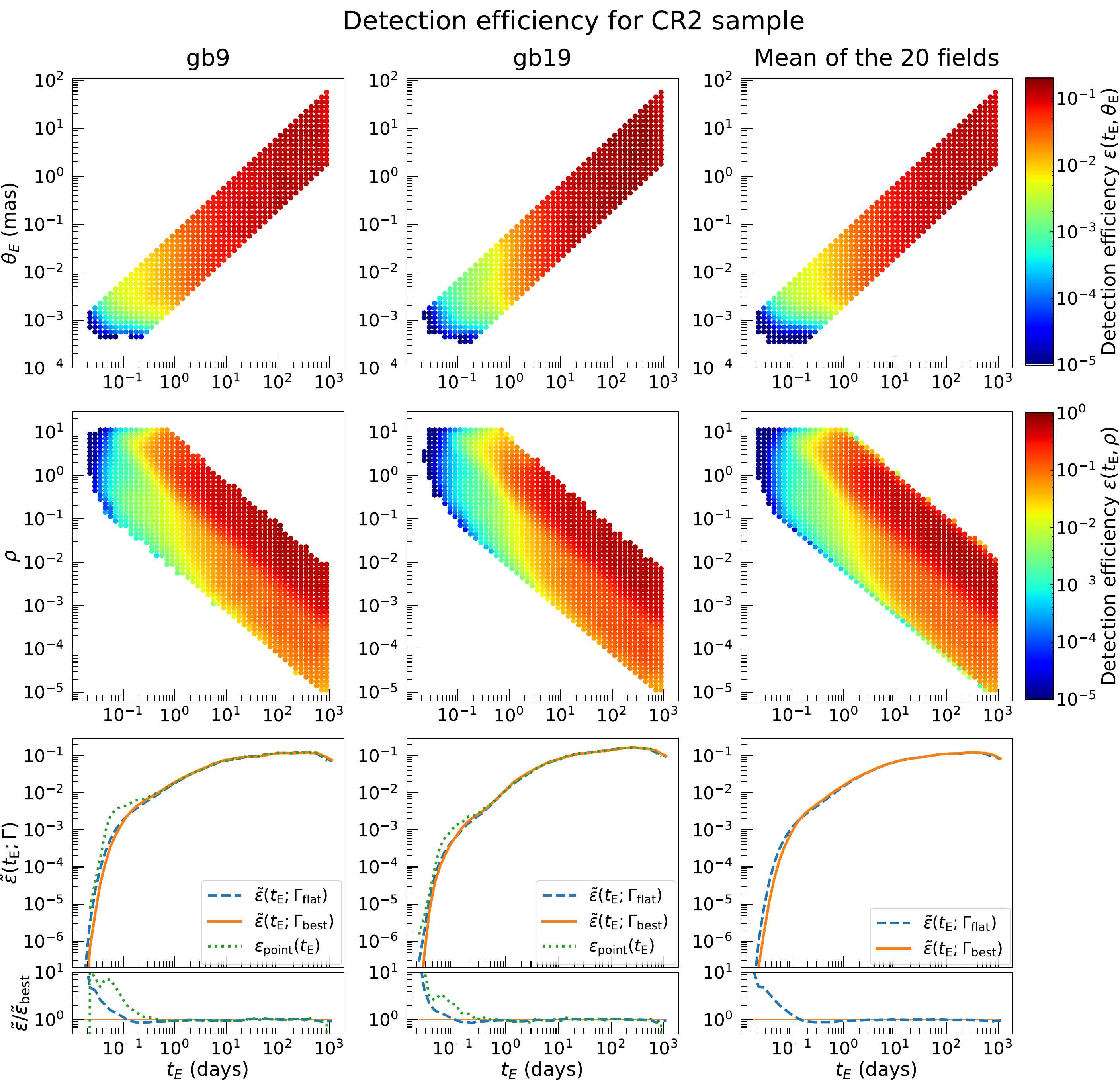}
\caption{ Two and one dimensional detection efficiencies including finite source measurements. The top row shows 
$\epsilon (t_{\rm E}, \theta_{\rm E})$, which is the detection efficiency as 
a function of the Einstein radius crossing time, $t_{\rm E}$, and angular Einstein radius $\theta_{\rm E}$. The middle row
shows the detection efficiency, $\epsilon (t_{\rm E},\rho)$, as a function of $t_{\rm E}$ and the finite source parameter $\rho$, and  the bottom row shows the one dimensional integrated detection efficiencies as a function of $t_{\rm E}$.
The orange solid, blue dashed, and green dotted curves in the bottom panels show $\tilde{\epsilon} (t_{\rm E} ; \Gamma)$ with 
the best-fit $\Gamma (t_{\rm E}, \theta_{\rm E})$ taken from \citetalias{sumi2023}, with $\Gamma (t_{\rm E}, \theta_{\rm E}) = $ const. in the simulated range, and with the PSPL assumption, respectively.  
The left, middle and right columns show the detection efficiencies for fields gb9, gb19, and the mean of all 20 fields
using criteria CR2. 
The simulation is limited in 0.8 $\leq \mu_{\rm rel} / ({\rm mas \, yr^{-1}})  \leq$ 20. Note that our selection criteria (Table \ref{tbl:criteria}) does not require $\theta_{\rm E}$ measurable.}
\end{center}
\label{fig:EFF_tEthE}
\end{figure*}

Because a likely range of the lens-source relative proper motion, $\mu_{\rm rel}=\theta_{\rm E}/t_{\rm E}$, is 0.8 mas/yr to 20 mas/yr 
regardless of $t_{\rm E}$ less than 100 days \citep[see Fig. 1 of][]{kos21b},
the angular Einstein radius values $\theta_{\rm E}$ are randomly drawn from a log uniform distribution between 
$\log (0.8 \, t_{\rm E}/{\rm 365.25 \, days}) \le \log (\theta_{\rm E}/{\rm mas}) \le  \log (20 \, t_{\rm E}/{\rm 365.25 \, days})$ depending on the assigned $t_{\rm E}$ (see Fig. \ref{fig:EFF_tEthE}).
Note that the detection efficiency for long timescale events has little dependence on $\theta_{\rm E}$, and thus the $\mu_{\rm rel}$ range taken here does not affect 
our results even if there is a non-negligible population of events with $\mu_{\rm rel} < 0.8$ mas/yr among events with $t_{\rm E} >$ 100 days.

The source angular radius $\theta_{*}$ is calculated from the assigned $I_{\rm s}$ by 
using the same procedure used in Section \ref{sec:FSPL}.
Then $\rho=\theta_{*} /\theta_{\rm E}$ is used for the finite source effect in the simulated events.

To embed the artificial events, we calculated the differences of the flux in each frame relative to that of the reference image, 
$\Delta F(t_i) = F(t_i) - F(t_{\rm ref})$. 
Here $F(t_i)$ and $F(t_{\rm ref})$ are model fluxes given by Eq. (\ref{eq:ft}) 
at the time when each frame $t_i$ and the reference images $t_{\rm ref}$ are taken, respectively.
The PSF  derived by DOPHOT on each subframe of the reference images are
convolved by the kernel to match to the seeing, scale, PSF shape variation on each observed subframe. 
Here we used same kernels which are derived in DIA process.
We added this convolved PSF scaled by $\Delta F(t_i)$ on all frames of the real difference images.
Then we reduced these simulated difference images with artificial events by using the same pipeline and 
``detect" the events through the same selection criteria as what used for the real events,
to calculate the detection efficiency as a function of $t_{\rm E}$ and $\theta_{\rm E}$  in each field.

The detection efficiency for events with $(t_{\rm E}, \theta_{\rm E})$ in $j$th field ($j =$1, 2, ... 22, but gb6 and gb22 aren't used) is calculated by
\begin{align}
\epsilon_j (t_{\rm E}, \theta_{\rm E}) =   \sum_{k \in j} \sum_{i = 1}^{N_{{\rm sim}, k}} w_{i, k} (t_{\rm E}, \theta_{\rm E}) X_{{\rm det}, i}, \label{eq:eps02D}
\end{align}
where $k$ denotes a subframe in $j$th field ($k =$ 1, 2, ..., 80, i.e., 10 chips $\times$ 8 subrames), 
$i$ denotes an artificial event in the subframe, and $N_{{\rm sim}, k}$ is the number of artificial events in the grid of $(t_{\rm E}, \theta_{\rm E})$.
$X_{{\rm det}, i}$ takes 1 when the $i$th event is detected and takes 0 when it is undetected. The weight for each event $w_{i, k} (t_{\rm E}, \theta_{\rm E})$ is given by
\begin{align}
\ \  w_{i, k} (t_{\rm E}, \theta_{\rm E}) = \frac{n_{{\rm RC}, k}^2 \, f_{{\rm LF}, k} (I_{s, i})}{\sum_{k \in j} \left(n_{{\rm RC}, k}^2 \sum_{i = 1}^{N_{{\rm sim}, k}}\, f_{{\rm LF}, k} (I_{s, i}) \right), } 
\end{align}
where $n_{{\rm RC}, k}$ is the number density of RCGs in the $k$th subfield, $f_{{\rm LF}, k} (I_{s, i})$ is the fraction of stars that has a source magnitude $I_{s, i}$ given by 
the luminosity function (LF) in $k$th subfield, and $n_{{\rm RC}, k}^2 \, f_{{\rm LF}, k} (I_{s, i})$ is thus proportional to the expected event rate.
Note that $n_{{\rm RC}, k}$ does not reflect the number of stars in the foreground or the far disk. We assumed that their contribution to the relative event rate among the fields is negligible.

The LF in $k$th subfield is given by using the combined luminosity function (LF)
from the OGLE-III photometry map \citep{uda11} and the HST data \citep{hol98}. This uses
the OGLE LF for bright stars and the HST LF for faint stars down to $I = 24$ mag.
This combined LF is calibrated to the extinction and Galactic bulge distance for
each subfield by using the position of RCG stars as a standard candle in the CMD. 


Figure \ref{fig:EFF_tEthE} shows the calculated detection efficiency when using the criteria CR2 in gb9 and gb19, in addition to the mean of all the 20 fields used. We picked these two fields here because the field gb9 is the highest cadence field while the gb19 is one of the lowest cadence fields but has MOA-9y-5919, the shortest event with $t_{\rm E} = 0.057\pm 0.016$ days in our sample.
In the top panels, the detection efficiencies as a function of $t_{\rm E}$ and $\theta_{\rm E}$, $\epsilon (t_{\rm E}, \theta_{\rm E})$, are shown.
At $t_{\rm E} \simgt 3$ days, dependence of the detection efficiency on $\theta_{\rm E}$ for a given $t_{\rm E}$ is not seen in the color map for all the three columns, but for $t_{\rm E} \simlt 1$ day, it is clearly seen.

In the middle row, the detection efficiency as a function of $t_{\rm E}$ and $\rho$ are shown.
The conversion from $\epsilon (t_{\rm E}, \theta_{\rm E})$ can be done via $\rho = \theta_*/\theta_{\rm E}$, where $\theta_*$ is calculated using the $I_s$ value for each artificial event as described in cut 3 in Section \ref{sec:select}.
Unlike $\epsilon (t_{\rm E}, \theta_{\rm E})$, $\epsilon (t_{\rm E}, \rho)$ has a dependence on $\rho$ for a given $t_{\rm E}$ even when $t_{\rm E}$ is long.
This is because the source tends to be brighter when $\rho$ is larger, and thus $\epsilon (t_{\rm E}, \rho)$ is high with a larger $\rho$ when $t_{\rm E}$ is long.
However, when $t_{\rm E}$ is short, $\epsilon (t_{\rm E}, \rho)$ gets smaller at a larger $\rho$ value, which is due to the finite source effect causing a suppression of magnification.
Note that the sharp cut in $\epsilon (t_{\rm E}, \rho)$ at $\rho = 10$ is because we assumed $\epsilon (t_{\rm E}, \theta_{\rm E})=0$ for $\rho > 10$. This is because the peak magnification with $\rho > 10$ is $\simlt$ 2\% and it is the limitation in the FSPL calculation algorithm by \citet{Bozza2018}\footnote{The latest version (v3.5) of VBBinaryLensing supports sources as big as $\rho$ = 100.}. Nevertheless, this is negligible for $\tilde{\epsilon} (t_{\rm E} ; \Gamma)$ because only $< 1\%$ of simulated events have $\rho > 10$ for $t_{\rm E} > 0.05$ days, which includes  all the events in our sample.

The bottom panels show the integrated detection efficiency as a function of $t_{\rm E}$, $\tilde{\epsilon} (t_{\rm E} ; \Gamma)$, which is discussed in Section \ref{sec:eff_tE} below.

\subsection{Integrated detection efficiency as a function of $t_{\rm E}$}
 \label{sec:eff_tE}
 
When finite source effects are important, the true detection efficiency is a function of $t_{\rm E}$ and $\theta_{\rm E}$,  
$\epsilon (t_{\rm E}, \theta_{\rm E})$, although it can also be described by other pairs of parameters, such as 
$(t_{\rm E}, \rho)$, that describe the same parameter space. However, all previous work in the field has considered
the detection efficiency for single lens events to be described by only a single parameter, $t_{\rm E}$, except for 
\citet{Gould2022} who proposed a detection efficiency model depending only on $\theta_{\rm E}$. The
integrated detection efficiency, $\epsilon(t_{\rm E};\Gamma)$, is defined in Eq.~(\ref{eq:eps_tE}), but this
equation includes a dependence of the event rate. Since the event rate depends on the FFP mass function, it
is problematic to try and use $\epsilon(t_{\rm E};\Gamma)$ to determine the FFP mass function. In the forward
Bayesian analysis of the FFP mass function that we present in \citetalias{sumi2023}, we avoid this problem by
integrating Eq.~(\ref{eq:eps_tE}) for every proposed mass function to separately 
determine $\tilde{\epsilon} (t_{\rm E} ; \Gamma)$ used in the calculation of the likelihood
function for the mass function parameters.

In this section, we probe the dependence of $\tilde{\epsilon} (t_{\rm E} ; \Gamma)$ on the FFP mass function
in order to investigate what circumstances might allow the dependence of $\tilde{\epsilon} (t_{\rm E} ; \Gamma)$ on the
event rate, $\Gamma (\theta_{\rm E} | t_{\rm E}) = \Gamma (t_{\rm E}, \theta_{\rm E}) / \Gamma (t_{\rm E})$,
to be ignored.
 
Since $t_{\rm E} \propto \sqrt{M}$, the event rate $\Gamma (t_{\rm E})$ can be separated from the mass function \citep{han96, weg17}
\begin{equation}
\Gamma (t_{\rm E}) = \int \gamma (t_{\rm E} M^{-1/2}) \Phi (M) \sqrt{M} \, dM, \label{eq:G_tE}
\end{equation}
where $\gamma (t_{\rm E})$ is the event rate for lenses with the mass $1\,M_\sun$ and $\Phi (M)$ is the present-day mass function. 
Similarly, the event rate $\Gamma (t_{\rm E}, \theta_{\rm E})$ is 
\begin{align}
 \Gamma (t_{\rm E}, \theta_{\rm E}) = \int \gamma (t_{\rm E} M^{-1/2}, \theta_{\rm E} M^{-1/2}) \Phi (M) \sqrt{M} dM. \label{eq:G_tEthE}
\end{align}
To calculate $\gamma (t_{\rm E})$ and $\gamma (t_{\rm E}, \theta_{\rm E})$, we use the stellar density and velocity distributions of the Galactic model from \citet{Koshimoto2021}.

With Eqs. (\ref{eq:G_tE}) and (\ref{eq:G_tEthE}), the calculation of $\tilde{\epsilon} (t_{\rm E} ; \Gamma)$ in Eq. (\ref{eq:eps_tE}) becomes a double integration over $M$ and $\theta_{\rm E}$.
To reduce computation time, we divide the integration over $\theta_{\rm E}$ into two parts,
\begin{align}
\tilde{\epsilon} (t_{\rm E} ; \Gamma) =  \int_{\theta_{\rm E} \leq \theta_{\rm E, th}}  \Gamma (\theta_{\rm E} &|  t_{\rm E})   \epsilon (t_{\rm E}, \theta_{\rm E}) d\theta_{\rm E}  \notag\\
 + \  & \epsilon_{\rm point}  (t_{\rm E}) \times \left( 1 - f_{\rm FS} (t_{\rm E} ; \Gamma) \right), \label{eq:eps_tE2}
\end{align}
where $\theta_{\rm E, th}$ is a $\theta_{\rm E}$ value above which $\epsilon (t_{\rm E}, \theta_{\rm E})$ is independent of $\theta_{\rm E}$, $\epsilon_{\rm point} (t_{\rm E})$ is the value of $\epsilon (t_{\rm E}, \theta_{\rm E})$ when $\theta_{\rm E} > \theta_{\rm E, th}$, and
\begin{align}
f_{\rm FS} (t_{\rm E} ; \Gamma) &\equiv \int_{\theta_{\rm E} \leq \theta_{\rm E, th}}  \Gamma ( \theta_{\rm E} |  t_{\rm E}) d \theta_{\rm E}
\end{align}
represents a fraction of events with $t_{\rm E}$ whose detectability can be affected by the finite source effect. We use $\theta_{\rm E, th} = 0.02 \, {\rm mas}$ determined based on the color map of $\epsilon (t_{\rm E}, \theta_{\rm E})$ in Fig. \ref{fig:EFF_tEthE}.

Using Eqs. (\ref{eq:G_tE}) and (\ref{eq:G_tEthE}) and switching the order of integrals over $M$ and $\theta_{\rm E}$,
\begin{align}
f_{\rm FS} (t_{\rm E} ; \Gamma) &= \int_{\theta_{\rm E} \leq \theta_{\rm E, th}}  \Gamma ( \theta_{\rm E} |  t_{\rm E}) d \theta_{\rm E} \notag\\
& = \  \frac{\int \eta_{\rm FS} (t_{\rm E},  M) \gamma (t_{\rm E} M^{-1/2}) \Phi (M) \sqrt{M} \, d M}{\int \gamma (t_{\rm E} M^{-1/2}) \Phi (M) \sqrt{M} \, d M},
\end{align}
where 
\begin{align}
\eta_{\rm FS} (t_{\rm E},  M) \equiv \int_{\theta_{\rm E} \leq \theta_{\rm E, th}} \gamma (\theta_{\rm E} M^{-1/2} |  t_{\rm E} M^{-1/2}) \, d \theta_{\rm E}
\end{align}
is a cumulative fraction of $\gamma (\theta_{\rm E} M^{-1/2} |  t_{\rm E} M^{-1/2})$ up to $\theta_{\rm E, th}$, and can be instantly calculated once the cumulative distribution of $\gamma (\theta_{\rm E} | t_{\rm E})$ is stored. This can be understood easier if considered in the ($\log t_{\rm E}$, $\log \theta_{\rm E}$) plane, where our actual calculations are performed. Let the cumulative distribution of $\gamma (\log \theta_{\rm E} | \log t_{\rm E})$ be $p (\log \theta_{\rm E} | \log t_{\rm E})$, then $\eta_{\rm FS} (\log t_{\rm E}, \log M)$ can be given by $p (\log \theta_{\rm E, th} -0.5 \log M | \log t_{\rm E} -0.5 \log M)$, i.e., just by shifting the offset of $-0.5 \log M$ on the ($\log t_{\rm E}$, $\log \theta_{\rm E}$) plane.
Thus, we can calculate $f_{\rm FS} (t_{\rm E} ; \Gamma)$ without a double integration using the $\gamma (\theta_{\rm E} | t_{\rm E})$ calculated by the Galactic model beforehand.
This makes the calculation of $\tilde{\epsilon} (t_{\rm E} ; \Gamma)$ faster because we can approximate Eq. (\ref{eq:eps_tE2}) as $\tilde{\epsilon} (t_{\rm E} ; \Gamma) \simeq \epsilon_{\rm point} (t_{\rm E})$ when $f_{\rm FS} (t_{\rm E} ; \Gamma) \ll 1$.
In the calculation of $\tilde{\epsilon} (t_{\rm E} ; \Gamma)$ for a proposed mass function during the fitting in \citetalias{sumi2023}, we applied the approximation when $f_{\rm FS} (t_{\rm E} ; \Gamma) < 0.01$, whereas we calculated both terms of Eq. (\ref{eq:eps_tE2}) when $f_{\rm FS} (t_{\rm E} ; \Gamma) \geq 0.01$.\footnote{
Note that 
Eq. (\ref{eq:eps_tE}) can be represented as 
\begin{align}
\tilde{\epsilon} (t_{\rm E} ; \Gamma) &= \int_{\theta_{\rm E}} \Gamma (\theta_{\rm E} |  t_{\rm E})   \epsilon (t_{\rm E}, \theta_{\rm E}) d\theta_{\rm E} \notag\\
&= \frac{\int \zeta (t_{\rm E},  M) \gamma (t_{\rm E} M^{-1/2}) \Phi (M) \sqrt{M} \, d M}{\int \gamma (t_{\rm E} M^{-1/2}) \Phi (M) \sqrt{M} \, d M},
\end{align}
where
\begin{align}
\zeta (t_{\rm E},  M) \equiv \int_{\theta_{\rm E}} \epsilon (t_{\rm E}, \theta_{\rm E}) \gamma (\theta_{\rm E} M^{-1/2} |  t_{\rm E} M^{-1/2}) \, d \theta_{\rm E}.
\end{align}
Because $\zeta (t_{\rm E},  M)$ is not dependent on the mass function $\Phi (M)$, we can avoid the double integration during the fitting by calculating $\zeta (t_{\rm E},  M)$ beforehand even when $f_{\rm FS} (t_{\rm E} ; \Gamma)$ is not negligible. However, we did not do this in \citetalias{sumi2023} because the computation was fast enough.}


The bottom panels in Fig. \ref{fig:EFF_tEthE} show the integrated detection efficiency as a function of $t_{\rm E}$, $\tilde{\epsilon} (t_{\rm E} ; \Gamma)$, calculated with the best-fit $\Gamma (t_{\rm E}, \theta_{\rm E})$ taken from \citetalias{sumi2023} (orange solid curves) and with $\Gamma (t_{\rm E}, \theta_{\rm E}) = $ const. in the simulated range (blue dashed curves).
We also show $\epsilon_{\rm point} (t_{\rm E})$ in green dotted curves for gb9 and gb19, where we do not consider the finite source effect. Note that the used best-fit $\Gamma (t_{\rm E}, \theta_{\rm E})$ is shown in Figure \ref{fig:Gamma} in Appendix \ref{sec:app}.

In each panel, all the curves agree when $t_{\rm E} \simgt 1$ day. However, they deviate from each other at $t_{\rm E} \simlt 1$ day, where the finite source effect is important, making $\epsilon (t_{\rm E}, \theta_{\rm E})$ dependent on $\theta_{\rm E}$ even when $t_{\rm E}$ is fixed. 
This demonstrates that consideration of both finite source effect and relative event rate is important for the calculation of $\tilde{\epsilon} (t_{\rm E} ; \Gamma)$ with $t_{\rm E} \simlt 1$ day. In Appendix \ref{sec:app}, we also show how $\tilde{\epsilon} (t_{\rm E} ; \Gamma)$ and $\Gamma (\theta_{\rm E} | t_{\rm E})$ depend on the mass function by changing the slope for the planetary mass range, $\alpha_4$.

As expected, the detection efficiency for short $t_{\rm E}$ events in gb9 is higher than in gb19, due to the
higher observing cadence for gb9. The mean detection efficiency for short events is between these two.

\begin{figure*}
\begin{center}
\includegraphics[width=16cm]{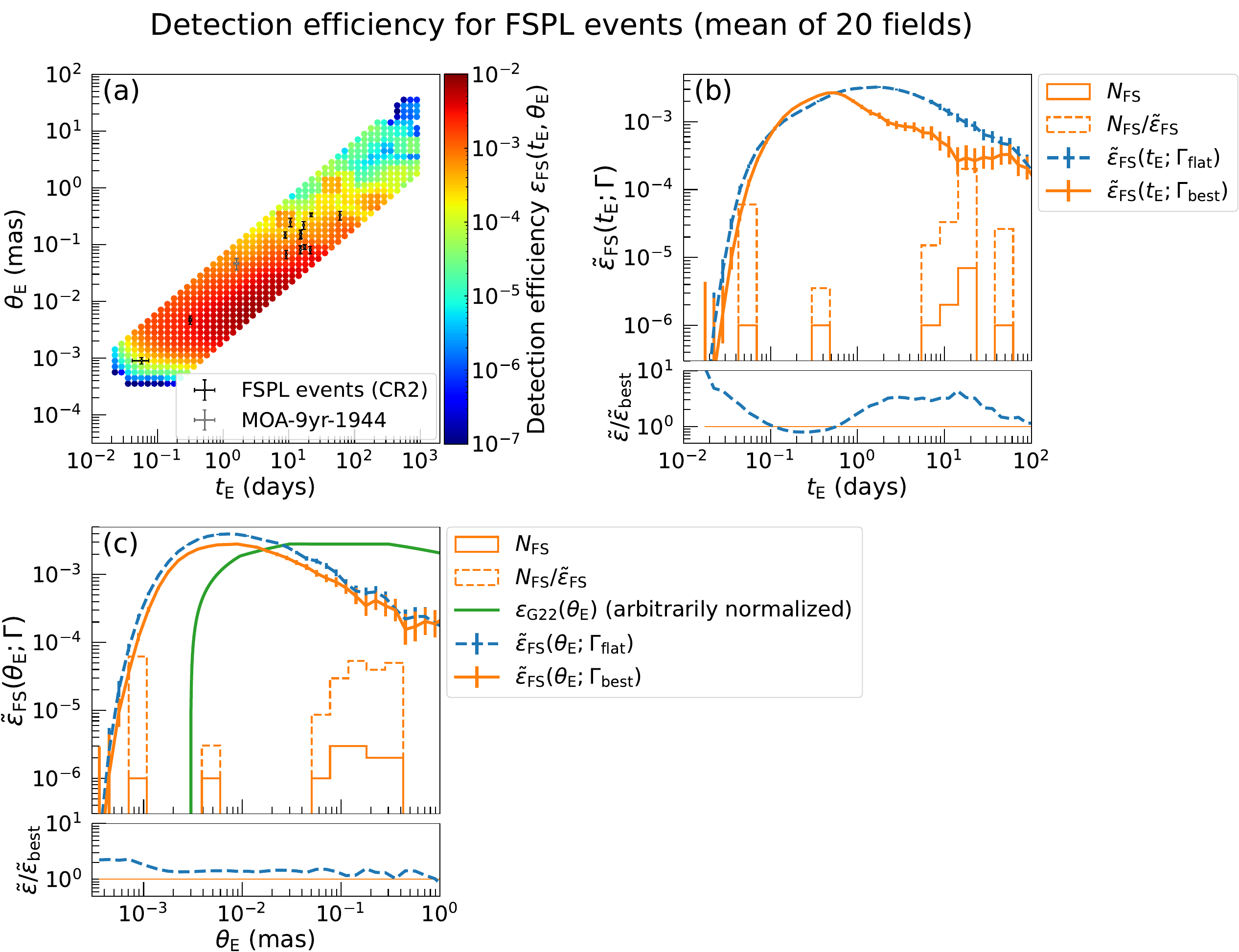}
\caption{
Detection efficiencies for FSPL events as functions of (a) the timescale $t_{\rm E}$ and the angular Einstein radius $\theta_{\rm E}$, (b) $t_{\rm E}$, and (c) $\theta_{\rm E}$, for the mean of all the 20 fields with the criteria CR2. In panels (b) and (c), the orange solid and blue dashed curves show integrated detection efficiencies with the best-fit $\Gamma (t_{\rm E}, \theta_{\rm E})$ taken from \citetalias{sumi2023} and with $\Gamma (t_{\rm E}, \theta_{\rm E}) = $ const. in the simulated range, respectively.
The solid orange histogram indicates the 13 FSPL events detected in our sample, where 10$^{-6}$ corresponds to one event. 
The green line in panel (c) shows the relative detection efficiency for the KMTNet giant source sample \citep{Gould2022}.}
\end{center}
\label{fig:EFF_FS}
\end{figure*}

\section{Detection Efficiency for FSPL events}
\label{sec:eff_thetaE}
In section \ref{sec:efficiency}, we showed when finite source effects are important, the detection efficiency is
a function of two variables, which can be either ($t_{\rm E}$, $\rho$) or ($t_{\rm E}$, $\theta_{\rm E}$). 
In sub-section \ref{sec:eff_tE},
we discussed the conversion of the two dimensional detection efficiency into a one dimensional function depending only
on $t_{\rm E}$, and we showed that the finite source effects generally do not have a significant effect on the detection efficiency
in our sample for events with $t_{\rm E} > 1\,$day. This depends somewhat on the angular size of the source stars in the sample.
For a sample of microlensing events with giant star sources, finite source effects are likely to affect the detection efficiencies for
events with $t_{\rm E} \simgt 1\,$day, but microlensing events discovered by 
the Roman Space Telescope exoplanet microlensing survey, the source stars will have a smaller average angular size, 
so the detection efficiency for events with $t_{\rm E} \simlt 1\,$day should be less dependent on finite source effects
than our MOA survey sample is.
In this section, we consider the calculation of detection efficiencies for event selection criteria that include a requirement
that $\theta_{\rm E}$ be measured, so we add a $\theta_{\rm E}$ measurement  criterion to our cut CR2.  
Although \citetalias{sumi2023} do not use this detection efficiency for FSPL events, it may
be important for the future studies focusing on FSPL events. It is also important to establish calculation methods 
and to see how the detection efficiency for FSPL events depends on $t_{\rm E}$, $\theta_{\rm E}$, and event rate.

Our criteria to declare that $\theta_{\rm E}$ is measured is the same as the criteria we use to decide if the FSPL fit 
result should be adopted during the cut-3 process in our event selection process. That is, the $\chi^2$ of the FSPL model must be improved by at least 20 over the PSPL model for events with $0.8 < \mu_{\rm rel} / {\rm mas \, yr^{-1}} < 0.9$, and improved by at least 50 over the PSPL model for events with $0.9 < \mu_{\rm rel} / {\rm mas \, yr^{-1}} < 20$. Only the 13 FSPL events in Table \ref{tbl:candlistFS1} remained in our sample after this selection.

We denote the detection efficiency for FSPL events as a function of $(t_{\rm E}, \theta_{\rm E})$ by $\epsilon_{\rm FS} (t_{\rm E}, \theta_{\rm E})$, and this can be calculated by Eq. (\ref{eq:eps02D}) with the new $\theta_{\rm E}$ measurement criteria.
The integrated detection efficiencies for FSPL events as a function of either $t_{\rm E}$ or $\theta_{\rm E}$ are both dependent on the event rate $\Gamma (t_{\rm E}, \theta_{\rm E})$. These single parameter, event rate dependent, integrated detection efficiencies are denoted by $\tilde{\epsilon}_{\rm FS} (t_{\rm E} ; \Gamma)$ and $\tilde{\epsilon}_{\rm FS} (\theta_{\rm E} ; \Gamma)$, respectively. These are given by
\begin{align}
\tilde{\epsilon}_{\rm FS} (t_{\rm E} ; \Gamma) &=  \int_{\theta_{\rm E}} \Gamma (\theta_{\rm E} | t_{\rm E})  \epsilon_{\rm FS} (t_{\rm E}, \theta_{\rm E}) d\theta_{\rm E},  \label{eq:epsFS_tE}
\end{align}
and
\begin{align}
\tilde{\epsilon}_{\rm FS} (\theta_{\rm E} ; \Gamma) &=  \int_{t_{\rm E}} \Gamma (t_{\rm E} | \theta_{\rm E})  \epsilon_{\rm FS} (t_{\rm E}, \theta_{\rm E}) dt_{\rm E}, \label{eq:epsFS_thE}
\end{align}
respectively. Note that these event rate dependent detection efficiencies cannot easily be used in a likelihood analysis
to determine the FFP mass function, because the event rate $\Gamma$ depends on the mass function. We deal with
this issue in \citetalias{sumi2023} by evaluating $\tilde{\epsilon}_{\rm FS} (t_{\rm E} ; \Gamma)$ separately for each FFP mass function considered
in our likelihood analysis.

Figures \ref{fig:EFF_FS}(a), (b), and (c) show $\epsilon_{\rm FS} (t_{\rm E}, \theta_{\rm E})$, $\tilde{\epsilon}_{\rm FS} (t_{\rm E} ; \Gamma)$, and $\tilde{\epsilon}_{\rm FS} (\theta_{\rm E} ; \Gamma)$, respectively.
These are noisier than the original detection efficiency shown in Figure \ref{fig:EFF_tEthE} especially at long $t_{\rm E}$ because at longer $t_{\rm E}$, the average $\rho$ value is smaller and FSPL events become less common, and only a small fraction of
our simulated events can be used to calculate the detection efficiencies with the $\theta_{\rm E}$ measurement.

In each of panels (b) and (c), the orange solid curve shows the integrated detection efficiency calculated using the best-fit event rate $\Gamma (t_{\rm E}, \theta_{\rm E})$ taken from \citetalias{sumi2023} while the blue dashed curve shows that using a 
simple, but unrealistic, model of a constant event rate $\Gamma (t_{\rm E}, \theta_{\rm E}) = {\rm const.}$.
At $t_{\rm E} < 0.5$ days, the difference between the two curves in panel (b) is similar to the one for $\tilde{\epsilon} (t_{\rm E} ; \Gamma)$ in Figure \ref{fig:EFF_tEthE}.
However, $\tilde{\epsilon}_{\rm FS} (t_{\rm E} ; \Gamma_{\rm best})$ is significantly smaller than $\tilde{\epsilon}_{\rm FS} (t_{\rm E} ; \Gamma_{\rm flat})$ for $1 < t_{\rm E}/{\rm days} < 100$. This is because, for a given $t_{\rm E}$ in the range $1 < t_{\rm E}/{\rm days} < 100$, $\Gamma_{\rm best} (\theta_{\rm E} | t_{\rm E})$ is significantly higher at the upper half of the simulated range of $\theta_{\rm E}$ than at the lower half (see Figure \ref{fig:Gamma} in Appendix \ref{sec:app}), while the detection efficiency is smaller at larger $\theta_{\rm E}$ as shown in Figure \ref{fig:EFF_FS}(a).

The green line in Figure \ref{fig:EFF_FS}(c) shows the relative detection efficiency used by \citet{Gould2022} for an analysis of 
a sample of the KMTNet microlensing events with giant source stars. It is unclear how this assumed detection efficiency
was determined, as \citet{Gould2022} present no discussion of this. In fact, the event selection criteria used by
\citet{Gould2022} includes both automated and manual light curve fitting, which could make a proper calculation of 
$\epsilon_{\rm FS} (t_{\rm E}, \theta_{\rm E})$ quite difficult.

Because the \citet{Gould2022} selection criteria differ significantly from our selection criteria with the 
added $\theta_{\rm E}$ measurement criteria, we should not expect the detection efficiencies for our analysis to match
the true detection efficiency for the \citet{Gould2022} analysis or their adopted detection efficiency. 
\citet{Gould2022} do not discuss their procedure to develop the detection efficiency that they adopted. They also do not
mention the dependence of the detection efficiency on the event rate, but we can consider how the detection efficiency
of our event selection method with the $\theta_{\rm E}$ measurement requirement depends on the assumed event rate.

Figure \ref{fig:EFF_FS}(c) shows that the peak sensitivity region of our survey to $\theta_{\rm E}$ is 
smaller than the detection efficiency adopted by \citet{Gould2022} would predict.
This is qualitatively consistent with the fact that the KMTNet sample contains only events with giant source star, 
whereas our sample contains many turn-off and main-sequence source stars, which have a smaller angular size. This
allows our sample to detect events with smaller $\theta_{\rm E}$ values which would diminish the magnification and
render giant source star events undetectable. However, Figure~\ref{fig:CMDLDall} shows that just over half of our
FSPL sample have sources in the giant branch, in the vicinity of the red clump. So, we might expect the peaks
for the measured MOA and adopted KMTNet detection efficiencies to be less than an order of magnitude, but
perhaps this is because of giant stars that are larger than red clump stars in the KMTNet sample.

The relationship between the sensitivity to low-mass planets and the source star angular size may be better
understood with a comparison of the source star angular radii, $\theta_*$, and angular Einstein radii, $\theta_{\rm E}$,
for the 8 FFP candidates with $\theta_{\rm E} < 10\,\mu$as measurements found by microlensing. These include 4 candidates
\citep{Mroz18,Mroz19b,Mroz20b,Mroz20c}
found by OGLE, 2 candidates found by KMTNet \citep{Kim2021,Ryu2021},
and the two candidates presented here, MOA-9y-770 and MOA-9y-5919. These events have angular Einstein radii 
values in the range $1.26\,\mu{\rm as} \leq \theta_* \leq 15.1\,\mu{\rm as}$, but it is only the two events,
OGLE-2016-BLG-1928 and MOA-9y-5919, with $\theta_* < 4\,\mu{\rm as}$, that are terrestrial planet candidates
with $\theta_{\rm E} < 2\,\mu{\rm as}$. MOA-9y-5919 has the smallest angular size of any of these FFP candidates,
with $\theta_* = 1.26 \pm 0.48\,\mu{\rm as}$, which is less than half of the angular source size for OGLE-2016-BLG-1928 
($\theta_* = 2.85 \pm 0.20\,\mu{\rm as}$). The MOA-9y-5919 source is a G7 turn-off star, as shown the CMD in 
Figure \ref{fig:CMDLDall} indicates. These results imply that 
it is important to investigate turn-off and main-sequence source events to study FFP population down to terrestrial masses
and to measure the FFP mass function.

Figure \ref{fig:EFF_FS}(c) also indicates that $\tilde{\epsilon}_{\rm FS} (\theta_{\rm E} ; \Gamma)$ also depends on 
$\Gamma (t_{\rm E}, \theta_{\rm E})$, since the blue and orange lines differ because they assume different event rate
functions, $\Gamma$. This is most clearly seen in the bottom panel of Figure \ref{fig:EFF_FS}(c).
However, for our modified selection criteria that requires a $\theta_{\rm E} $ measurement, we find 
that the dependence of $\tilde{\epsilon}_{\rm FS} (\theta_{\rm E} ; \Gamma)$ on the mass function is 
smaller than that of $\tilde{\epsilon}_{\rm FS} (t_{\rm E} ; \Gamma)$, as shown in Appendix \ref{sec:app}.

The dashed orange histograms in panels (b) and (c) show $t_{\rm E}$ and $\theta_{\rm E}$ distributions, respectively, corrected by the detection efficiency with the best-fit event rate. The far left bin's value is an order of magnitude higher than the second left bin's value, which implies that terrestrial mass objects are about ten times more common than Neptune mass objects. This is consistent with the conclusion in \citetalias{sumi2023}, where we do a more detailed likelihood analysis using our sample and discuss the FFP population further.

\section{Discussion and conclusions}
\label{sec:discussionAndSummary}

We conducted a systematic offline analysis of a 9-year subset of the MOA-II survey towards the Galactic bulge.
We found 6,111 microlensing candidates in which 3,554 or 3,535 events have been 
selected with criteria sets CR1 or CR2, respectively, to be used in a statistical analysis.
Among these selected events, we found 12 very short ($t_{\rm E}<1$ day) events. 

Among the 12 short events, we found 2 FSPL events,
MOA-9y-770 ($t_{\rm E}=0.315\pm 0.017$ days), and
MOA-9y-5919 ($t_{\rm E}=0.057\pm 0.016$ days).
These events have very small angular Einstein radii of 
$\theta_{\rm E}= 4.73 \pm 0.75$\,$\mu$as and
$0.90 \pm 0.14$\,$\mu$as, respectively.
Our Bayesian analysis using a Galactic model and information of 
observed $t_{\rm E}$ and $\theta_{\rm E}$ infer their masses as 
 $22.3^{+42.2}_{-17.4}$ $M_\earth$ and
$0.75^{+1.23}_{-0.46}$ $M_\earth$, i.e., likely a Neptune mass and a terrestrial mass objects, respectively.

There were 7 known FFP candidates with $\theta_{\rm E}$ measurements and our discoveries increased the sample to 9 in total.
Among these 7 known FFP candidates, only OGLE-2016-BLG-1928L \citep{Mroz20b} has a terrestrial mass.
MOA-9y-5919L is the second terrestrial mass FFP candidate.
This discovery confirmed the existence of a terrestrial mass FFP population, and a simple comparison using a detection efficiency corrected histogram in Figure \ref{fig:EFF_FS} indicates that terrestrial mass objects like MOA-9y-5919L are about ten times more common than Neptune mass objects like MOA-9y-770. A more detailed analysis and discussion on the FFP population are presented in our companion paper, \citetalias{sumi2023}.

Compared with our detection rate of one terrestrial mass object out of two short FSPL events, a relatively small number of low mass FFP candidates have been found to date.
This is partly because FFPs have been mainly sought in events with giant or super-giant source stars.
Giant or super-giant source stars have an advantage to detect the finite source effect because of 
their large angular source radii. On the other hand, a large source radius tends to suppress 
the maximum event magnification.
It is important to search for FSPL events in sub-giant and dwarf source stars to detect events with small $\theta_{\rm E}$, i.e., low mass lenses.

We developed a new method for calculating the detection efficiency of the survey
by taking the influence of the finite source effect into account for the first time.
When finite source effects are important, as is generally the case for FFP analyses, then
the detection efficiency, $\epsilon (t_{\rm E}, \theta_{\rm E})$,
becomes a function of both $t_{\rm E}$ and $\theta_{\rm E}$. If one wishes to define a single variable detection
efficiency as a function of either $t_{\rm E}$ or $\theta_{\rm E}$ , then one must integrate over the other variable, 
yielding the integrated detection efficiency as a function of $t_{\rm E} $, $\tilde{\epsilon} (t_{\rm E} ; \Gamma)$,
or $\theta_{\rm E} $, $\tilde{\epsilon} (\theta_{\rm E} ; \Gamma)$. These expressions include the ``$; \Gamma$" term to
indicate that they depend on the event rate, $\Gamma (t_{\rm E}, \theta_{\rm E})$, because the distribution of $\theta_{\rm E}$ 
values for fixed $t_{\rm E}$ and the distribution of $t_{\rm E}$ values for fixed $\theta_{\rm E}$ both depend on the 
event rate.
If a single variable integrated detection efficiency function (either $\tilde{\epsilon} (t_{\rm E} ; \Gamma)$,
or $\tilde{\epsilon} (\theta_{\rm E} ; \Gamma)$) is used in an analysis of FFP mass function models,
then the dependence of the event rate on the FFP mass function requires that $\tilde{\epsilon} (t_{\rm E} ; \Gamma)$,
or $\tilde{\epsilon} (\theta_{\rm E} ; \Gamma)$
must be reevaluated for every mass function
that is considered. Alternatively, the more straightforward procedure of directly using the two-dimensional detection 
efficiency, $\epsilon (t_{\rm E}, \theta_{\rm E})$ can be used.
In the case of our final sample consisting of both PSPL and FSPL events, this method is important for constraining the low mass end of the mass function by modeling the
short tail at $t_{\rm E}<0.5$ days, where the finite source effect is more important, as is done in \citetalias{sumi2023}.
In the case of a sample consisting of only FSPL events, we show that the integrated detection efficiency depends on the event rate $\Gamma (t_{\rm E}, \theta_{\rm E})$ over a wide range of $t_{\rm E}$ or $\theta_{\rm E}$.
Our method will also be useful for the analysis of the survey by the Roman Space telescope, 
which expects to detect many more low mass FFP candidates \citep{ben02, joh20}.

\acknowledgments
NK was supported by the JSPS overseas research fellowship.
The MOA project is supported by JSPS KAKENHI Grant Number JP24253004, JP26247023, JP23340064, JP15H00781, JP16H06287, JP17H02871 and JP22H00153.
DPB acknowledges support from NASA grants 80NSSC20K0886 and 80NSSC18K0793.
The work of DPB, AB, SIS, and AV was supported by NASA under award number 80GSFC21M0002.

\appendix

\section{Dependence of integrated detection efficiency on the mass function} \label{sec:app}

\begin{figure*}
\begin{center}
\includegraphics[width=15cm]{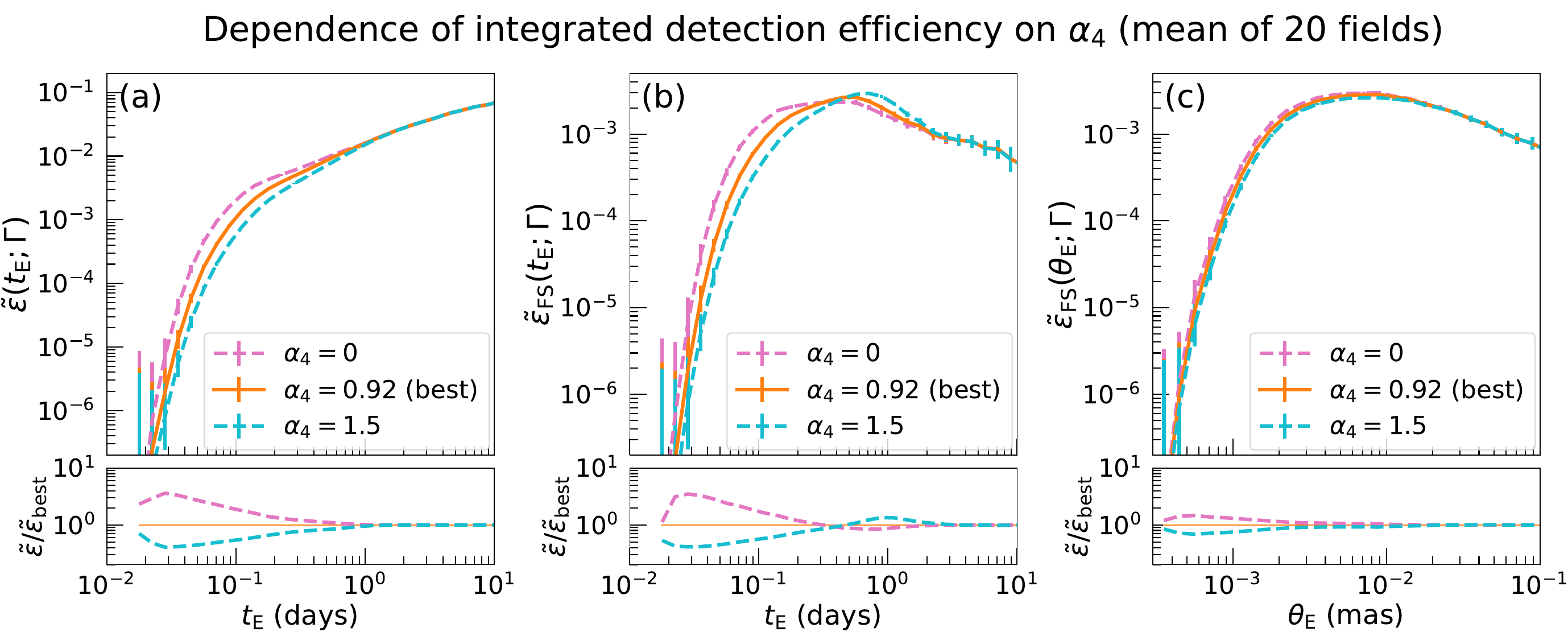}
\caption{
Dependence of integrated detection efficiency on the mass function for three different slopes in the planetary mass range, 
$\alpha_4 = 0,\ 0.92$, and 1.5 for different event selection criteria and different variables, $t_{\rm E}$ and $\theta_{\rm E}$.
Panels (a) and (b) show the detection efficiency as a function of $t_{\rm E}$ after integrating over $\theta_{\rm E}$ for our
original selection criteria CR2, which yields $\tilde{\epsilon} (t_{\rm E} ; \Gamma)$ in panel (a) and for our modified criteria, which
requires a $\theta_{\rm E}$ measurement, yielding $\tilde{\epsilon}_{\rm FS} (t_{\rm E} ; \Gamma)$ in panel (b). Panel (c) shows
$\tilde{\epsilon}_{\rm FS} (\theta_{\rm E} ; \Gamma)$, which is the integrated detection efficiency as a function of $\theta_{\rm E}$ using the
modified selection criteria that requires a $\theta_{\rm E}$ measurement.
In each panel, the orange solid, magenta dashed, and cyan dashed curves show integrated detection efficiencies with the event rate $\Gamma (t_{\rm E}, \theta_{\rm E})$ calculated with the mass functions with $\alpha_4 = 0.92$ (consistent with the best-fit in \citetalias{sumi2023}), $\alpha_4 = 0$, and $\alpha_4 = 1.5$.}
\end{center}
\label{fig:eff_alpha4}
\end{figure*}

\begin{figure*}
\begin{center}
\includegraphics[width=15cm]{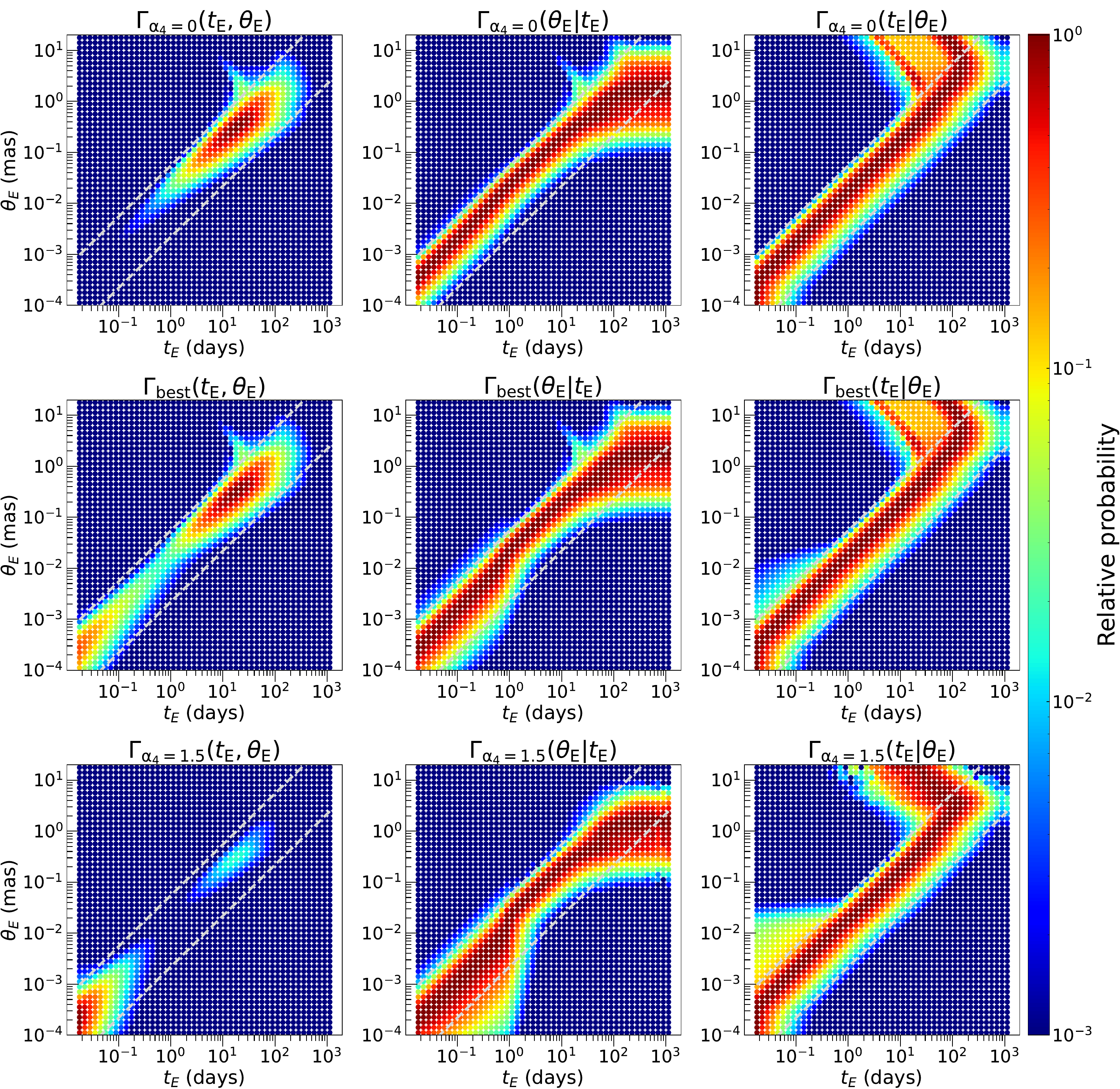}
\caption{
Relative probability distributions of event rate, $\Gamma (t_{\rm E}, \theta_{\rm E})$ (left), $\Gamma (\theta_{\rm E} | t_{\rm E})$ (middle), and $\Gamma (t_{\rm E} | \theta_{\rm E})$ (right), calculated using the Galactic model of \citet{Koshimoto2021} but with the modified broken power law given by Eq. (\ref{eq:MF}), where $\alpha_4 = 0$, $\alpha_4 = 0.92$ (best),  and $\alpha_4 = 1.5$ are used in the top, middle, and bottom panels, respectively.
Lower and upper dashed gray lines in each panel indicate $\mu_{\rm rel} = 0.8$ mas/yr and $\mu_{\rm rel} = 20$ mas/yr, respectively.
Note that while the distributions in long $t_{\rm E}$ and/or large $\theta_{\rm E}$ should be the same across all $\alpha_4$ values, they appear different because we used Monte Carlo simulations, which generated fewer high-mass lens events with larger $\alpha_4$ values.}
\end{center}
\label{fig:Gamma}
\end{figure*}

We argued when the finite source effect is considered, the detection efficiency depends on two variables, $(t_{\rm E}, \rho)$ or $(t_{\rm E}, \theta_{\rm E})$. When it is the case, a single variable detection efficiency, which is referred to as an integrated detection efficiency, as a function of either $t_{\rm E}$ or $\theta_{\rm E}$ depends on the event rate, $\Gamma (t_{\rm E}, \theta_{\rm E})$, which has a dependency on the mass function.
Figure \ref{fig:eff_alpha4} shows how much integrated detection efficiencies depend on the mass function slope in the planetary mass range, $\alpha_4$, which is defined in Eq. (\ref{eq:MF}). Panels (a), (b), and (c) show the integrated detection efficiencies for the CR2 sample as a function of $t_{\rm E}$, $\tilde{\epsilon} (t_{\rm E} ; \Gamma)$, for the FSPL sample as a function of $t_{\rm E}$, $\tilde{\epsilon}_{\rm FS} (t_{\rm E} ; \Gamma)$, and for the FSPL sample as a function of $\theta_{\rm E}$, $\tilde{\epsilon}_{\rm FS} (\theta_{\rm E} ; \Gamma)$, respectively. 
We use $\alpha_4 = 0$, $\alpha_4 = 0.92$, and $\alpha_4 = 1.5$ to plot the dashed magenta, solid orange, and dashed cyan lines in each panel, where $\alpha_4 = 0$ is a common assumption when we do not have any prior knowledge, $\alpha_4 = 0.92$ corresponds to the tentative best-fit model used for the Bayesian analysis in Section \ref{sec:shortFSPL}, and $\alpha_4 = 1.5$ is a possible value within the uncertainty given by \citetalias{sumi2023}.
The event rate, $\Gamma (t_{\rm E}, \theta_{\rm E})$, used to calculate these efficiency curves are shown in Figure \ref{fig:Gamma}.

Figure \ref{fig:eff_alpha4} shows that both $\tilde{\epsilon} (t_{\rm E} ; \Gamma)$ and $\tilde{\epsilon}_{\rm FS} (t_{\rm E} ; \Gamma)$ exhibit a similar trend with respect to variations in $\alpha_4$. Notably, a difference of up to an order of magnitude can be seen between the two efficiency curves at $t_{\rm E} \sim 0.03$ days when comparing the cases of $\alpha_4 = 0$ and $\alpha_4 = 1.5$. 
In contrast, $\tilde{\epsilon} (\theta_{\rm E} ; \Gamma)$ shows less dependence on the variation of $\alpha_4$, with a difference of up to a factor of 2.5 at $\theta_{\rm E} \sim 5 \times 10^{-4}$ mas, and almost no difference at $\theta_{\rm E} \simgt 3 \times 10^{-3}$ mas. 

To understand what causes these different dependencies on $\alpha_4$, we plot the $\Gamma (\theta_{\rm E} | t_{\rm E})$ and $\Gamma (t_{\rm E} | \theta_{\rm E})$ distributions in the middle and right rows in Figure \ref{fig:Gamma}.
These distributions show that the sensitivity to variations in $\alpha_4$ is greater for $\Gamma (\theta_{\rm E} | t_{\rm E})$ than for $\Gamma (t_{\rm E} | \theta_{\rm E})$. 
As $\alpha_4$ increases, the average lens mass decreases.
This results in smaller $\theta_{\rm E}$ values for a given $t_{\rm E}$, which is reflected in the figure as a higher value of $\Gamma (\theta_{\rm E} | t_{\rm E})$ at lower $\theta_{\rm E}$ values for larger $\alpha_4$.
On the other hand, when the average lens mass is lower, the average lens distance, $D_{\rm L}$, gets closer for a given $\theta_{\rm E}$.
This results in a change in the distribution of $\mu_{\rm rel}$ (or equivalently $t_{\rm E}$ for a given $\theta_{\rm E}$). 
However, as the $\mu_{\rm rel}$ distribution is relatively insensitive to changes in the lens distance $D_{\rm L}$ \citep[e.g., Fig. 2 of][]{zhu17}, the $\Gamma (t_{\rm E} | \theta_{\rm E})$ distribution does not change significantly with variations in $\alpha_4$.
An exception is the increase of events with extremely close $D_{\rm L}$, which leads to the appearance of very fast relative proper motion events in the $\mu_{\rm rel} > 20$ mas/yr range in the $\Gamma_{\rm best} (t_{\rm E} | \theta_{\rm E})$ and $\Gamma_{\rm \alpha_4 = 1.5} (t_{\rm E} | \theta_{\rm E})$ panels. However, such events are still rare even with a large $\alpha_4$ value, and do not significantly contribute to the overall distribution.

The dependence of the integrated detection efficiency as a function of $t_{\rm E}$ on $\alpha_4$ is determined by $\Gamma (\theta_{\rm E} | t_{\rm E})$ as shown in Eqs. (\ref{eq:eps_tE}) and (\ref{eq:epsFS_tE}), making $\tilde{\epsilon} (t_{\rm E} ; \Gamma)$ and $\tilde{\epsilon}_{\rm FS} (t_{\rm E} ; \Gamma)$ sensitive to variations in $\alpha_4$.
On the other hand, the dependence of the integrated detection efficiency as a function of $\theta_{\rm E}$ on $\alpha_4$ is determined by $\Gamma (t_{\rm E} | \theta_{\rm E})$ as shown in Eq. (\ref{eq:epsFS_thE}), making $\tilde{\epsilon}_{\rm FS} (\theta_{\rm E} ; \Gamma)$ less sensitive to variations in $\alpha_4$.

\begin{thebibliography}{}
\bibitem[Akeson et al.(2013)]{ake13} Akeson, R.~L., Chen, X., Ciardi, D., et al.\ 2013, \pasp, 125, 989. doi:10.1086/672273
\bibitem[Alard(2000)]{ala00} Alard, C.\ 2000, \aaps, 144, 363. doi:10.1051/aas:2000214
\bibitem[Alard \& Lupton(1998)]{ala98} Alard, C. \& Lupton, R.~H.\ 1998, \apj, 503, 325. doi:10.1086/305984
\bibitem[An et al.(2002)]{an02} An, J.~H., Albrow, M.~D., Beaulieu, J.-P., et al.\ 2002, \apj, 572, 521. doi:10.1086/340191
\bibitem[Bennett(2008)]{bennett_rev} Bennett, D.P, 2008, in Exoplanets, ed. J. Mason (Berlin: Springer), 47
\bibitem[Bennett(2010)]{bennett-himag} Bennett, D.P.\ 2010, \apj, 716, 1408
\bibitem[Bennett et al.(2008)]{ben08} Bennett, D.~P., Bond, I.~A., Udalski, A., et al.\ 2008, \apj, 684, 663. doi:10.1086/589940
\bibitem[Bennett \& Khavinson(2014)]{rhie_obit} Bennett, D.~P. \& Khavinson, D.\ 2014, Physics Today, 67, 64. doi:10.1063/PT.3.2318
\bibitem[Bennett \& Rhie(1996)]{bennett96}Bennett, D.P. \& Rhie, S.H.\ 1996, \apj, 472, 660
\bibitem[Bennett \& Rhie(2002)]{ben02} Bennett, D.~P. \& Rhie, S.~H.\ 2002, \apj, 574, 985. doi:10.1086/340977
\bibitem[Bennett et al.(2012)]{bennett2012} Bennett, D.~P., Sumi, T., Bond, I. A., et~al.\ 2012, \apj, 757, 119
\bibitem[Bensby et al.(2011)]{Bensby2011} Bensby, T.,  Ad\'{e}n, D., Mel\'{e}ndez, J., et al. 2011, A\&A, 533, A134
\bibitem[Bensby et al.(2013)]{Bensby2013} Bensby, T., Yee, J. C., Feltzing, S., et al. 2013, A\&A, 549, A147
\bibitem[Bond et al.(2001)]{bon01} Bond, I.~A., Abe, F., Dodd, R.~J., et al.\ 2001, \mnras, 327, 868. doi:10.1046/j.1365-8711.2001.04776.x
\bibitem[Boyajian et al.(2014)]{Boyajian2014}Boyajian, T. S., van Belle, G., \& von Braun, K. 2014, AJ, 147, 47
\bibitem[Bozza et al.(2018)]{Bozza2018} Bozza, V., Bachelet, E., Bartoli{\'c}, F., et al.\ 2018, \mnras, 479, 5157. doi:10.1093/mnras/sty1791
\bibitem[Claret \& Bloemen(2011)]{Claret2011} Claret, A. \& Bloemen, S., 2011, A\&A, 529, A75
\bibitem[Fukui et al.(2015)]{Fukui2015} Fukui, A., Gould, A., Sumi, T.,  et al. 2015, \apj, 809, 74
\bibitem[Gaudi(2012)]{gaudi_rev} Gaudi, B.~S.\ 2012, \araa, 50, 411 
\bibitem[Gould(1992)]{gou92}Gould, A. 1992, ApJ, 392, 442
\bibitem[Gould(1994a)]{gou94a}Gould, A. 1994a, ApJ, 421, L71 
\bibitem[Gould et al.(2022)]{Gould2022} Gould, A., Jung, Y.~K., Hwang, K.-H., et al.\ 2022, Journal of Korean Astronomical Society, 55, 173. doi:10.5303/JKAS.2022.55.5.173
\bibitem[Han \& Gould(1996)]{han96} Han, C. \& Gould, A.\ 1996, \apj, 467, 540. doi:10.1086/177631
\bibitem[Hirao et al.(2020)]{Hirao2020} Hirao, Y., Bennett, D.~P., Ryu, Y.-H., et al.\ 2020, \aj, 160, 74. doi:10.3847/1538-3881/ab9ac3
\bibitem[Holtzman et al.(1998)]{hol98} Holtzman, J.~A., Watson, A.~M., Baum, W.~A., et al.\ 1998, \aj, 115, 1946. doi:10.1086/300336
\bibitem[James(1994)]{James1994} James F., 1994, MINUIT reference Manual, https://root.cern.ch/download/minuit.pdf
\bibitem[Johnson et al.(2020)]{joh20} Johnson, S.~A., Penny, M., Gaudi, B.~S., et al.\ 2020, \aj, 160, 123. doi:10.3847/1538-3881/aba75b
\bibitem[Kim et al.(2010)]{kmtnet2010} Kim, S.-L., Park, B.-G., Lee, C.-U., et al.\ 2010, \procspie, 7733, 77333F
\bibitem[Kim et al.(2016)]{kim16} Kim, S.-L., Lee, C.-U., Park, B.-G., et al.\ 2016, Journal of Korean Astronomical Society, 49, 37. doi:10.5303/JKAS.2016.49.1.37
\bibitem[Kim et al.(2021)]{Kim2021} Kim, H.-W., Hwang, K.-H., Gould, A., et al.\ 2021, \aj, 162, 15
\bibitem[Kondo et al.(2019)]{Kondo2019} Kondo, I., Sumi, T., Bennett, D.~P., et al.\ 2019, \aj, 158, 224. doi:10.3847/1538-3881/ab4e9e
\bibitem[Koshimoto et al.(2021)]{Koshimoto2021} Koshimoto, N., Baba, J., \& Bennett, D.~P.\ 2021, \apj, 917, 78. doi:10.3847/1538-4357/ac07a8
\bibitem[Koshimoto et al.(2021b)]{kos21b} Koshimoto, N., Bennett, D.~P., Suzuki, D., et al.\ 2021b, \apjl, 918, L8. doi:10.3847/2041-8213/ac17ec
\bibitem[Koshimoto \& Ranc(2021)]{kosran21} Koshimoto, N. \& Ranc, C.\ 2021, Zenodo
\bibitem[Lam et al.(2022)]{lam22} Lam, C.~Y., Lu, J.~R., Udalski, A., et al.\ 2022, \apjl, 933, L23. doi:10.3847/2041-8213/ac7442
\bibitem[Mao \& Paczy\'{n}ski(1991)]{mao1991}Mao, S., \& Paczy\'{n}ski, B. 1991, ApJ, 374, L37
\bibitem[Metropolis et al.(1953)]{metrop} Metropolis, N., Rosenbluth, A.~W., Rosenbluth, M.~N., Teller, A.~H., \& Teller, E.\ 1953, \jcp, 21, 1087
\bibitem[Mr\'{o}z et al.(2017)]{Mroz17} Mr\'{o}z, P. Udalski A., Skowron, J.,  et al. 2017, Nature, 548, 183
\bibitem[Mr\'{o}z et al.(2018)]{Mroz18} Mr\'{o}z, P. Y.-H.Ryu, Skowron, J.,  et al. 2018, \apj, 155, 121
\bibitem[Mr{\'o}z et al.(2019)]{Mroz19b} Mr{\'o}z, P., Udalski, A., Bennett, D.~P., et al.\ 2019, \aap, 622, A201. doi:10.1051/0004-6361/201834557
\bibitem[Mr\'{o}z et al.(2020a)]{Mroz20b} Mr\'{o}z, P., Poleski, R. \& Gould, A. 2020, \apj, 903, 11
\bibitem[Mr\'{o}z et al.(2020b)]{Mroz20c} Mr\'{o}z, P., Poleski, R., Han, C. 2020, \aj, 159, 262
\bibitem[Mr{\'o}z et al.(2022)]{mro22} Mr{\'o}z, P., Udalski, A., \& Gould, A.\ 2022, \apjl, 937, L24. doi:10.3847/2041-8213/ac90bb
\bibitem[Nataf et al.(2016)]{Nataf2016}Nataf, D. M., et al. 2016, \mnras, 456, 2692
\bibitem[Nemiroff \& Wickramasinghe(1994)]{nem94} Nemiroff, R. J. \& Wickramasinghe, W. A. D. T. 1994, ApJ, 424, 21
\bibitem[Niikura et al.(2019a)]{Niikura2019a} Niikura, H., Takada, M., Yasuda, N., et al.\ 2019, Nature Astronomy, 3, 524. doi:10.1038/s41550-019-0723-1
\bibitem[Niikura et al.(2019b)]{Niikura2019b} Niikura, H., Takada, M., Yokoyama, S., et al.\ 2019, \prd, 99, 083503. doi:10.1103/PhysRevD.99.083503
\bibitem[Paczy\'{n}ski(1986)]{pac86}Paczy\'{n}ski, B.  1986, ApJ, 304, 1 
\bibitem[Ryu et al.(2021)]{Ryu2021} Ryu, Y.-H., Mr{\'o}z, P., Gould, A., et al.\ 2021, \aj, 161, 126. doi:10.3847/1538-3881/abd55f
\bibitem[Sahu et al.(2022)]{sah22} Sahu, K.~C., Anderson, J., Casertano, S., et al.\ 2022, \apj, 933, 83. doi:10.3847/1538-4357/ac739e
\bibitem[Sako et al.(2008)]{sako2008}Sako, T., et al. 2008, Experimental Astronomy, 22, 51
\bibitem[Schechter, Mateo \& Saha(1993)]{sch93}Schechter, L., Mateo, M., \& Saha, A., 1993, PASP, 105, 1342S
\bibitem[Smith et al.(2002)]{smi02} Smith, M.~C., Mao, S., \& Wo{\'z}niak, P.\ 2002, \mnras, 332, 962. doi:10.1046/j.1365-8711.2002.05427.x
\bibitem[Stetson(1987)]{daophot}Stetson, P.~B. DAOPHOT - A computer program for crowded-field stellar photometry. \pasp {\ \bf 99}, 191-222 (1987)
\bibitem[Sumi et al.(2003)]{sumi03} Sumi, T., Abe, F., Bond, I.~A., et al.\ 2003, \apj, 591, 204. doi:10.1086/375212
\bibitem[Sumi et al.(2010)]{sumi2010} Sumi, T., Bennett, D.~P., Bond, I.~A., et al.\ 2010, \apj, 710, 1641. doi:10.1088/0004-637X/710/2/1641
\bibitem[Sumi et al.(2011)]{sumi2011} Sumi, T., Kamiya, K., Bennett, D.~P., et al.\ 2011, \nat, 473, 349. doi:10.1038/nature10092
\bibitem[Sumi et al.(2013)]{sumi2013} Sumi, T., Bennett, D.~P., Bond, I.~A., et al.\ 2013, \apj, 778, 150. doi:10.1088/0004-637X/778/2/150
\bibitem[Sumi et al.(2016)]{sumi2016a} Sumi, T., Udalski, A., Bennett, D.~P., et al.\ 2016, \apj, 825, 112. doi:10.3847/0004-637X/825/2/112
\bibitem[Sumi et al. (2023)]{sumi2023} Sumi, T., Koshimoto, N., Bennett, D.~P., et al.\ 2023, submitted (S23)
\bibitem[Suzuki et al.(2016)]{suzuki2016} Suzuki, D., Bennett, D.~P., Sumi, T., et al.\ 2016, \apj, 833, 145. doi:10.3847/1538-4357/833/2/145
\bibitem[Szyma{\'n}ski et al.(2011)]{uda11} Szyma{\'n}ski, M.~K., Udalski, A., Soszy{\'n}ski, I., et al.\ 2011, \actaa, 61, 83
\bibitem[Tomaney \& Crotts(1996)]{tom96} Tomaney, A.~B. \& Crotts, A.~P.~S.\ 1996, \aj, 112, 2872. doi:10.1086/118228
\bibitem[Udalski et al.(1994)]{uda94}Udalski, A. et al. 1994, Acta Astronomica, 44, 165
\bibitem[Udalski et al.(2015)]{uda15} Udalski, A., Szyma{\'n}ski, M.~K., \& Szyma{\'n}ski, G.\ 2015, \actaa, 65, 1
\bibitem[Wegg et al.(2017)]{weg17} Wegg, C., Gerhard, O., \& Portail, M.\ 2017, \apjl, 843, L5. doi:10.3847/2041-8213/aa794e
\bibitem[Witt \& Mao(1994)]{wit94}Witt, H. J. \& Mao, S. 1994, ApJ, 430, 505
\bibitem[Wo\'{z}niak(2000)]{wozniak2000} Wo\'{z}niak P. R., 2000, Acta Astronomica, 50, 421
\bibitem[Jung et al.(2020)]{YounKil2020} Jung, Y.~K., Gould, A., Udalski, A., et al.\ 2020, \aj, 160, 148. doi:10.3847/1538-3881/abacc8
\bibitem[Zhu et al.(2017)]{zhu17} Zhu, W., Udalski, A., Novati, S.~C., et al.\ 2017, \aj, 154, 210. doi:10.3847/1538-3881/aa8ef1
\end{thebibliography}
\end{document}